\documentclass[11pt, draftclsnofoot, peerreview, onecolumn]{IEEEtran}

\usepackage{bbm}
\usepackage{dsfont}
\usepackage{lineno}
\usepackage{amssymb}
\usepackage{amsmath}
\usepackage{amsthm}
\usepackage{epsfig}
\usepackage{graphicx}
\usepackage{graphics}
\usepackage{float}
\usepackage{subfigure}
\usepackage{multirow}
\usepackage{color}
\usepackage{tcolorbox}
\usepackage{makeidx}
\usepackage{xspace}
\usepackage{wrapfig}
\usepackage{lipsum}
\usepackage{mathtools,amsmath}
\usepackage{cuted}
\usepackage{cite}
\usepackage{amsfonts}
\usepackage{textcomp}
\usepackage{enumerate}
\usepackage{bm}
\newcommand{\diag}{\mathop{\mathrm{diag}}}
\usepackage{textcomp}
\usepackage{enumerate}
\usepackage{algpseudocode}
\usepackage{algorithm}
\usepackage{mathtools}
 \usepackage{relsize}
 \usepackage{url}
 \usepackage{perpage}

\DeclarePairedDelimiter{\ceil}{\lceil}{\rceil}

\makeindex
\theoremstyle{plain}
\newtheorem{Theorem}{Theorem}

\theoremstyle{definition}
\newtheorem{Definition}{Definition}
\newtheorem{Corollary}{Corollary}
\newtheorem{Lemma}{Lemma}

\newtheorem{remark}{Remark}
\newtheorem{Proof of Lemma}{Proof of Lemma}
\theoremstyle{remark}
\newtheorem{example}{Example}

\makeatletter

\renewcommand\@endtheorem{\vvv@endmarker\endtrivlist\@endpefalse}
\newcommand\vvv@endmarker{%
  {\unskip\nobreak\hfil\penalty50
  \hskip2em\vadjust{}\nobreak\hfil\openbox
  \parfillskip=0pt \finalhyphendemerits=0 \par
  \penalty 10000 \parskip=0pt\noindent}\ignorespaces}
\makeatother

\def \mod {\text{{\normalfont mod}}}

\def \supp {\text{supp}}
\def \Supp {\text{Supp}}

\def \T {\mathcal{T}}

\def \I {\mathcal{I}}
\def \J {\mathcal{J}}

\def \one \mathbf{1}

\def \min {\mathrm{min}}
\def \rank {\mathrm{rank}}
\def \minimize{\mathrm{minimize}}
\DeclareMathOperator{\tr}{tr}
\MakePerPage{footnote}
\allowdisplaybreaks

\def\BibTeX{{\rm B\kern-.05em{\sc i\kern-.025em b}\kern-.08em
    T\kern-.1667em\lower.7ex\hbox{E}\kern-.125emX}}

\cleardoublepage
\begin{document}
  
 \title{ Tessellated Distributed  Computing
      \thanks{%
     This work was supported by the European Research Council (ERC)
through the EU Horizon 2020 Research and Innovation Program under Grant
725929 (Project DUALITY), the ERC-PoC project LIGHT (Grant 101101031), as well as by the Huawei France-funded Chair towards
Future Wireless Networks. The authors are with the Communication Systems Department at
EURECOM, 450 Route des Chappes, 06410 Sophia Antipolis, France (email:
khalesi@eurecom.fr; elia@eurecom.fr).
}}
    \author{\IEEEauthorblockN{Ali Khalesi and Petros Elia}}
    \maketitle
\begin{abstract}
The work considers the $N$-server distributed computing scenario with $K$ users requesting functions that are linearly-decomposable over an arbitrary basis of $L$ real (potentially non-linear) subfunctions. In our problem, the aim is for each user to receive their function outputs, allowing for reduced reconstruction error (distortion) $\epsilon$, reduced computing cost ($\gamma$; the fraction of subfunctions each server must compute), and reduced communication cost ($\delta$; the fraction of users each server must connect to).
For any given set of $K$ requested functions --- which is here represented by a coefficient matrix $\mathbf {F} \in \mathbb{R}^{K \times L}$ --- our problem is made equivalent to the open problem of sparse matrix factorization that seeks --- for a given parameter $T$, representing the number of shots for each server --- to minimize the reconstruction distortion $\frac{1}{KL}\|\mathbf {F} - \mathbf{D}\mathbf{E}\|^2_{F}$ overall $\delta$-sparse and $\gamma$-sparse  matrices $\mathbf{D}\in \mathbb{R}^{K \times NT}$ and  $\mathbf{E} \in \mathbb{R}^{NT  \times L}$. With these matrices respectively defining which servers compute each subfunction, and which users connect to each server, we here design our $\mathbf{D},\mathbf{E}$ by designing tessellated-based and SVD-based fixed support matrix factorization methods that first split $\mathbf{F}$ into properly sized and carefully positioned submatrices, which we then approximate and then decompose into properly designed submatrices of $\mathbf{D}$ and $\mathbf{E}$. For the zero-error case and under basic dimensionality assumptions, the work reveals achievable computation-vs-communication corner points $(\gamma,\delta)$ which, for various cases, are proven optimal over a large class of $\mathbf{D},\mathbf{E}$ by means of a novel 
tessellations-based converse. Subsequently, for large $N$, and under basic statistical assumptions on $\mathbf{F}$, the average achievable error $\epsilon$ is concisely expressed using the incomplete first moment of the standard Marchenko-Pastur distribution, where this performance is shown to be optimal over a large class of $\mathbf{D}$ and $\mathbf{E}$. In the end, the work also reveals that the overall achieved gains over baseline methods are unbounded.
\end{abstract}

\section{introduction}
We are currently witnessing a growing need for novel parallel computing techniques that efficiently offload computations across multiple distributed computing servers\cite{dean2008mapreduce,zaharia2010spark}. To address this urgent need, a plethora of works has proposed novel methods that address various elements of distributed computing, such as scalability~\cite{li2017scalable,haddadpour2019trading,yang2020coded,charalambides2021numerically,soleymani2021analog,Brunero1,Parrinello1,Reza1,Brunero2}, privacy and security~\cite{Nezhad2023,soleymani2020distributed,khalesi2021capacity,soleymani2020privacy,soleymani2021list,bitar2022adaptive,yang2021coded,yu2020coded,ehteram2023trainedmpc,AVEST1}, as well as latency and straggler mitigation~\cite{raviv2020gradient,lee2017speeding,egger2022efficient,kai1,yu2020straggler,yu2017polynomial,jia2021cross}, to mention just a few. For a detailed survey of such related works, the reader is referred to~\cite{ng2020survey,CIT-103,Parrinello5}. \nocite{Brunero1,Brunero2,Lampiris}
In addition to the above elements, the celebrated computation-vs-communication relationship stands at the very core of distributed computing as a fundamental principle with profound ramifications. This principle appears as a limiting factor in a variety of distributed computing scenarios~\cite{verbraeken2020survey,ulukus2022private,wang2018fundamental,li2017fundamental,wan2022cache,yu2017polynomial2,dutta2019optimal,reisizadeh2021codedreduce,woolsey2021new,woolsey2021coded,woolsey2021practical,egger2022efficient,chen2021distributed} where indeed communication and computation are often the two intertwined bottlenecks that most heavily define the overall performance.

Another important factor pertains to computational accuracy and the ability to recover desired functions with reduced error or distortion. There is indeed a variety of techniques dedicated to increasing accuracy (cf.~\cite{wang2021price,ozfatura2021coded,Malak,woodruff2014sketching,jahani2021codedsketch,RaviApproximated,CharalambidesApproximated,RamchandranApproximated,StarkApproximated,ZhuRamchandranApproximated,TayyebehBerrutMaddah-Ali,CadambeApproximated,KianiStarkApproximated,NarayananKrishna,RashmiApproximated,ZhuJinggeApproximatedLearning,JahaniNezhadApproximated,MaddahAliApproximated}), such as for instance the \emph{sketching technique} \cite{woodruff2014sketching,MaddahAliApproximated,RamchandranApproximated} which can for example utilize a randomized linear algebraic approach to compute an approximation of the multiplication of two massive matrices (often by approximating input matrices by multiplying them with a random matrix having certain properties). Another related approach can be traced in successive approximation coding techniques (cf.~\cite{KianiStarkApproximated}) which make it possible to tradeoff accuracy and speed, allowing for better approximations and increased accuracy over time. 

This triptych between accuracy, communication costs and computation costs, lies at the center of distributed computing. We here explore this triptych for a pertinent setting of multi-user distributed computing.

\subsection{Multi-User Linearly-Decomposable Distributed Computing}\label{System-Model} 
We focus on the very broad and arguably practical setting of~\emph{multi-user distributed computation of linearly-decomposable real functions}, which captures several classes of computing problems that include distributed gradient coding problems~\cite{tandon2017gradient, ye2018communication,raviv2020gradient,halbawi2018improving}, the distributed linear-transform computation problem~\cite{dutta2016short,wang2018fundamental}, the distributed matrix multiplication or
the distributed multivariate polynomial computation problems~\cite{ramamoorthy2019universally,das2019distributed,haddadpour2018codes,lee2017speeding,yu2017polynomial,wang2018coded,yu2020straggler,dutta2019optimal,ramamoorthy2020straggler,jia2021cross}, as well as a plethora of distributed computing problems of training large-scale machine learning algorithms and deep neural networks with massive data~\cite{verbraeken2020survey}. These constitute a broad collection of problems where both computation and communication costs are crucial~\cite{zinkevich2010parallelized,chilimbi2014project}.

\begin{figure}
      \centering
\includegraphics[scale=0.8]{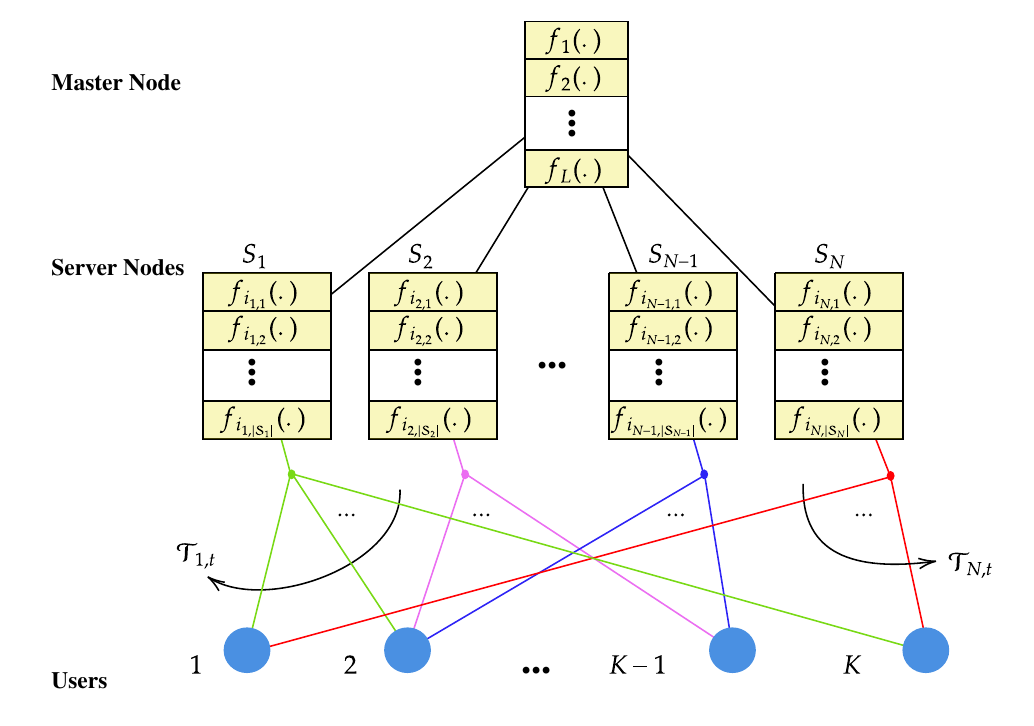}
      \caption{The $K$-user, $N$-server, $T$-shot setting. Each server $n$ computes the subfunctions in $\mathcal{S}_n=\{f_{i_{n,1}}(.),f_{i_{n,2}}(.),\hdots , f_{i_{n,|\mathcal{S}_n|}}(.)\}$ and communicates to users in $\mathcal{T}_{n,t}$, under computational constraint $|\mathcal{S}_n|\leq \Gamma\leq L$ and communication constraint $|\mathcal{T}_n|\leq \Delta \leq K$, yielding a system with normalized constraints $\gamma = \frac{\Gamma}{L}, \delta = \frac{\Delta}{K}$ and with an error constraint $\epsilon = \frac{\mathcal{E}}{KL}$, where $\gamma,\delta, \epsilon\in[0,1]$.
      }
      \label{Fig: System Model}
  \end{figure}

Our setting, as depicted in~Fig.~\ref{Fig: System Model}, initially considers a \emph{master node} that coordinates, in three phases, a set of $N$ distributed \emph{servers} that compute functions requested by the $K$ \emph{users}. During the initial \emph{demand phase}, each user $k \in \{1,2,\hdots, K\}$ independently requests the computed output of a single real function $F_{k}(.)$. Under the real-valued\footnote{The real-valued exposition entails a variety of advantages over finite-field approaches~\cite{kai1,kai2,kai3,khalesi4}, such as accuracy advantages stemming from using real-valued fixed point data representations, as well as advantages regarding computation overflows, quantization errors and scalability barriers~\cite{soleymani2020privacy,makkonen2022analog,soleymani2021analog}.} linear decomposability assumption\footnote{This naturally incorporates linearly separable functions (see for example~\cite{kai1}) where each $F_{k}(.)$, taking $L$ subfunctions as input, can be written as a linear combination of $L$ \emph{univariate} subfunctions. In our work, these subfunctions need not be univariate.}, these functions take the basic form
\begin{align} \label{linearlySep1}
    F_{k}(.) = \sum^{L}_{\ell=1} f_{k,\ell} f_{\ell}(.) = \sum^{L}_{\ell=1} f_{k,\ell} W_\ell
\end{align}
where $f_\ell(\cdot)$ denotes a \emph{(basis or component) subfunction}, where $f_{k,\ell}$ denotes a real-valued basis coefficient, and where $W_\ell = f_\ell(x), x \in \mathcal{D},$ denotes the real-valued \emph{output file} of $f_\ell(\cdot)$ for a multi-variate input $x$ from any domain set $\mathcal{D}$. 

Subsequently, during the \emph{computing phase}, the master assigns to each server $n \in [N]$, a set of subfunctions $\mathcal{S}_{n} \subseteq [L]$ 

to compute locally\footnote{We will interchangeably use $\mathcal{S}_{n}$ to describe sets of indices of subfunctions, as well as the subfunctions themselves.} in order to generate the corresponding
$W_\ell$, and then during the \emph{communication phase}, each server $n$ forms signals
\begin{align}
  z_{n,t}\triangleq \sum_{\ell \in [L]} e_{n,\ell,t} W_\ell,\:\: n\in [N],\: t \in [T] \label{EncodedFiles}
\end{align}
as dictated by the \emph{encoding coefficients} $e_{n,\ell,t} \in \mathbb{R}, n\in [N], t\in [T],\ell\in [L]$, and subsequently proceeds to transmit $z_{n,t}$ during time-slot $t=1,2,\dots,T$, to a subset of users $\T_{n,t} \subseteq [K]$, via a dedicated error-free broadcast channel. Finally, during the decoding part of the last phase, each user $k$ linearly combines its received signals to get
\begin{align}
    F'_{k} \triangleq \sum_{n \in [N],t\in[T]} d_{k,n,t} z_{n,t} \label{DecedFiles}
\end{align}
as dictated by the \emph{decoding coefficients} $d_{k,n,t} \in \mathbb{R}, n\in [N], t\in [T], k\in [K]$. Naturally $d_{k,n,t} =0,\forall k \notin \mathcal{T}_{n,t},$ simply because user $k \in [K]$ does not receive any symbol from server $n \in [N]$ during time $t \in [T]$. Note that both encoding and decoding coefficients are determined by the master node after the demand phase, and thus are dependent on the functions requested, but remain independent of the instance of the \emph{input} to the requested functions.

In case of allowable error, we consider 
 \begin{align}
     \mathcal{E} = \sum^{K}_{k=1}{|F'_k - F_k|}^2, \: \mathcal{E} \in \mathbb{R},\:\: \: \forall k \in [K] \label{Decoding-Criteria}
 \end{align}
to be the Euclidean distortion during function retrieval. Furthermore, for $\mathcal{T}_{n} = \cup^{T}_{t=1} \mathcal{T}_{n,t}$, we consider the computation and communication costs
\begin{align}  \label{eq:GammaDelta}
    \Gamma \triangleq  \max_{n \in [N]} |\mathcal{S}_{n}|, \ \ \Delta \triangleq \max_{n  \in  [N]} |\mathcal{T}_{n}|  
\end{align}
respectively representing the maximum number of subfunctions to be locally computed at any server\footnote{Our focus directly on the cost of computing the component subfunctions $f_\ell(\cdot)$ stems from the point of view that these subfunctions typically capture computationally intensive (and generally non-linear) tasks which would dominate, in terms of load, the remaining easier linear manipulations at servers and users during encoding and decoding. We also note that our constraints of $\Gamma$ and $\Delta$ are strict constraints in the sense that they must hold for any instance of the problem.}, and the maximum number of users that a server can communicate to. After normalization, we here consider the normalized costs
     \begin{align}\label{Nonasymptotic-constatnts}
       \epsilon \triangleq \frac{\mathcal{E}}{KL}, \ \ \ \ \gamma \triangleq \frac{\Gamma}{L}, \ \ \ \ \delta \triangleq \frac{\Delta}{K}.
     \end{align}
The three parameters are  bounded\footnote{In brief, $\gamma$ is the fraction of subfunctions that must be computed locally, and $\delta$ is the fraction of available links to be activated. Having $\gamma=1$ corresponds to the centralized scenario of having to locally calculate all subfunctions, while $\delta =1$ matches an extreme parallelized  scenario that activates all available communication links. We will discuss later in Section \ref{AsymptoticResults} the justification of having $\epsilon \leq 1$.}  between $0$ and $1$.

Another two parameters of interest are 
\begin{align} \label{eq:zetaKappa}    
\zeta \triangleq \frac{\Delta}{L}, \ \ \ \kappa \triangleq \frac{K}{L}
\end{align}
 where $\zeta$ normalizes the number of activated communication links by the number of subfunctions, while $\kappa $ reflects the dimensionality (`fatness') of matrix $\mathbf{F}$ and thus will define the statistical behavior of the singular values of $\mathbf{F}$ which will be crucial in the lossy case where we allow for function reconstruction error.

In a system defined by $K,N$ and $L$, our goal is to find schemes that can recover any set of desired functions, with the best possible decoding refinement $\epsilon$, and the smallest possible computation and communication loads $\gamma, \delta$. To do so, we must carefully decide which subfunctions each server computes, and which combinations of computed outputs each server sends to which users.  
Having to serve many users with fewer servers naturally places a burden on the system (suggesting higher $\gamma,\delta, \epsilon$), bringing to the fore the concept of the \emph{system rate} 
\begin{align}
    R \triangleq \frac{K}{N}\label{Rate1}
\end{align}
and the corresponding \emph{system capacity} $C$ representing the supremum of all rates.

\subsection{Connection to Sparse Matrix Factorization, and Related Works}
Toward analysing our distributed computing problem, we can see from~\eqref{linearlySep1} that the desired functions are fully represented by a matrix $\mathbf{F}\in \mathbb{R}^{K \times L}$ of the aforementioned coefficients $f_{k,\ell}$. With $\mathbf{F}$ in place, we must decide on the computation-assignment and communication (encoding and decoding) protocol. As we have seen in~\cite{khalesi4,khalesi3}, for the error-free case of $\epsilon = 0$, this task is equivalent --- directly from \eqref{EncodedFiles},\eqref{DecedFiles} and \eqref{Decoding-Criteria} --- to solving a (sparse) matrix factorization problem of the form 
\begin{equation}
    \mathbf{D}\mathbf{E}=\mathbf{F} \label{eq:DEF1}
\end{equation} where, as we will specify later on, the $NT\times L$ \emph{computing matrix} $\mathbf{E}$ holds the coefficients $e_{n,\ell,t}$ from~\eqref{EncodedFiles}, while the $K\times NT$ \emph{communication matrix} $\mathbf{D}$ holds the decoding coefficients $d_{k,n,t}$ from~\eqref{DecedFiles}.
As one can suspect, a sparser $\mathbf{E}$ reflects a lower $\gamma$ and a sparser $\mathbf{D}$ a lower $\delta$, at the cost though of a potentially higher $\epsilon$. 
Focusing on functions over finite fields, the work in~\cite{khalesi4}, after making the connection between distributed computing and the above factorization problem, employed from coding theory the class of covering codes and a new class of \emph{partial covering codes}, in order to derive bounds on the optimal communication and computation costs for the \emph{error-free case}. In brief, by choosing $\mathbf{D}$ to be the sparse parity-check matrix of a (shown to exist) sparse partial covering code, each column of $\mathbf{E}$ was subsequently produced to be the coset leader from syndrome decoding (with the syndrome being) the corresponding column\footnote{Thus for example, the first column of $\mathbf{E}$ is the coset leader to the coset corresponding to the syndrome described by the first column of $\mathbf{F}$ and by the code whose parity check matrix is $\mathbf{D}$.} of $\mathbf{F}$. This allowed for reduced communication and computation costs, where for example in the single shot scenario with a $q$-ary finite field, the (somewhat different from here) normalized optimal computation cost $\gamma\in (0,1]$ was bounded as a function of the $q$-ary entropy function $H_q$ to be in the range $\gamma \in [H_q^{-1}(\frac{\log_q(L)}{N}), H_q^{-1}(\frac{K}{N})]$.

A first exposition of the \emph{real-valued} variant of our computing problem, again for the error-free case of $\epsilon = 0$, can be found in~\cite{khalesi5}, which reformulated the equivalent sparse matrix factorization problem $\mathbf{D}\mathbf{E}=\mathbf{F}$ into the well-known compressed sensing problem $\mathbf{A}\mathbf{x}=\mathbf{y}$
which seeks to efficiently identify unique sparse solutions to an under-determined system of equations\footnote{This reformulation identifies the observed vector $\mathbf{y}$ with the vectorized $\mathbf{F}$, the sparse solution $\mathbf{x}$ with the vectorized $\mathbf{E}$, and the alphabet matrix $\mathbf{A}$ with the Kronecker product of the communication matrix $\mathbf{D}$ with an identity matrix.}. This reformulation allowed for conditional bounds on $\gamma$ of the form $\gamma \leq -\frac{1}{r} \frac{K}{N} W^{-1}_{1}(\frac{-2K}{erN})$, where though these bounds\footnote{Here $W_1(.)$ is the first branch of the Lambert function, while $r$ calibrates the statistical distribution of $\mathbf{D}$.} remained loose and  conditional, for two main reasons. The first reason stems from the fact that the focus of the compressed sensing machinery is mainly on the search efficiency and uniqueness of the sparse solutions, rather than on the level of sparsity itself\footnote{Search efficiency and uniqueness are not fundamental to our distributed computing problem. For example, what a compressed sensing exposition of our problem effectively shows is that, under the assumption that the sparsest solution for $\mathbf{E}$ has sparsity-level not more than the above $\gamma = -\frac{1}{r} \frac{K}{N} W^{-1}_{1}(\frac{-2K}{erN})$, and under the additional assumption that this solution is unique, then --- with high probability, in the limit of large $N$ --- there is an $l_1$-minimization approach that will efficiently find this sparsest unique solution. For us, the efficiency of identifying $\mathbf{E}$ is of secondary importance, and the possibility of having another equally sparse $\mathbf{E}$ is not an issue.}. The second reason is that, while in our computing setting our communication matrix $\mathbf{D}$ must be a function of $\mathbf{F}$, compressed sensing places its focus on designing $\mathbf{A}$ in a manner that is oblivious to the instance of $\mathbf{y}$ (which corresponds to our $\mathbf{F}$). These mismatches are part of what our work here addresses, allowing us to directly explore the fundamental principles of our computing problem.

\subsection{New Connection Between Distributed Computing, Fixed Support Matrix Factorization, and Tessellations}   

As we will see almost directly from~\eqref{EncodedFiles}, \eqref{DecedFiles} and \eqref{Decoding-Criteria},~\eqref{eq:DEF1}, and also from Lemma~\ref{Lemma-Epsilon-normalization}, solving our distributed computing problem will be equivalent to solving the approximate matrix factorization problem
\begin{align}
\hat{\mathcal{E}}=\underset{\mathbf{D,E}}{\min}&\:\: \| \mathbf{D}\mathbf{E} - \mathbf{F} \|^2_{F} \label{eq:minDe-F2} 
\end{align}
under dimensionality constraints posed by $K,NT, L$, and under sparsity constraints on $\mathbf{D}$ and $\mathbf{E}$ posed by $\delta$ and $\gamma$ respectively. These sparsity constraints will be described in detail later on.

This problem encompasses the problem of compressed sensing, and it is known to be NP hard~\cite{gribonval2010dictionary}. In general, finding the optimal solution $(\hat{\mathbf{D}},\hat{\mathbf{E}}) = \underset{\mathbf{D,E}}{\arg\min} \:\: \| \mathbf{D}\mathbf{E} - \mathbf{F} \|^2_{F}$ to \eqref{eq:minDe-F2}, under the aforementioned dimensionality and sparsity constraints, requires an infeasible coverage of the entire space of solutions. Otherwise, establishing optimality of an algorithmic solution, generally requires establishing uniqueness of that solution, which is challenging \cite{zheng2021identifiability, zheng2023efficient}. Furthermore, to date, little is known in terms of clear guarantees on the optimal error performance $\hat{\mathcal{E}}$, for any given $\mathbf{F}$ and any given dimensionality and sparsity constraints on $\mathbf{D,E}$.

Recently, the work in~\cite{le2023spurious} explored the problem of \emph{Fixed Support (sparse) Matrix Factorization} (FSMF), which --- under the equivalent dimensionality and sparsity constraints of the unbounded problem of \eqref{eq:minDe-F2} --- seeks to find
\begin{align}
\hat{\mathcal{E}}_{\mathcal{I},\mathcal{J}}=\underset{\mathbf{D,E}}{\min}&\:\: \| \mathbf{D}\mathbf{E} - \mathbf{F} \|^2_{F}\label{FSMF1}\\ 
\text{Subject to:}& \:\:\supp(\mathbf{D}) \subseteq \mathcal{I}, \supp(\mathbf{E}) \subseteq \mathcal{J} \nonumber
\end{align}
where $\mathcal{I} \subseteq [K]\times[NT]$ and $\mathcal{J} \subseteq [NT]\times[L]$ 
respectively define the support constraint $\supp(\mathbf{D}) $ and $\supp(\mathbf{E})$ of $\mathbf{D}$ and $\mathbf{E}$, where such support constraint entails that $\mathbf{D}(i,j) =0, \ \forall (i,j) \notin \mathcal{I}$ and $\mathbf{E}(i,j) =0, \ \forall (i,j) \notin \mathcal{J}$. 
FSMF remains a broad\footnote{For connections between the FSMF problem with \emph{Low-rank matrix approximation} \cite{eckart1936approximation}, \emph{LU decomposition}\cite{van1996matrix}, \emph{ Butterfly structure and fast transforms}\cite{dao2019learning}, Hierarchical $\mathcal{H}-$matrices \cite{hackbusch1999sparse} and \emph{matrix completion} \cite{johnson1990matrix}, the reader may read~\cite{le2023spurious}.} and challenging problem, partly because, as argued in~\cite{le2023spurious}, it does not directly accept the existing algorithms from the unconstrained problem in~\eqref{eq:minDe-F2}. 

After showing that ill-conditioned supports may lead certain algorithms \cite{le2023spurious} to converge to local minima (referred to as `spurious local valleys'), the same work in~\cite{le2023spurious} revealed that for some specific $\mathcal{I,J}$,  
some algorithms can provably converge to the corresponding $\hat{\mathcal{E}}_{\mathcal{I},\mathcal{J}}$ which is shown to be unique but, clearly, optimal only~\emph{within the space of} $\mathbf{D,E}$ \emph{defined by the specific} support $\mathcal{I,J}$. The work in~\cite{le2023spurious} placed some of its focus on a particular class of ``disjoint" supports, corresponding to the class of those supports $\mathcal{I},\mathcal{J}$ that (as we will clarify later on) map onto disjoint
regions of $\mathbf{F}$. The finding in~\cite{le2023spurious} is that such ``disjoint" $\mathcal{I},\mathcal{J}$ render~\eqref{FSMF1} tractable. 

Naturally, depending on $\mathcal{I},\mathcal{J}$, even such optimal ``support-limited" solutions in~\eqref{FSMF1} can have unbounded gaps $\hat{\mathcal{E}}_{\mathcal{I},\mathcal{J}} - \hat{\mathcal{E}}$ to the global optimal $\hat{\mathcal{E}}$ from~\eqref{eq:minDe-F2}, as we simply do not know how badly the performance deteriorates by limiting the search within the specific fixed-support set of matrices. 

In summary, to date, in terms of explicit solutions to the matrix factorization problem in \eqref{eq:minDe-F2}, little is known in terms of designing good supports, while in terms of optimality guarantees, these are restricted to within the specific problem in~\eqref{FSMF1} where the search is for a given specific support. To date, little is known about explicitly characterizing a desired error performance $\hat{\mathcal{E}}$ under any desired sparsity constraints.  

\paragraph{Our contribution to the sparse matrix factorization problem}
We begin by clarifying that the progress that we have made in this domain is limited to the sparse matrix factorization problem where the sparsity constraints are on the columns of $\mathbf{D}$ and the rows of $\mathbf{E}$. To be clear, we are placing a constraint that no column of $\mathbf{D}$ has a fraction of non-zero elements that exceeds $\delta$, and no row of $\mathbf{E}$ has a fraction of non-zero elements that exceeds $\gamma$. With this in place, in terms of designs, for the lossless case of $\epsilon = 0$, our contribution is to identify supports that are optimal over a very large class of $\mathbf{D,E}$, and to explicitly identify the conditional optimal $(\hat{\mathbf{D}},\hat{\mathbf{E}})$ of the problem in~\eqref{eq:minDe-F2} restricted to the aforementioned very large class of $\mathbf{D,E}$. Under our constraints, these are the sparsest supports that guarantee lossless reconstruction of any $\mathbf{F}$.
On the other hand, when $\mathcal{E}>0$, we can identify --- under some additional uniformity assumptions --- the best possible support among all conceivable disjoint supports, in that no other disjoint support can be sparser. Hence, in terms of achievable schemes for $\mathbf{D,E}$, our contribution is to first make the connection between the approach in~\cite{le2023spurious} and {the combinatorics analysis of tiling }~\cite{ardila2010tilings} which guarantees that our schemes satisfy a crucial covering condition, and to then explicitly design easy to represent SVD-based schemes for $\mathbf{D,E}$ for any $\mathbf{F}$. 

In terms of guarantees on performance, and under our disjoint assumption, we provide clear expressions on the minimum possible sparsity $\gamma,\delta$ that guarantees an $\epsilon =0$. Now for the lossy case of $\epsilon >0$, we are providing upper bounds on the optimal $\epsilon$ under any given sparsity constraint (or equivalently, lower bounds on the sparsity and dimensionality constraints, for any given $\epsilon>0$). These bounds on the optimal $\epsilon$ are provided after we employ the asymptotic setting of scaling $K,N,L$, after we accept some uniformity assumptions on the weights of the supports of $\mathbf{D}$ and $\mathbf{E}$ (reflecting a uniformity on the load of the servers), and they are provided in the stochastic sense by averaging over the ensemble of possible $\mathbf{F}$, under some basic statistical assumptions. These, to the best of our knowledge, are the first explicit characterizations that identify or bound the optimal performance of the sparse matrix factorization problem, again under our assumptions.

Our ability to handle the stochastic problem, particularly benefits from being able to reduce the overall factorization problem into a sequence of combinatorially-designed\footnote{As we note here, the approach of \cite{le2023spurious} is a special case of tiling \cite{ardila2010tilings}, focusing on disjoint tiles.}, SVD-resolved low-rank matrix approximation problems (\!\!\cite{le2023spurious}), whose simplicity opens up to the benefits from existing powerful results from random matrix theory~\cite{tao2012topics}. This same problem of using tessellations to allow for matrix decomposition with statistically good approximations, is a very interesting problem because as we will see, the shape of the tiles affects the goodness of the approximation that each tile offers to particular parts of the matrix $\mathbf{F}$.  This is an interesting connection which, to the best of our knowledge, appears here for the first time.

\paragraph{Summary of our contributions on the problem of multi-user distributed computing of linearly-decomposable functions}
Having made the connection between matrix factorization and distributed computing, we here identify the FSMF problem to be key in the resolution of our real-valued multi-user distributed computing problem, for which we provide the following results.

We first consider the lossless case of $\mathcal{E} = 0$, and after we design an achievable scheme using  novel concepts and algorithms introduced in \cite{le2023spurious} and a converse using combinatorial tiling  arguments~\cite{ardila2010tilings}, Theorem~\ref{Achievability-Converse} establishes the exact system capacity $C =  \frac{K}{N_{opt}}$ where 
\begin{align}
N_{opt} &=\frac{\min(\Delta, \Gamma) }{T}  \lfloor \frac{K}{\Delta}\rfloor  \lfloor \frac{L}{\Gamma}\rfloor + 
    \frac{\min(\textrm{{\normalfont mod}}(K,\Delta),\lfloor \frac{L}{\Gamma} \rfloor)}{T}  \lfloor \frac{L}{\Gamma} \rfloor \\ &+ 
    \frac{\min(\mod(L,\Gamma) , \lfloor \frac{K}{\Delta}  \rfloor)}{T}\lfloor \frac{K}{\Delta} \rfloor  + \frac{ \min(\mod(K,\Delta), \mod(L, \Gamma))}{T}
    \label{eq:Nopt1}
\end{align} 
where this exact optimality is established for the case of $T \geq \min(\Delta,\Gamma)$ as well as the case of $\Delta \geq \Gamma, \Delta |K, T | \Gamma$ as well as the case of $\Gamma \geq \Delta, \Gamma | K, T  | \Delta$, or $\Delta | K , \Gamma | L, T | \min(\Delta,\Gamma)$. For the remaining cases (corresponding to a subset of the cases where $T < \min(\Delta,\Gamma)$), our achievable scheme is shown to suffer from only a constant gap to the optimal. 
In terms of design, many of the above cases are of particular interest because the tessellation patterns that we must construct, must accommodate for tiles of various sizes and shapes. 
In terms of insight, we can highlight for example the simplified case of \eqref{Zero-error-capacity}, which applies to the relatively broad setting of $\delta^{-1}, \gamma^{-1} \in \mathbb{N}$ (corresponding to having $\Delta | K , \Gamma | L$), where 
the capacity now takes the insightful form
\begin{align}
       C=
    \begin{cases}
     T \max(\zeta, \gamma),\:\:& \text{if } T\: |\: L\:  \min(\zeta, \gamma) \\
     L \zeta\gamma,\:\: & \text{if } T > L\min(\zeta, \gamma)
    \end{cases}
    \end{align} 
    revealing that for the first case, the optimal communication-vs-computation points $(\gamma,\delta)$, are    $(\frac{K}{NT},\frac{T}{K})$ and $(\frac{T}{N},\frac{L}{NT})$, while for the other case the tradeoff takes the form \[\gamma\delta = \frac{1}{N}.\] 

    Subsequently, for the lossy case, we consider the large-$N$ scaling regime with an average error guarantee $\epsilon$, averaged over matrices $\mathbf{F}$ and over the subfunctions' outputs.  Employing a similar achievable scheme as in the error-free case, paired with a standard truncated-SVD low-rank approximation approach, and under the assumption that the elements of $\mathbf{F}$ and the output of subfunctions are \emph{i.i.d} with zero mean and unit variance, we can bound the average optimal error by the expression
        \[ \Phi_{\text{MP},\lambda}(t,r)= \int_{r}^{t} x f_{\text{MP},\lambda}(x) dx\] where $\Phi_{\text{MP},\lambda}(t,r) $ is the incomplete first moment of the standard Marchenko-Pastur distribution with parameters $\lambda = \frac{\delta K}{\gamma L} = \frac{\Delta}{\Gamma}, r=(1 - \sqrt{\lambda})^2$, where $t$ is the solution to $F_{MP,\lambda}(t)= 1- T \frac{\gamma N}{K}$, and where $F_{MP,\lambda}(.),f_{MP,\lambda}(.)$ are the CDF and PDF of the same distribution. The scheme that provides the above bound, employs tessellations after consideration that the size and shape of the tiles, alters the statistics of the SVD approximations of the parts of the matrices that each tile corresponds to. As it turns out, this scheme and the above performance, is optimal over all choices of $\mathbf{D}$ and $\mathbf{E}$ whose supports satisfy the balanced disjoint support assumption (cf. Definition~\ref{disjointBalancedSupportAssumption}).

\subsection{Paper Organization and Notations}\label{NOTATIONS}
  The rest of the paper is organized as follows. Section~\ref{Formulating} formulates the system model for the setting of multi-user distributed computing of linearly-decomposable functions. 
  Section~\ref{Non-Asymptotics} addresses the error-free case, providing schemes and converses that lead to Theorem~\ref{Achievability-Converse}. Subsequently, Section~\ref{AsymptoticResults} addresses the lossy-computation case in the asymptotic setting, providing schemes and converses that lead to Theorem~\ref{asymptotic-capacity}. Section~\ref{Discussion} discusses some of the results, and then various appendix sections host some of our proofs as well as a small primer on matrix approximation.

\emph{Notations}:
We define $[n] \triangleq \{1,2,\hdots , n\}$.
For matrices $\mathbf{A}$ and $\mathbf{B}$, $[\mathbf{A},\mathbf{B}]$ indicates the horizontal concatenation of the two matrices.
For any matrix $\mathbf{X} \in \mathbb{R}^{m \times n}$, then $\mathbf{X}(i,j),\: i \in [m],\: j \in [n]$, represents the entry in the $i$th row and $j$th column, while $\mathbf{X}(i,:),\: i \in [m]$, represents the $i$th row, and $\mathbf{X}(:,j),\: j \in [n]$ represents the $j$th  column of $\mathbf{X}$. For two index sets $\I\subseteq [m], \J\subseteq[n]$, then $\mathbf{X}(\I,\J)$ represents the submatrix comprised of the rows in $\I$ and columns in $\J$. We will use $\omega(\mathbf{X})$ to represent the number of nonzero elements of some matrix (or vector) $\mathbf{X}$. We will use $\text{supp}(\mathbf{x}^{\intercal})$ to represent the support of some vector $\mathbf{x}^{\intercal} \in \mathbb{R}^{n}$, describing the set of indices that can accept non-zero elements. We will also use $\supp(\mathbf{X})$ to represent the support of some matrix $\mathbf{X} \in \mathbb{R}^{m \times n}$, describing the set of two-dimensional indices $(i,j) \in [m] \times [n]$ that can accept non-zero elements. Furthermore, $\mathbb{I}(\mathcal{X},\mathcal{Y}) \in \mathbb{R}^{m \times n}$ represents a matrix that all of its elements are zero except the elements, $\mathbb{I}(i,j), i \in \mathcal{X}, j \in \mathcal{Y}$. $\Supp (\mathbf{A}) \in \{0,1\}^{m \times n}, \mathbf{A} \in \mathbb{R}^{m \times n}$, is a binary matrix representing the locations of non-zero elements of $\mathbf{A}$. $\|\mathbf{x}\|_{0}$ denotes the zero-norm of a vector $\mathbf{x} \in \mathbb{R}^{n}$. For two binary matrices $\mathbf{I}, \mathbf{J} \in \{0,1\}^{m \times n}$, then $ (\mathbf{I} \cup \mathbf{J})(i,j) \triangleq \mathbf{I}(i,j)\vee \mathbf{J}(i,j), (\mathbf{I} \cap \mathbf{J})(i,j)\triangleq \mathbf{I}(i,j)\wedge \mathbf{J}(i,j), \mathbf{I}' (i,j) \triangleq \neg \mathbf{I}(i,j), \mathbf{I} \backslash \mathbf{J}(i,j) =  \mathbf{I} \cap \neg \mathbf{J},$ where $\vee,\wedge $ and $\neg$ are the logical ``OR'', ``AND''  and ``negation'' operators, respectively. Vector $\mathbf{1}_n$ defines the all-one  $n$-dimensional real column vector, and $\mathbf{0}_m$ defines the all-zero $m$-dimensional real column vector. The operation $\odot $ represents a Hadamard product, while $\|\mathbf{A}\|_{F} = \sqrt{\sum^{m}_{i=1} \sum^{n}_{j=1} \mathbf{A}^{2}(i,j)}$ represents the Frobenius norm of the matrix  $\mathbf{A}$. Furthermore, for $x 
 \in \mathbb{R}$, then $\lceil x \rceil, \lfloor x\rfloor $ respectively represent the ceiling and floor functions on $x$.  Additionally, $\mod(a,b),\; a ,b \in \mathbb{N}$ is the remainder of division of $a$ by $b$. Finally we use $\Phi_{\text{MP},\lambda}(t,r) \triangleq \int_{r}^{t} x f_{\text{MP},\lambda}(x) dx$ to denote the \emph{incomplete first moment} of the Marchenko–Pastur distribution with ratio $\lambda$.

\section{Problem Formulation}\label{Formulating}
We here describe in detail the main parameters of our model, establish the clear link between our distributed computing problem and sparse matrix factorization, define matrices $\mathbf{D,E,F}$ and the various metrics, rigorously establish the link between the distributed computing problem and the factorization in~\eqref{eq:DEF1} (see also~\eqref{matrixDecomposition}) for the error-free case, as well as rigorously present the equivalence of our lossy distributed computing problem to the approximate matrix factorization problem corresponding to~\eqref{eq:minDe-F2}. Let us consider
\begin{align}
      \mathbf{f}&\triangleq [F_1,F_2,\hdots,F_K]^{\intercal} \in \mathbb{R}^{K}, \label{function-vectors-0}\\
      \mathbf{f}_k &\triangleq [f_{k,1},f_{k,2},\hdots,f_{k,L}]^{\intercal} \in \mathbb{R}^{L},\: k \in [K],\label{function-vectors-1}\\
    \mathbf{w}&\triangleq [W_{1},W_{2},\hdots,W_{L}]^{\intercal} \in \mathbb{R}^{L}\label{message-vectors-1}
\end{align}
where $\mathbf{f}$ represents the vector of desired function outputs $F_{k}$ from~\eqref{linearlySep1}, where $\mathbf{f}_k$ represents the vector of function coefficients $f_{k,\ell}$ from~\eqref{linearlySep1} for the function requested by user $k$, and where $\mathbf{w}$ denotes the vector of output files $W_\ell = f_\ell(\cdot)$ again from~\eqref{linearlySep1}. Then recalling the encoding coefficients $e_{n,\ell,t}$ and transmitted signals $z_{n,t}$ from~\eqref{EncodedFiles}, as well as the decoding coefficients $d_{k,n,t}$ and decoded functions $F'_{k}$ from~\eqref{DecedFiles}, we have
   \begin{align}
    \mathbf{e}_{n,t} &\triangleq [e_{n,1,t},e_{n,2,t},\hdots, e_{n,L,t}]^\intercal \in \mathbb{R}^{L},\: n \in [N],\: t \in [T], \label{encoding-vectors-per-shot}\\
       \mathbf{z}_{n} &\triangleq [z_{n,1},z_{n,2},\hdots, z_{n,T}]^ \intercal \in \mathbb{R}^{T},\: n \in [N],\\
        \mathbf{E}_{n} &\triangleq [\mathbf{e}_{n,1},\mathbf{e}_{n,2},\hdots, \mathbf{e}_{n,T}]^\intercal \in \mathbb{R}^{T \times L},\: n \in [N], \label{encoding-vectors}\\
       \mathbf{d}_{k,n} &\triangleq [d_{k,n,1},d_{k,n,2},\hdots, d_{k,n,T}]^ \intercal \in \mathbb{R}^{T},\: k \in [K], n \in [N], \label{decoding-vectors-per-shot-1}\\
        \mathbf{d}_k &\triangleq [\mathbf{d}_{k,1}^{\intercal},\mathbf{d}_{k,2}^{\intercal},\hdots, \mathbf{d}_{k,N}^{\intercal}]^ \intercal \in \mathbb{R}^{N \times T},\: k \in [K]\label{decoding-vectors}
\end{align}
and thus  from \eqref{function-vectors-0}, we have the output vector taking the form
\begin{align}
    \mathbf{f} & =[\mathbf{f}_1,\mathbf{f}_2,\hdots,\mathbf{f}_K]^{\intercal} \mathbf{w}\label{Functions}
\end{align}
as well as the transmitted vector by server $n$ taking the form
\begin{align}
    \mathbf{z}_n &= \mathbf{E}_n \mathbf{w} =[\mathbf{e}_{n,1}, \mathbf{e}_{n,2}, \hdots,\mathbf{e}_{n,T}]^\intercal \mathbf{w}.\label{EncodedCashedData}
\end{align}
This allows us to form the matrices 
\begin{align}
      \mathbf{F}& \triangleq [\mathbf{f}_1,\mathbf{f}_2,\hdots,\mathbf{f}_K]^{\intercal} \in \mathbb{R}^{K \times L},\label{Demand-Matrix-1}\\
        \mathbf{E} &\triangleq [\mathbf{E}_{1}^{\intercal},\mathbf{E}_{2}^{\intercal},\hdots, \mathbf{E}^{\intercal}_{N}]^\intercal \in \mathbb{R}^{NT \times L},\label{EncodingMatrix}\\
    \mathbf{D} &\triangleq [\mathbf{d}_1,\mathbf{d}_2, \hdots , \mathbf{d}_K]^{\intercal} \in \mathbb{R}^{K \times NT} \label{DecodingMatrix}
\end{align}
where $\mathbf{F}$ represents the $K\times L$ matrix of all function coefficients across all the users, where $\mathbf{E}$ represents the aforementioned $NT\times L$ \emph{computing matrix} capturing the computing and linear-encoding tasks of servers in each shot, and where $\mathbf{D}$ represents the $K\times NT$ \emph{communication matrix} capturing the communication protocol and the linear decoding task done by each user. Thus, we clarify here, that all presented results, schemes and converses, assume our multi-user linearly-decomposable distributed computing problem, and assume, in their entirety, linear encoding at the servers, and decoding at the receivers.

To see the transition to the matrix factorization problem, we first note that from~\eqref{EncodedFiles} and~\eqref{Functions} we have that the overall transmitted vector $\mathbf{z} \triangleq [\mathbf{z}_{1}^{\intercal},\mathbf{z}^{\intercal}_{2},\hdots, \mathbf{z}^{\intercal}_{N}]^ \intercal \in \mathbb{R}^{N \times T}$ takes the form
\begin{align}
    \mathbf{z} =[\mathbf{E}^{\intercal}_{1}, \mathbf{E}^{\intercal}_{2}, \hdots,\mathbf{E}^{\intercal}_{N}]^\intercal \mathbf{w}  = \mathbf{E}   \mathbf{w}  \label{EncodedCashedData-1}
\end{align}
and then that given the decoding from~\eqref{DecedFiles}, each retrieved function takes the form
\begin{align}
    F'_k= \mathbf{d}_{k}^{\intercal} \mathbf{z} \label{DecodedData-one}
\end{align}
thus resulting in the vector of all retrieved functions taking the form $\mathbf{f}' =[\mathbf{d}_1,\mathbf{d}_2,\hdots,\mathbf{d}_K]^{\intercal} \mathbf{z}$. 
Our aim is to minimize the recovery error 
\begin{align}
\mathcal{E} \triangleq \|\mathbf{f}' - \mathbf{f} \|^2    \label{Recovery-Error}
\end{align}
solving the following problem
\begin{align}
  \underset{ \mathbf{f}'}{\minimize}\: \|\mathbf{f}' - \mathbf{f} \|_2 = \underset{\mathbf{D}, \mathbf{E}}{\minimize}\: \| \mathbf{D}\mathbf{E}\mathbf{w} - \mathbf{F}\mathbf{w}\|_2 =  \underset{\mathbf{D}, \mathbf{E}}{\minimize}\| (\mathbf{D}\mathbf{E} - \mathbf{F})\mathbf{w}\|_2. \label{MainEquationWithW}
\end{align}

Directly from the above, we see that for the error-free case of $\mathcal{E}  = 0$, resolving our distributed computing problem requires that $\mathbf{F}$ be decomposed as 
\begin{align}
   \mathbf{F} = \mathbf{DE} \label{matrixDecomposition}
\end{align}
as seen in~\eqref{eq:DEF1}. 

For the lossy case, directly from upcoming Lemma~\ref{Lemma-Epsilon-normalization}, and under the assumption of statistical independence of $\mathbf{w}$ from  $\mathbf{D,E,F}$, as well as from the fact that $\mathbf{D},\mathbf{E}$ are deterministic functions of $\mathbf{F}$, and from the fact that the elements of $\mathbf{w}$ are \emph{i.i.d} with unit variance, we have that 

\begin{align}\label{eq:Err1}
\epsilon \triangleq \frac{\mathbb{E}_{\:{ \mathbf{F},\mathbf{w}}}\{\mathcal{E} \}}{{KL}} = \frac{\mathbb{E}_{\:{\mathbf{F},\mathbf{w}}}\{\| (\mathbf{D}\mathbf{E} - \mathbf{F})\mathbf{w}\|^{2}_2 \}}{{KL}} = \frac{\mathbb{E}_{\mathbf{F}}\{\| \mathbf{D}\mathbf{E} -\mathbf{F}\|^{2}_{F}\}}{KL}
\end{align}
which holds for any scheme that deterministically derives $\mathbf{D,E}$. This brings to the fore the aforementioned (cf.~\eqref{eq:minDe-F2}) minimization $\hat{\mathcal{E}}=\underset{\mathbf{D,E}}{\min} \:\: \| \mathbf{D}\mathbf{E} - \mathbf{F} \|^2_{F}$, and our aim will be to bound 
\begin{align}\label{eq:Err2}
\hat{\epsilon} \triangleq \frac{\mathbb{E}_{\:{\mathbf{F},\mathbf{w}}}\{\underset{\mathbf{D,E}}{\min} \:\: \| \mathbf{D}\mathbf{E} - \mathbf{F} \|^2_{F} \}}{{KL}} 
\end{align}
which matches the function reconstruction error 
\begin{align}\label{eq:Err3}
\hat{\epsilon} = \frac{\mathbb{E}_{\:{\mathbf{F},\mathbf{w}}}\{\underset{\mathbf{D,E}}{\min} \:\:  \sum_{k=1}^{K} | F_k-F^{'}_k|^2 \}}{{KL}}. 
\end{align}

In terms of the corresponding connection to the sparsity of $\mathbf{D}$ and $\mathbf{E}$, we recall from \eqref{eq:GammaDelta} our metrics $\Gamma \triangleq  \max_{n \in [N]} |\mathcal{S}_{n}|$ and $\Delta \triangleq \max_{n  \in  [N]} |\mathcal{T}_{n} |$, which directly from~\eqref{decoding-vectors}--\eqref{EncodedCashedData} and from~\eqref{Demand-Matrix-1}--\eqref{DecodingMatrix}, imply that
\begin{align}
\max_{n \in [N]} |\cup^{T}_{t=1}\supp(\mathbf{D}(:,(n-1)T +t)) | \leq \Delta    \label{Communication-cost-condition}
\end{align}
and that
\begin{align}
     \max_{n \in [N]} |\cup^{T}_{t=1}\text{supp}(\mathbf{E}((n-1)T +t,:))| \leq \Gamma \label{Computation-cost-condition}
 \end{align}
and thus we see how the normalized costs $ \delta = \frac{\Delta}{K},  \gamma = \frac{\Gamma}{L}$ from~\eqref{Nonasymptotic-constatnts} form the upper bound on the fraction of non-zero elements of the columns of $\mathbf{D}$ and rows of $\mathbf{E}$, respectively.

Finally, from~\eqref{Rate1} we recall the system rate $R = \frac{K}{N}$, the corresponding system capacity $C$ representing the supremum of all rates that allow for error-free function reconstruction, and the two parameters of interest $\zeta = \frac{\Delta}{L}$ and $\kappa = \frac{K}{L}$.

Some additional definitions and assumptions on the stochastic aspects of our problem, will be described in Section~\ref{AsymptoticResults}.  

\section{Lossless Distributed Computing of Linearly-Decomposable Functions}\label{Non-Asymptotics}
We proceed with the main results for the error-free (lossless) case. We recall the distributed computing setting, which involves $K$ users, $N$ servers, $T$ communication slots, computational and communication costs $\Gamma\leq L, \ \Delta\leq K$ respectively, normalized costs $\gamma,\delta\in[0,1]$, a communication-related parameter $\zeta = \frac{\Delta}{L}$, and a system capacity corresponding to the maximum ratio $K/N$ that achieves lossless reconstruction of the functions.  We first present the result without any restriction on the dimensions, while we recall that the optimality requires that each submatrix of $\mathbf{F}$ has full rank, which is a condition that is readily justified in our real-function setting of interest here. All the results hold under the assumption that $NT\geq L$ and $NT\geq K$. The following main result will hold under the basic \emph{disjoint support assumption} on the matrices $\mathbf{D,E}$, where this assumption will be clarified in Definition~\ref{disjointSupportAssumption} immediately after the theorem below. The proofs will follow from an achievable scheme and from a converse that will be presented later on. 
\begin{Theorem} \label{Achievability-Converse}
The optimal achievable rate of the lossless $K,N,T,\Gamma,\Delta$ distributed computing setting takes the form $C = K/N_{\text{opt}}$, where 
\begin{align}
    &\lceil\frac{\min(\Delta, \Gamma) }{T} \rceil \lfloor \frac{K}{\Delta}\rfloor  \lfloor \frac{L}{\Gamma}\rfloor + 
    \lceil\frac{\min(\textrm{{\normalfont mod}}(K,\Delta), \Gamma)}{T}\rceil  \lfloor \frac{L}{\Gamma} \rfloor \nonumber\\ &+ 
    \lceil\frac{\min(\mod(L,\Gamma) , {\Delta})}{T}\rceil\lfloor \frac{K}{\Delta} \rfloor + \lceil\frac{ \min(\mod(K,\Delta), \mod(L,\Gamma))}{T}\rceil\label{eq-upper-N}\\
&\geq N_{opt} \geq \frac{KL}{T \max(\Gamma,\Delta)} \label{eq-upper-lower-N}
\end{align}
and the bounds exactly meet (and thus the achievable scheme is exactly optimal) for $T \geq \min(\Delta,\Gamma)$ as well as for $\Delta \geq \Gamma, \Delta |K, T | \Gamma$ as well as for $\Gamma \geq \Delta, \Gamma | L, T  | \Delta$. Else the achievable capacity is within a constant gap from the optimal. Finally, when $\delta^{-1}, \gamma^{-1} \in \mathbb{N}$ (corresponding to the case of $\Delta | K , \Gamma | L, T | \min(\Delta,\Gamma)$), the capacity takes the form 
    \begin{align}\label{Zero-error-capacity}
     C = 
    \begin{cases} 
     T \max(\zeta, \gamma),\:\:& \text{if } T\: |\: L\:  \min(\zeta, \gamma), \\
     L \zeta\gamma,\:\: & \text{if } T > L\min(\zeta, \gamma).
    \end{cases}
    \end{align} 
\end{Theorem}
\begin{proof}
The proof can be found in Appendix~\ref{Achievability} which describes the achievability of the corresponding decomposition $\mathbf{DE} = \mathbf{F}$, and how this decomposition is translated to our distributed computing setting with the corresponding communication and computation cost constraints. The converse can be found in Appendix~\ref{Converse}, and so is the proof of exact optimality and order optimality of the achievable scheme. 
\end{proof}
The above assumption on optimality, is clarified in the following definition. 
\begin{Definition}\label{disjointSupportAssumption}[Disjoint Support Assumption]
We say that two matrices $\mathbf{D}\in \mathbb{R}^{K\times NT},\mathbf{E}\in \mathbb{R}^{NT\times L}$, accept the \emph{disjoint  support assumption} if and only if for any two columns $\mathbf{D}(:,i),\mathbf{D}(:,i'), i, i' \in [NT]$ of $\mathbf{D}$ and the respective two rows $\mathbf{E}(i,:),\mathbf{E}(i',:)$ of $\mathbf{E}$, then $\text{supp}(\mathbf{D}(:,i)\mathbf{E}(i,:)) = \text{supp}(\mathbf{D}(:,i')\mathbf{E}(i',:) )$ or $\text{supp}(\mathbf{D}(:,{i})\mathbf{E}(i,:)) \cap \text{supp}(\mathbf{D}(:,{i'}),\mathbf{E}({i'},:)) = \emptyset$. 
\end{Definition}

To offer some insight on the result of our theorem above, let us provide here a small example that pertains to the simpler single-shot scenario. 

\begin{example}\label{single-shot-example-simple}
Let us see a basic example of a multi-user linearly-decomposable problem. Here, we have $N$ servers tasked with computing functions for $K = 6$ users. These functions are linear combinations of $L = 10$ subfunctions. We will focus on the simplified scenario with $T = 1$ (single-shot communication)\footnote{In the single shot scenario, a server broadcasts a single linear combination of the output files to its connected users. For $T>1$, a server can broadcast multiple linear combinations of the output files, not necessarily to the same set of users. The communication cost $\Delta$ measures the union of all the activated links throughout the $T$ shots.}. Our budget allows for a maximum of $\Delta = 3$ communication links per server, equivalent to $\delta = \frac{\Delta}{K} = \frac{3}{6}$, and a per-server computational cost of $\Gamma=5$ (as defined in equation~\eqref{eq:GammaDelta}) computed subfunctions, corresponding to $\gamma = \frac{\Gamma}{L}=\frac{5}{10}$. With this in place, we seek the minimum number of servers needed to guarantee lossless reconstruction $(\mathcal{E} = 0)$ of the desired functions at the users, and for this we directly use~\eqref{Zero-error-capacity} to conclude that we need $N = 12$ servers. To tackle this challenge of reconstruction, we need to construct two key matrices based on $\mathbf{F}$:
\begin{enumerate}
    \item The $(NT\times L) = (12\times 10)$ computing matrix $\mathbf{E}$, which specifies which subfunctions each server computes. Each row of $\mathbf{E}$ should have at most $\Gamma = 5$ non-zero elements.
    \item The $(K \times NT) = (6 \times 12)$ communication matrix $\mathbf{D}$, which determines where each server communicates to and how each user collects data from different servers. Each column of $\mathbf{D}$ should have at most $\Delta = 3$ non-zero elements.
\end{enumerate}
These matrices originate from the decomposition $\mathbf{F} = \mathbf{DE}$. Recalling from equation~\eqref{Demand-Matrix-1} that the matrix $\mathbf{F}$ representing the requested functions, is of size $ K\times L = 6 \times 10$, and considering $\Gamma = 5$, which corresponds to $\gamma = 1/2$, the solution is as follows:
    \begin{figure}
    \centering
   \[\begin{array}{lr}
        \includegraphics[scale=0.70]{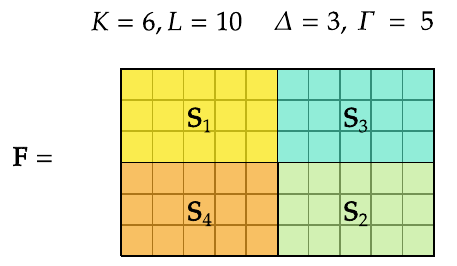}  & \hspace{20pt}\includegraphics[scale=0.55]{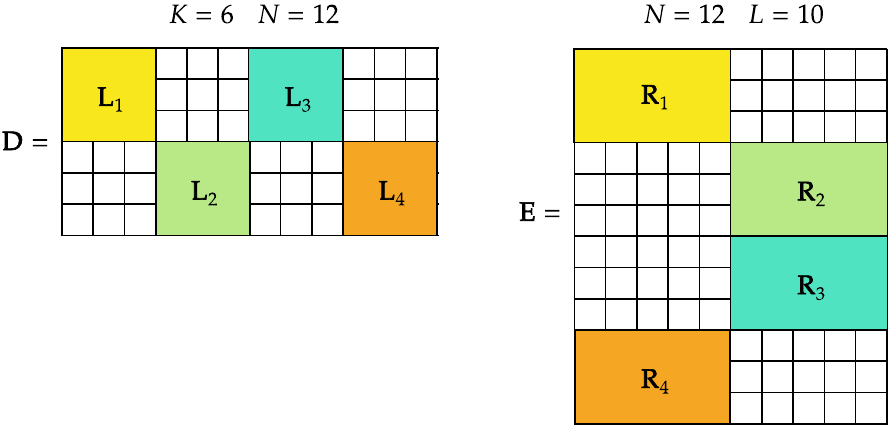}
   \end{array} \]
    \caption{Corresponding to Example~\ref{single-shot-example-simple}, this figure illustrates the partitioning of $\mathbf{F}$ into 4 tiles of size $ (\Delta \times \Gamma) = (3\times 5)$, and also illustrates the sparse tiling of $\mathbf{D}$ and $\mathbf{E}$ with tiles $\mathbf{L}_j$ and $\mathbf{R}_j$ respectively, resulting in the full tiling of $\mathbf{F} = \mathbf{DE}$ which is covered by the four $\mathbf{S}_j = \mathbf{L}_j  \mathbf{R}_j, j \in [4]$  (see Figure~\ref{ex1}), guaranteeing sparsity $\delta =\gamma =  \frac{1}{2}$ for $\mathbf{D}$ and $\mathbf{E}$ respectively, thus satisfying the per-server communication and computing constraints, while yielding lossless reconstruction of $\mathbf{F}$ and thus of the desired functions.}
    \label{ex1}
\end{figure}

\begin{enumerate}
    \item Initially we partition $\mathbf{F}$ into $\frac{K}{\Delta} \frac{L}{\Gamma} = 2\cdot 2 = 4$ disjoint $3\times 5$ submatrices $\mathbf{S}_{j}\in \mathbb{R}^{\Delta \times \Gamma} = \mathbb{R}^{3 \times 5}$ (we will refer to these submatrices as "tiles"), where $j$ ranges from $1$ to $4$. This is illustrated in Figure~\ref{ex1}. 
    \item  Then using the standard matrix decomposition form (see later in~\eqref{Complete-SVD}), we SVD-decompose each $\mathbf{S}_j$ into $\mathbf{S}_j = \mathbf{L}_j \mathbf{R}_j$, where $\mathbf{L}_j \in \mathbb{R}^{ 3 \times 3},  \mathbf{R}_j \in \mathbb{R}^{ 3 \times  5}$ for all $j \in [4]$, noting that such full decomposition is possible since the maximum rank of each $\mathbf{S}_j$ is $\min(\Delta, \Gamma)=3$. 
    \item Then we construct $\mathbf{D} \in \mathbb{R}^{6  \times 12}$ and $\mathbf{E} \in \mathbb{R}^{12 \times 10}$ by tiling them with $\mathbf{L}_j$ and $\mathbf{R}_j$ respectively, in the manner clearly illustrated in Figure~\ref{ex1}. Thus, for example, the upper left $3\times 3$ submatrix of $\mathbf{D}$ is equal to $\mathbf{L}_1$, the $3\times 3$ tile to the right of that is zero, while the lower left corner of $\mathbf{E}$ is equal to $\mathbf{R}_4$.  
    \end{enumerate}
\end{example}

The above example offers a glimpse, albeit partly illustrative, of the general principle behind creating our achievable scheme. In brief, for the simpler case corresponding to~\eqref{Zero-error-capacity}, we begin by splitting our $K\times L$ matrix $\mathbf{F}$, into $\gamma^{-1}\delta^{-1}$ submatrices of size $\Delta \times \Gamma$, and we SVD-decompose (using the matrix decomposition form) each submatrix into the $\mathbf{L}_j,j\in[4]$ part that becomes a tile of $\mathbf{D}$, and into the $\mathbf{R}_j, j \in [4]$ part that becomes a tile of $\mathbf{E}$. The tile placement must respect the sparsity constraints from $\Gamma,\Delta$ and must yield $\mathbf{DE} = \mathbf{F}$.  Regarding the required number of servers, the general rule (at least for the case of $T=1$) is that $N$ is simply the number of submatrices, multiplied with the rank of each submatrix\footnote{Having a larger $T$ augments the span of the transmitted signals, thus allowing for fewer required servers.}. In our example, we had $4$ submatrices, each of rank $3$, thus we employed $12$ servers, which later turns out to be optimal. 

\begin{figure}
        \centering
 \includegraphics[scale=0.65]{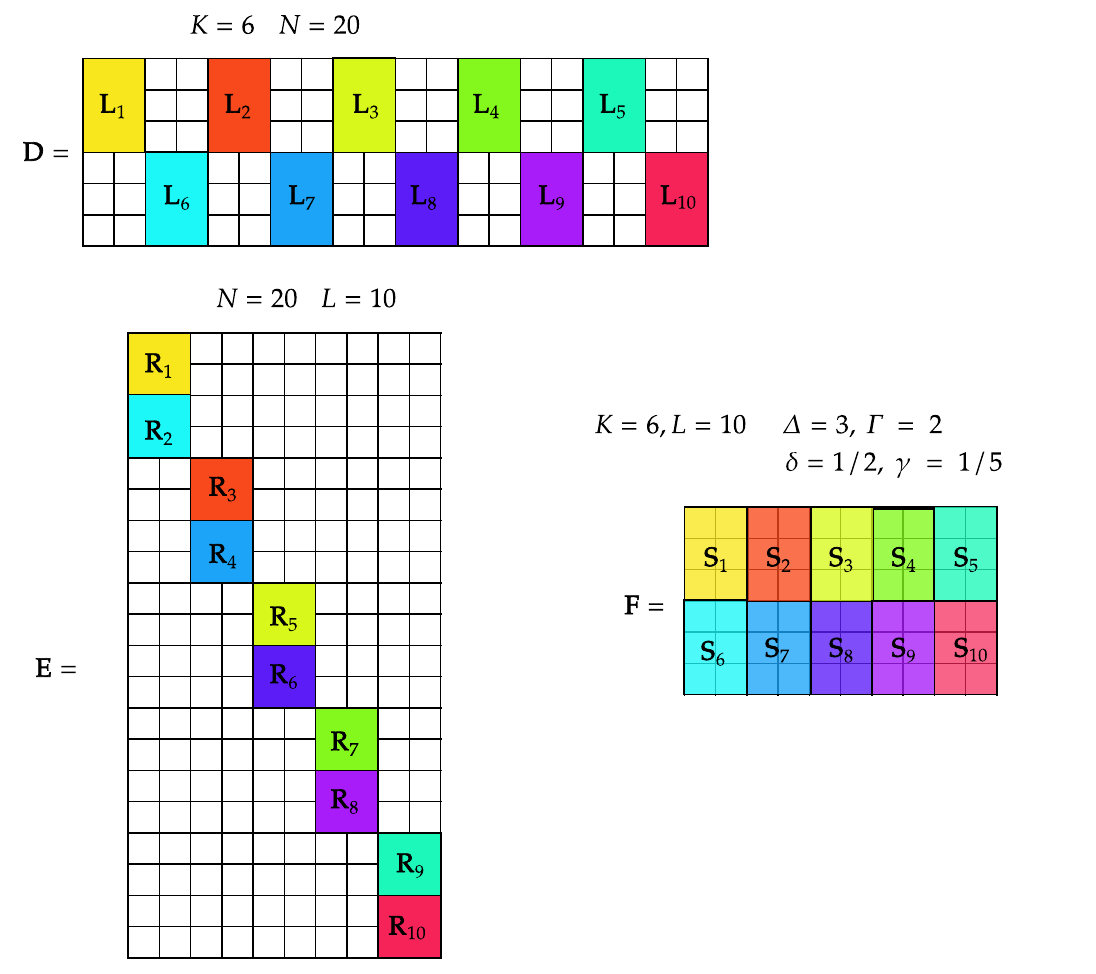}
        \caption{ A problem setting with the same $K=6,L=10,\
        \Delta=3$ and $\mathcal{E}=0$ as in Example \ref{ex1}, but a smaller computation  cost $\Gamma =2$ corresponding to $\gamma = 1/5$.  The number of servers used now for zero-error function recovery increases from  $12$ to $20$.  }
        \label{ADD2}
    \end{figure}

On the other hand, once we reduce the computation and communication capabilities of each server, more servers may be required. This is illustrated in the following example. 
\begin{example} \label{ex:exampleDelta3}
For the same setting as in Example~\ref{single-shot-example-simple}, again for $K = 6,L = 10,T=1$, and again for $\Delta = 3$ (corresponding to $\delta = \frac{1}{2}$), if we wish to substantially reduce the load on each server to only having to compute $\Gamma =2$ subfunctions (corresponding now to $\gamma = \frac{1}{5}$), then the minimum number of servers $N$ is now $20$ (cf.~\eqref{eq-upper-N}), and the corresponding tessellation pattern is presented in Figure~\ref{ADD2}, where we see $10$ submatrices of rank $2$.  
\end{example}

\begin{figure}
        \centering
 \includegraphics[scale=0.7]{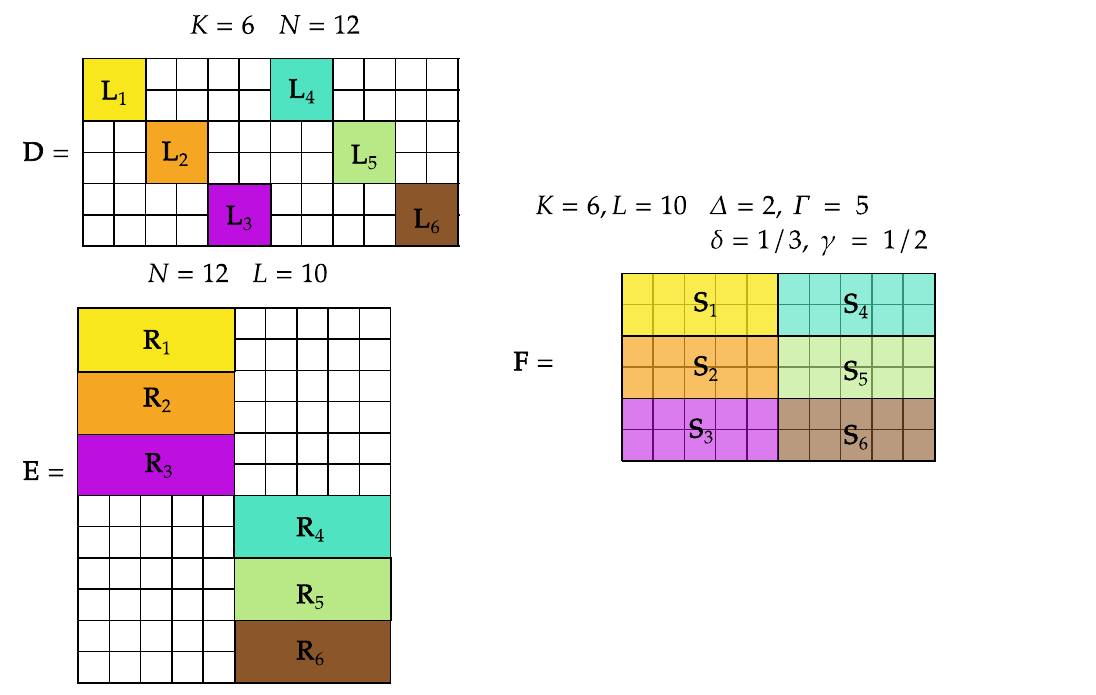}
        \caption{Pertaining to Example~\ref{ex:exampleDelta2} with $K = 6,L = 10,T=1,\Gamma = 5$ and an optimal number of $N=12$ servers, the new tessellation pattern allows for a reduced $\Delta =2$ reflecting a reduction from $\delta = 1/2$ to $\delta = 1/3$. 
        }
        \label{ADD1}
    \end{figure}

At the same time, some reductions in $\gamma,\delta$ may come for free. As it turns out, there can be multiple tessellation patterns resulting in the same required value of $N$, but depending on the size and placement of the tiles, such patterns could correspond to different $\Gamma$ and $\Delta$. To illustrate, consider the following example.
\begin{example} \label{ex:exampleDelta2}
In a lossless computing setting similar to that in Example~\ref{single-shot-example-simple}, we consider again $N=12$ servers, $K = 6$ users, $L = 10$ subfunctions and $T=1$. As in Example~\ref{single-shot-example-simple}, we maintain a computational cost of $\Gamma = 5$, but now we see that the tessellation pattern in Figure~\ref{ADD1} allows for a reduced $\Delta=2$ corresponding to $\delta = 1/3$. 
\end{example}

Such reductions in the communication and computation costs can continue, up to a point, and then again more servers may be required to provide lossless function reconstruction, as suggested by Example~\ref{ex:exampleDelta3}. The following corollary addresses this aspect, by providing the optimal communication-vs-computation tradeoff for a broad setting. We recall our usual conditions that $NT\geq L $ and $NT\geq K$.

\begin{figure}
        \centering
 \includegraphics[scale=0.8]{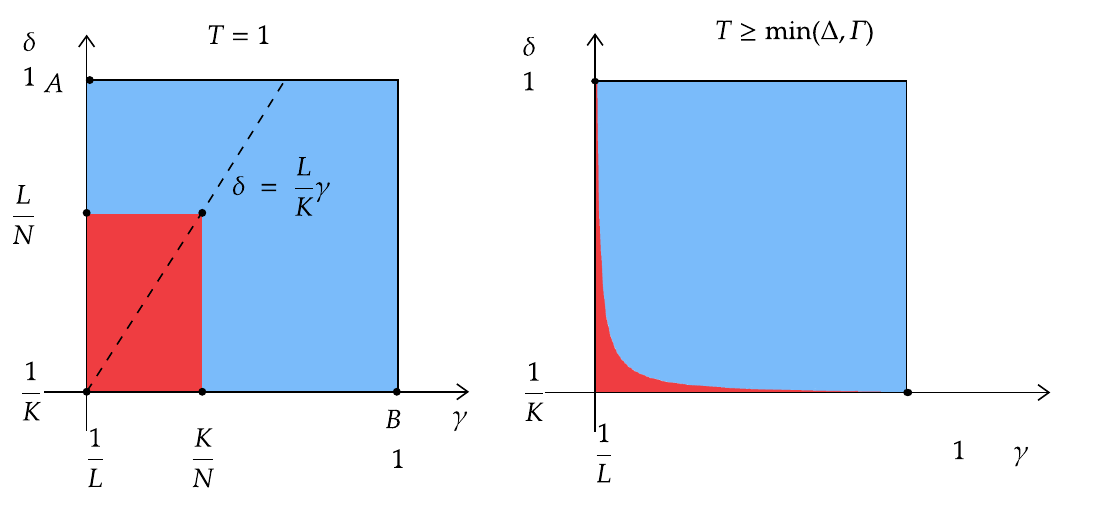}
        \caption{On the right we see the optimal performance for $T\geq \min(\Delta,\Gamma)$, which contrasts the blue achievable region with the red provably non-achievable region. On the left, we illustrate for the simple single-shot case, the two optimal points $(\gamma = \frac{K}{N},\delta= \frac{1}{K})$ and $(\gamma=\frac{1}{L},\delta = \frac{L}{N})$, which are compared to the operating points $A=(\gamma=1,\delta=1/K)$ and $B=
        (\gamma = 1/L,\delta=1)$ of two conceivable baseline schemes. Point $A=(\frac{1}{L},1)$ is that of a baseline fully-centralized scheme where servers $n\in[K]$ must compute all subfunctions (the rest are assigned no functions), while point $B=(1,\frac{1}{K})$ corresponds to a fully-parallelized baseline scheme where each server only computes one subfunction output and sends it, by necessity, to all users. The two points correspond to the trivial decompositions $\mathbf{F}=[\mathbf{I}_{K} \ \ \mathbf{0}_{(K, K-N)}] \cdot [\mathbf{F}^{\intercal} \ \ \mathbf{0}_{(L, N-K)} ]^{\intercal}$ and  $\mathbf{F} = [\mathbf{F} \ \  \mathbf{0}_{(K,N-L)}] \cdot [\mathbf{I}_{L}^{} \ \ \mathbf{0}^{}_{(L,N-L)}]$ respectively.  
        } 
    \label{gain-plot}
    \end{figure}

\begin{Corollary} \label{cor:TradeofLossLess}
In the $(K,N,T,\gamma,\delta)$ lossless distributed computing setting with $\delta^{-1}, \gamma^{-1} \in \mathbb{N}$, the optimal communication-vs-computation relationship takes the form
    \begin{align}\label{Tradeoff1}
     \gamma\delta = \frac{1}{N},  \  \  \text{for } T > L\min(\zeta, \gamma)
    \end{align} 
    whereas when $T\: |\: L\:  \min(\zeta, \gamma)$, the $(\gamma,\delta)$ operating  points 
    \begin{align}\label{Tradeoff2}
    (\frac{K}{NT},\frac{T}{K}) \  \text{ and } (\frac{T}{L},\frac{L}{NT})
    \end{align} 
    are again optimal.
\end{Corollary}
\begin{proof} The proof is direct from~\eqref{Zero-error-capacity}.
\end{proof}
The results of the above corollary are illustrated in Figure~\ref{gain-plot}.
As we have seen, when $T\geq \max(\Delta,\Gamma)$ then the above achievable rate is declared to be exactly optimal. Similarly when $T< \max(\Delta,\Gamma)$ then if $\Delta \geq \Gamma, \Delta |K, T | \Gamma$ or if $\Gamma \geq \Delta, \Gamma | L, T  | \Delta$, then again the achieved performance is exactly optimal. 
We now focus on the remaining cases (which are the remaining cases that abide to $T< \max(\Delta,\Gamma)$), for which we will show that the performance is at most a factor of $8$ from our converse. 
\begin{Corollary}\label{Gap}
For the case where $T< \max(\Delta,\Gamma)$, the achievable rate $R$ in \eqref{eq-upper-N} is at most a multiplicative factor of $8$ from the optimal.   
\end{Corollary}
\begin{proof}
    The proof can be found in the Appendix~\ref{Converse}. 
\end{proof}

\section{Lossy Distributed Computing of Linearly-Decomposable Functions}\label{AsymptoticResults}
We now examine the scenario of lossy function reconstruction, where it is possible for the reconstruction error to surpass zero. Our primary goal is to constrain and quantify the error that occurs when the available system resources, represented by $\gamma, \delta, N$, are insufficient for achieving lossless reconstruction.

Let us recall that our aim is to approach, under the available $\gamma,\delta,N,T,$ the minimum (cf.~\eqref{eq:Err2},\eqref{eq:Err3}) 
\begin{align}
\hat{\epsilon} = 
\frac{1}{KL} \mathbb{E}_{\:{\mathbf{F},\mathbf{w}}}\{\underset{\mathbf{D,E}}{\min} \:\:  \sum_{k=1}^{K} | F_k-F^{'}_k|^2 \} = \frac{1}{KL}\mathbb{E}_{\:{\mathbf{F},\mathbf{w}}}\{\underset{\mathbf{D,E}}{\min} \:\: \| \mathbf{D}\mathbf{E} - \mathbf{F} \|^2_{F} \} \label{RDefinedE_F}
\end{align}
which means that our distributed computing challenge ultimately reduces to the problem of sparse matrix factorization, as seen below
\begin{align} \label{ErrorNonStatistical}
\underset{\mathbf{D,E}}{\min}&\:\: \| \mathbf{D}\mathbf{E} - \mathbf{F} \|^2_{F}.
\end{align}

Solving this problem optimally has been a persistent challenge, often resisting a straightforward characterization of the most efficient performance~\cite{zheng2023efficient}. Consequently, we will turn to asymptotic scaling,  with parameters of interest that grow with $N$, as well as adopt a statistical approach, where instead of offering guarantees for each individual matrix $\mathbf{F}$, we will provide assurances across the ensemble of matrices $\mathbf{F}$ under certain basic assumptions.
Adopting the specific vector-wise metrics of sparsity $\gamma,\delta$, our aim will be to bound the average optimal error $\hat{\epsilon} = \frac{1}{KL} \underset{\mathbf{\mathbf{F},\mathbf{w}}}{\mathbb{E}} \{\underset{\mathbf{D,E}}{\min} \:\: \| \mathbf{D}\mathbf{E} - \mathbf{F} \|^2_{F}\}$, under the assumptions that the entries of $\mathbf{w}$ from~\eqref{message-vectors-1} and of $\mathbf{F}$, are i.i.d with zero mean and unit variance.

We proceed directly with the main result. We recall that $\Phi_{\text{MP},\lambda}(t,r), F_{MP,\lambda}(.)$ and $f_{MP,\lambda}(.)$ are respectively the incomplete first moment, the CDF and PDF of the standard Marchenko-Pastur distribution. We also recall that we operate under the assumption that $NT\geq L,NT\geq K$.

 \begin{figure}
        \centering
 \includegraphics[scale=0.8]{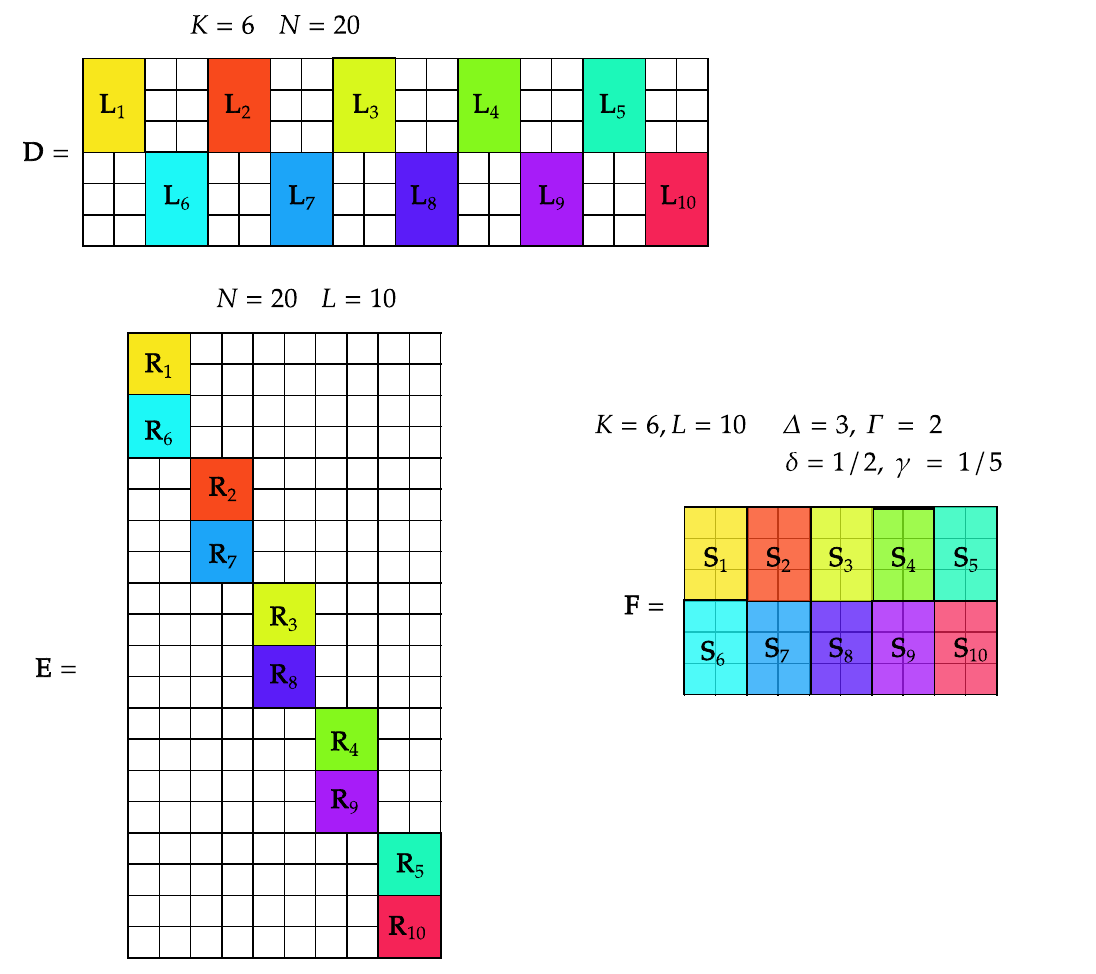}
        \caption{We consider $K=6,L=10,\
        \Delta=3$ (corresponding to $\delta=1/2$), $\Gamma =2$ (corresponding to $\gamma = 1/5$), and  $N=10$. We note that $\mathcal{E}= \sum^{10}
_{i=1}\sigma_{2, i}^2$, where $\sigma_{2, i}$ as the second singular value of $\mathbf{S}_i$ in decreasing order (i.e., the smallest singular value). In this setting, $\mathbf{L}_i \mathbf{R}_i$ is the best rank-$1$ approximation to submatrix $\mathbf{S}_i$, and is provided by the truncated SVD as described in the appendix Section~\ref{Basic-Concepts-and-Definitions}. Compared to Example~\ref{ex1} (Figure~\ref{ADD2}), we now have $\Gamma =2$ (corresponding to $\gamma = 1/5$) and a smaller number of servers $N=10$, at the expense though of having to tolerate a certain error in  our function recovery.}
        \label{ADD3}
    \end{figure}

The following main result will hold under a stronger disjoint \emph{balanced} support assumption on the matrices $\mathbf{D,E}$, where this assumption will be clarified in Definition~\ref{disjointBalancedSupportAssumption} right after the following theorem.

\begin{Theorem} \label{asymptotic-capacity} In the limit of large $N$ and constant $\delta, \gamma,\kappa, R$, the average optimal reconstruction error is bounded as 
\begin{align}
\hat{\epsilon} \leq \Phi_{\text{MP},\lambda}(t,r)=\int_{r}^{t} x f_{\text{MP},\lambda}(x) dx 
\end{align}
where $\lambda = \frac{\delta K}{\gamma L} = \frac{\Delta}{\Gamma}, r=(1 - \sqrt{\lambda})^2$, and where $t$ is the solution to $F_{MP,\lambda}(t)= 1- T \frac{\gamma N}{K}$. 
Furthermore, the bound is tight and the corresponding performance is optimal under the constraint that $\mathbf{D}$ and $\mathbf{E}$ satisfy the disjoint balanced  support assumption.
\end{Theorem}
\begin{proof}
The achievable part of the proof is based on the general scheme described in Appendix~\ref{ProofOfTheorem2}, while the analysis part draws from the properties of the Marchenko–Pastur distribution law, and is again found in Appendix~\ref{ProofOfTheorem2}. The converse utilizes basic arguments from the combinatorial tiling  literature~\cite{adams2022tiling}, after accounting for additional covering requirements, again as described in Appendix~\ref{ProofOfTheorem2}. 
\end{proof}
The above assumption on optimality is clarified in the following definition. 
\begin{Definition}\label{disjointBalancedSupportAssumption}[Disjoint Balanced Support Assumption]
We say that two matrices $\mathbf{D}\in \mathbb{R}^{K\times NT},\mathbf{E}\in \mathbb{R}^{NT\times L}$ accept the \emph{disjoint balanced support assumption} if an only if they accept the disjoint support assumption (cf. Definition~\ref{disjointSupportAssumption}) and additionally it holds that $\|\mathbf{D}(:,i)\|_{0} = \|\mathbf{D}(:,j)\|_{0}$, $\|\mathbf{E}(i,:)\|_{0} = \|\mathbf{E}(j,:)\|_{0}$, $\forall i,j \in [NT]$. 
\end{Definition}
The uniformity assumption reflects a uniformity in the computational and communication capabilities across the servers. 
Before proceeding to the next sections that present the schemes and the proofs, we provide here a simple example for our case of lossy function reconstruction.  
\begin{example}\label{Example4}
We consider a multi-user distributed computing scenario which, as in Example~\ref{single-shot-example-simple}, entails $K=6$ users, $T=1$ shots, $L=10$ subfunctions, and a per-server computation and communication cost defined by $\Gamma =5$ and $\Delta =3$ respectively. 
Unlike Example~\ref{single-shot-example-simple} that considered $\frac{KL}{\Delta \Gamma} \min(\Delta,\Gamma) = 12$ servers and error-free reconstruction, now we consider $N=4$ servers, thus forcing  a reconstruction error. For this setting, we describe how to construct our matrices $\mathbf{D} \in \mathbb{R}^{6  \times 4}$ and $\mathbf{E} \in \mathbb{R}^{4 \times 10}$, after receiving $\mathbf{F}$. Useful in this description will be Figure~\ref{ex2233}, which illustrates the tessellation pattern used to tile each non-zero submatrix of $\mathbf{D}$ and $\mathbf{E}$ with $\mathbf{L}_j$ and $\mathbf{R}_j$ respectively. 
The scheme entails three main steps.
\begin{enumerate} 
    \item 
In the first step in our example, we partition $\mathbf{F}$ into $4$ submatrices $\mathbf{F}_{1},\mathbf{F}_{2},\mathbf{F}_{3},\mathbf{F}_{4}$ of size $3\times 5$, as shown in Figure~\ref{ex1}. We will approximate these respectively by submatrices $\mathbf{S}_{1},\mathbf{S}_{2},\mathbf{S}_{3},\mathbf{S}_{4}$, such that $\mathbf{S}_1$ will approximate the upper left $3\times 5$ submatrix of $\mathbf{F}$, $\mathbf{S}_2$ the lower right $3\times 5$ submatrix, and so on.

    \item 
    As a second step, to create each $\mathbf{S}_j$, we start by SVD-decomposing each submatrix $\mathbf{F}_j$. As there are only $4$ servers and four submatrices, we remove the two less dominant singular directions of each decomposition, preserving only each dominant singular direction. Based on this rank-$1$ SVD approximation (see appendix Section~\ref{Basic-Concepts-and-Definitions} for more details), each $\mathbf{S}_j$  takes the form $\mathbf{S}_j = \mathbf{L}_j \mathbf{R}_j $ where now these dominant directions $\mathbf{L}_j \in \mathbb{R}^{ 3 \times 1} $ and $\mathbf{R}_j \in \mathbb{R}^{ 1 \times  5}, j \in [4]$, will serve as the tiles of our designed $\mathbf{D}$ and $\mathbf{E}$ respectively. 
    \item  In the final third step, we decide where to place the tiles  $\mathbf{L}_j, \mathbf{R}_j$ in $\mathbf{D}$ and $\mathbf{E}$ respectively. To do this, we employ the tessellation pattern in Figure~\ref{ex2233}. 
    For example $\mathbf{L}_1$ and $\mathbf{L}_4$ are respectively the upper left and lower right $3\times 1$ vectors in $\mathbf{D}$, while $\mathbf{R}_1$ and $\mathbf{R}_4$ are respectively the upper left and lower left $1\times 5$ vectors in $\mathbf{E}$.   
    \end{enumerate}

\begin{figure}
    \centering
   \[\begin{array}{lr}
        \includegraphics[scale=0.70]{ex1b.pdf}  & \hspace{20pt}\includegraphics[scale=0.7]{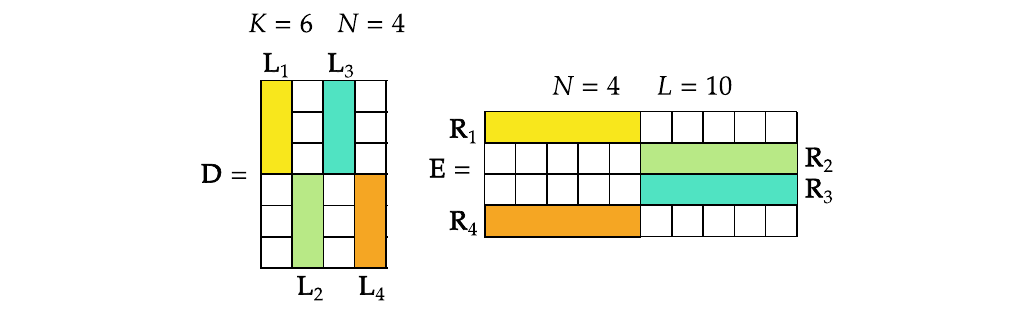}
   \end{array} \]
    \caption{Corresponding to Example~\ref{Example4}, this figure illustrates the tessellation pattern used to design $\mathbf{D}$ and $\mathbf{E}$ for a system with $K=6$ users, $T=1$ shots, $L=10$ subfunctions, $\Gamma =5$ and $\Delta =3$, but with only $N=4$ servers. The reduced number of servers forces SVD-based approximations which entail lossy function reconstruction.}
    \label{ex2233}
\end{figure}
Then in our example, in order to verify that the communication and computational costs at each server are not violated, we simply note that each column of $\mathbf{D}$ only has $\Delta =3$ non-zero elements, and each row of $\mathbf{E}$ only has $\Gamma = 5$ non-zero elements. Finally, the overall approximation error for $\mathbf{F}$ takes the form
   \begin{align}\label{eq:exampleError}
       \sum_{j=1}^4\|\mathbf{F}_j - \mathbf{L}_j \mathbf{R}_j\|^2=\sum_{j=1}^4 \sqrt{\sigma^{2}_{2,j} + \sigma^{2}_{3,j} }
       \end{align}
       where $\sigma_{i,j}, i \in [3], j \in [4]$ are the singular values of $\mathbf{F}_j$ in descending order. 
   \end{example}
   
\begin{remark}
    The above gives an example of the employed scheme for designing $\mathbf{D}$ and $\mathbf{E}$, and the expression in~\eqref{eq:exampleError} gives an example of the corresponding error performance. The scheme is described for any dimension in Section~\ref{AsymptoticResults}, and the error expression takes the form
      \begin{align}\label{eq:exampleErrorGeneral}
       \mathcal{E}_{\mathbf{F}} = \sum_{j=1}^n\|\mathbf{F}_j - \mathbf{L}_j \mathbf{R}_j\|^2=\sum_{j=1}^n \sqrt{\sigma^{2}_{q+1,j} + \sigma^{2}_{q+2,j} + \cdots  }
       \end{align}
where $n$ describes the number of submatrices that we have divided $\mathbf{F}$ into, and where $q$ simply represents the truncation depth of the SVD, where we keep only the $q$ most dominant dimensions of each submatrix $\mathbf{F}_j$.    
In terms of guarantees, we clarify as follows. As is probably clear, an exact (non-truncated) SVD approach corresponds to the error-free scenario of $\mathcal{E}_{\mathbf{F}} = 0$, for which we have indeed proven optimality under the assumptions of full rank $\mathbf{F}$ and disjoint supports, where our optimality results reveal that  $\mathcal{E}_{\mathbf{F}} = 0$ is achieved with the least amount of  $\gamma,\delta,N$ resources, compared to any other conceivable scheme. On the other hand, for the lossy case, we note that while the scheme applies and can be used for the finite dimensional case, we also recall that in terms of evaluating statistics in closed form and for providing optimality guarantees, we needed to revert to the asymptotic and stochastic setting, for which the scheme (and the corresponding averaging $\mathbb{E}[\mathcal{E}_{\mathbf{F}}]$ of~\eqref{eq:exampleErrorGeneral}, as rigorously described in \eqref{RDefinedE_F}) is optimal under the assumption of disjoint balanced supports. 
    On the other hand, for the non-asymptotic and non-stochastic lossy regime (where guarantees are required for any fixed-sized  $\mathbf{F}$), we clarify that the general expression in~\eqref{eq:exampleErrorGeneral} (seen more rigorously in \eqref{Error-Sigma}), offers a valid upper bound (directly from~\eqref{eq:Err2}) on the optimal error of our distributed computing problem, under an assumption of normalized outputs $\|\mathbf{w}\|^2 = 1$. Finally, in the asymptotic but non-stochastic (non-average) setting --- under commonly employed assumptions, and under a converging $\|\mathbf{w}\|^2 = 1$ --- the above general expression in~\eqref{eq:exampleErrorGeneral}, serves as an upper bound on the optimal error in the distributed computing problem, for any specific large $\mathbf{F}$. 
\end{remark}    
 
\section{Discussion and Conclusions} \label{Discussion}

In this work, we investigated the fundamental limits of multi-user distributed computing of real-valued  linearly-decomposable functions. In addressing this problem, we have made clear connections to the problem of fixed support matrix factorization, tessellation theory, as well as have established an interesting connection between the problem of distributed computing, and the statistical properties of large matrices. 
Under a basic disjoint support assumption, the error-free system capacity $C =  \frac{K}{N_{opt}}$ in Theorem~\ref{Achievability-Converse} revealed the optimal computational and communication resources $\gamma,\delta,N$ required to accommodate a certain number of users and subfunctions. The same result yields a simple relationship between computational complexity and communication load, as this is described in Corollary~\ref{cor:TradeofLossLess}. The derived performance is proven optimal over a sizeable class of schemes.  

For the lossy case, after transitioning to the equivalent problem of approximate matrix factorization, we employed asymptotics and a stochastic metric, allowing us to provide a clear bound on the optimal normalized reconstruction error as a function of the matrix sparsity as it reflects our communication and computational resources.  This bound is in fact tight under the assumption of uniform and disjoint supports, and it is an outcome of our schemes that employ tiling  techniques, together with truncated SVD decompositions\footnote{The derived solutions in Theorem~\ref{Achievability-Converse}  and Theorem~\ref{asymptotic-capacity} entail at most $\lceil \frac{K}{\Delta}\rceil \lceil \frac{L}{\Gamma}\rceil$ SVD decompositions, and a corresponding additional (unaccounted for) complexity of $\lceil \frac{K}{\Delta}\rceil \lceil \frac{L}{\Gamma}\rceil O(\Delta \Gamma^2) = O(KL \Gamma)= O(KL^2)$. Our paper operates under the assumption that such costs are small compared to the costs of evaluating the $L$ subfunctions that may often be non-linear.}.

 \begin{figure}
        \centering
        \includegraphics[scale=0.35]{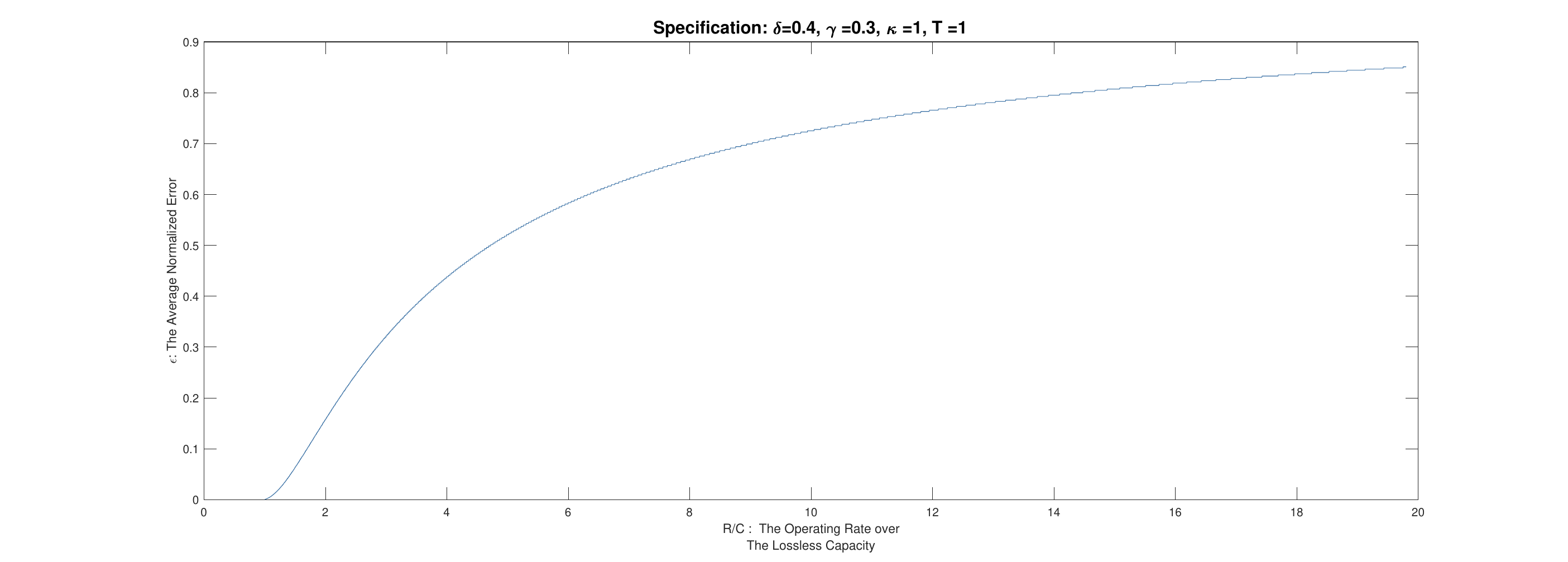}
        \caption{The $y$-axis represents the conditionally optimal average error $\epsilon$ derived in Theorem~\ref{asymptotic-capacity} for $\delta= 0.4,\gamma = 0.3,  \kappa=1$ and $T=1$. The $x$-axis describes the ratio of the operating rate to the error-free or lossless system capacity, i.e. describes how many times higher is the system rate from the error-free system capacity.} 
        \label{FigureErrorVSServer}
    \end{figure}
One of the interesting outcomes of the work is an analytical handle on how we can tradeoff our system rate (corresponding to our server resources or the clients we serve) with the function reconstruction error. Figure~\ref{FigureErrorVSServer} offers some understanding on how lossy reconstruction is affected by having either too few servers or too many users. Starting from a lossless scenario where the operating ratio $K/N$ matches the error-free system capacity (corresponding to the value of 1 on the $x$-axis), we see how the error increases as we either add more users or as we remove servers. For example, when the ratio $R/C$ between the operating rate and the error-free capacity, is around $2$ (after doubling the users or halving the servers) then we expect an error corresponding to $\epsilon = 0.15$. On the other hand, if we could accept $\epsilon = 0.3$, then --- compared to the error-free scenario --- we could triple the number of users or equivalently reduce our servers to about one third.

\begin{figure}
        \centering
        \includegraphics[scale=0.34]{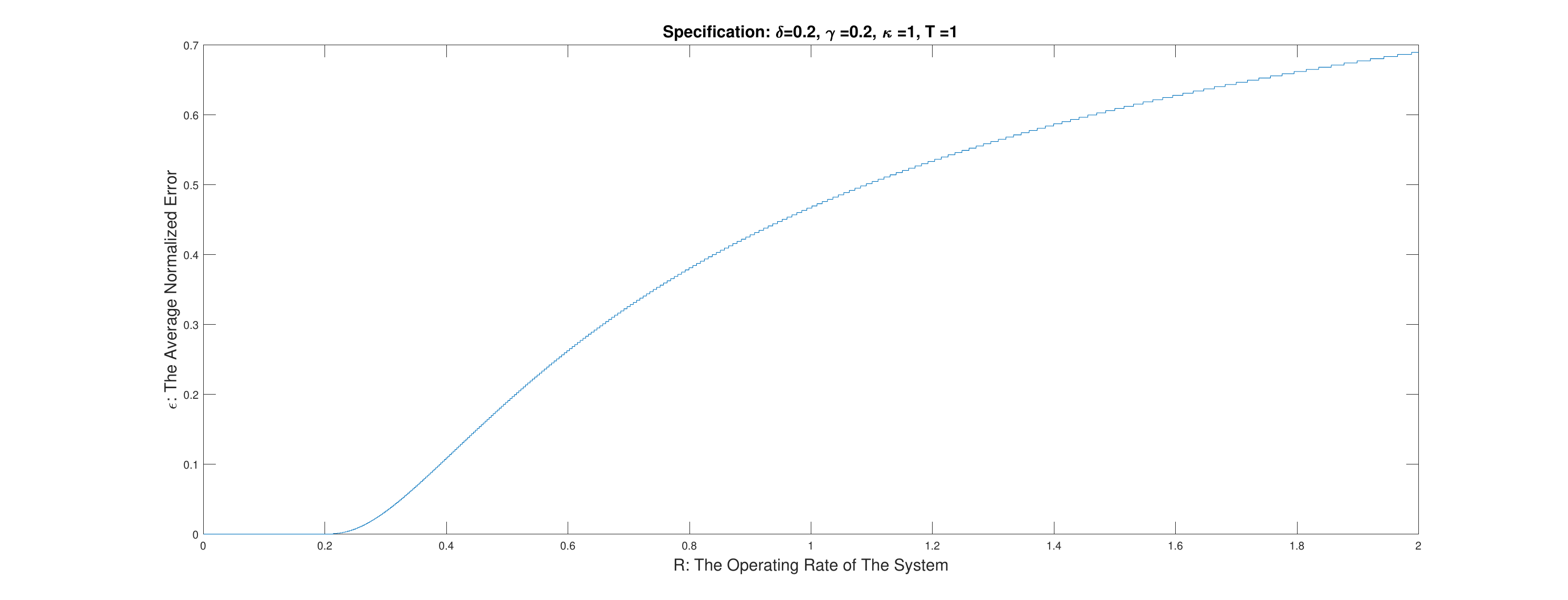}
        \caption{The $y$-axis plots $\epsilon$ from Theorem~\ref{asymptotic-capacity} for $\delta= 0.2,\gamma = 0.2,  \kappa=1$ and $T=1$, while the $x$-axis represents the rate $K/N$.}\label{ErrorVSRate}
\end{figure}

\begin{figure}\label{FigureErrorVSOthers}
        \centering
        \includegraphics[scale=0.54]{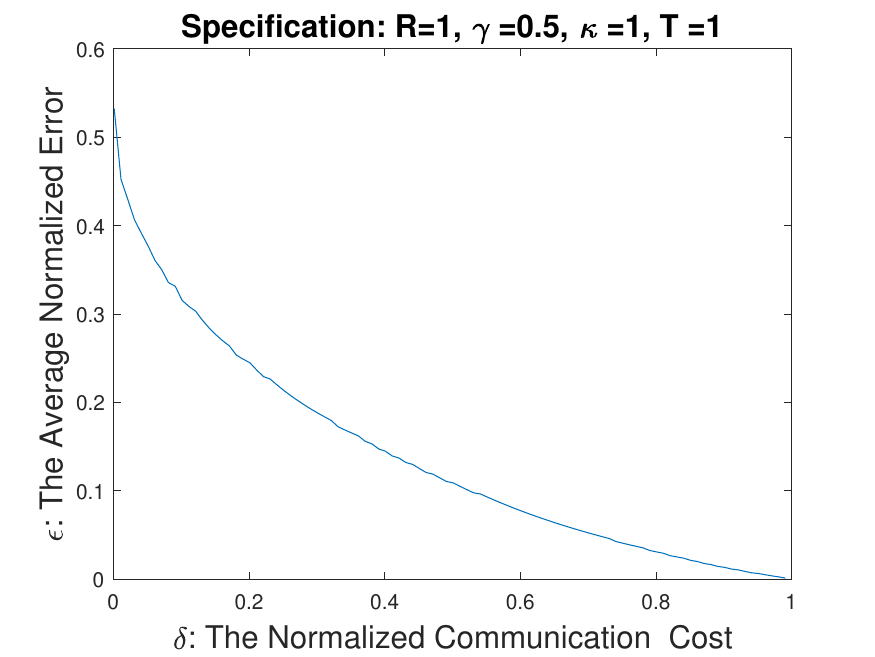}
         \includegraphics[scale=0.54]{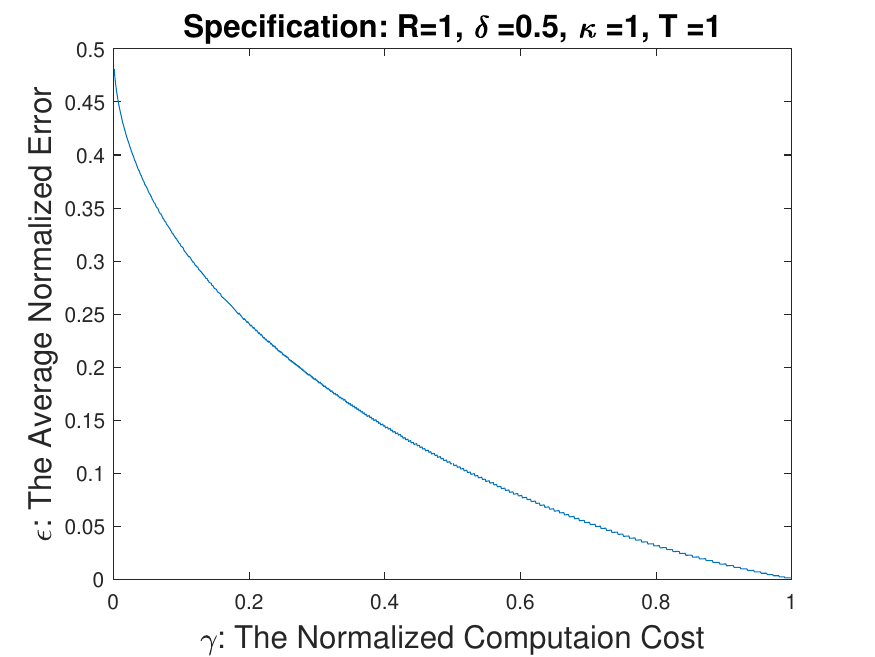}
         \includegraphics[scale=0.54]{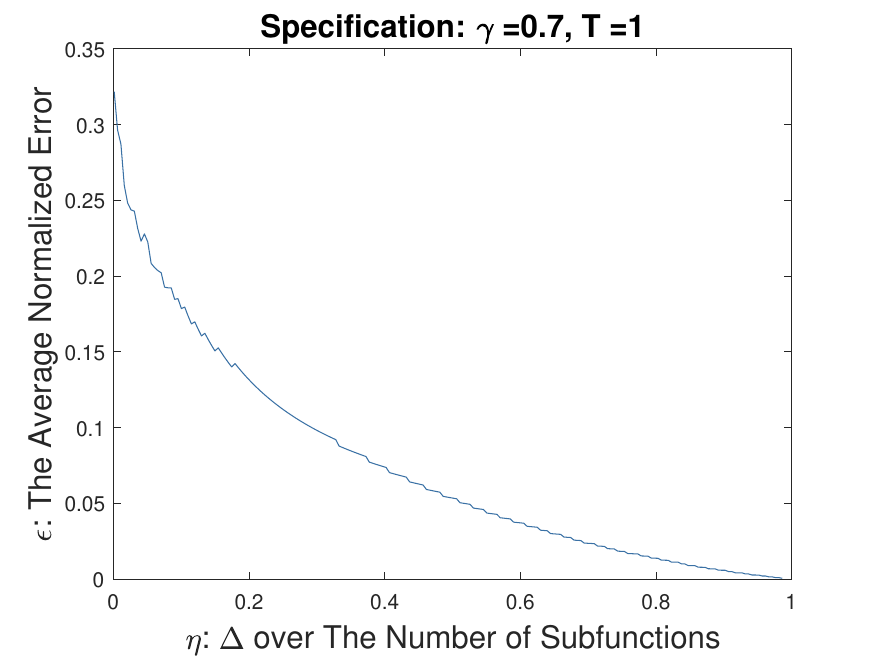}
        \includegraphics[scale=0.54]{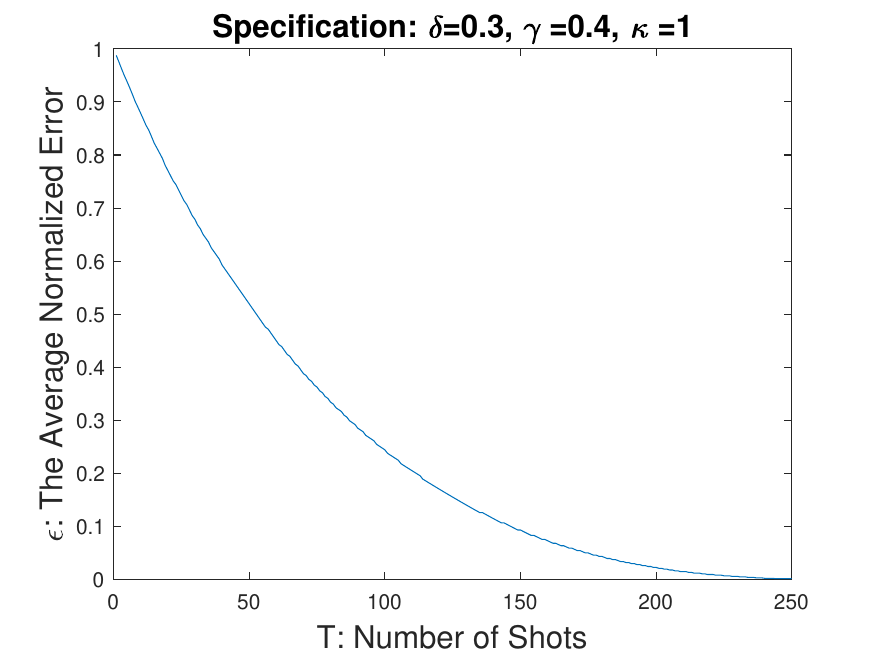}
        \caption{Plotting the error-effect of $
        \delta,\gamma,\eta,T$ (Theorem~\ref{asymptotic-capacity}) in various settings. The $y$ axis corresponds to $\epsilon$ (cf.~\eqref{eq:Err2}).}
        \label{ex22222}
    \end{figure}

Similarly Figure~\ref{ErrorVSRate}, for the same setting, plots the error as a function of the operating rate (unnormalized), for a slightly different setting where now $\delta=0.2, \gamma=0.2, \kappa=1, T=1$.  In this scenario, where the error-free capacity is merely $C_{0}= 0.2$ (cf. \eqref{Zero-error-capacity}), an aggressive increase in the rate to $R=2$ (2 times more users than servers, and $10 = \frac{R}{C_0}$ times more users than in the error-free case) yields a generally unmanageable $\epsilon= 0.68$, etc. 
Finally, Figure~\ref{ex22222} reflects the effect of the per-server computation resources $\gamma$, the per-server communication resources $\delta$, the effect of $T$ and the effect of  $\eta$ (cf.~\eqref{eq:zetaKappa}) where a higher $\eta$ implies more communication resources per subfunction. 

\begin{figure}
        \centering
        \includegraphics[scale=0.43]{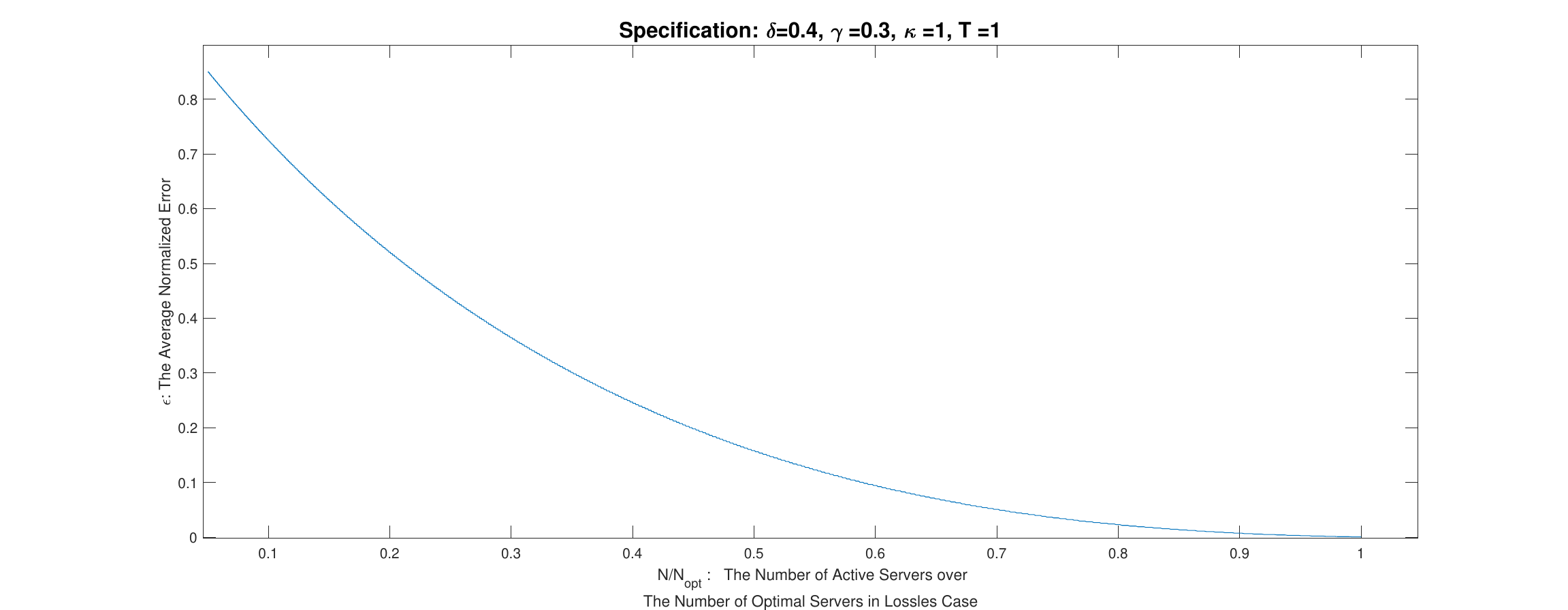}
        \caption{The $y$-axis represents the conditionally optimal average error $\epsilon$ derived in Theorem~\ref{asymptotic-capacity} for $\delta= 0.4,\gamma = 0.3,  \kappa=1$ and $T=1$. The $x$-axis describes the number of active servers over the number of optimal serves in the lossless scheme.} 
        \label{FigurServers}
    \end{figure}

One interesting outcome of our work is that substantial computational and communication savings can be harvested with only a modest cost in function reconstruction error. This has to do with the nature of the singular value distribution of larger matrices (corresponding to many users and many basis subfunctions). Due to this nature, in principle, removing computational resources will indeed introduce error, but will do so in a manner that is initially slow. Such error power is initially proportional to the smallest singular value of a large $\mathbf{F}$, and then with more computational/communication savings, it becomes proportional to the sum of the few smallest singular values, and so on. Due to the statistical nature of larger matrices, this error accumulates slowly, thus considerable computational savings can be obtained, with a relatively modest function reconstruction error (See Figure~\ref{FigurServers}). 

For future work, we can consider various related scenarios such as the scenario where  some subfunctions contribute more to the overall function than other subfunctions do. Another interesting direction could involve a pre-defined and fixed communication topology, which would then entail a support of $\mathbf{D}$ that is fixed and independent of $\mathbf{F}$, or similarly a scenario where each server can only compute a predefined subset of subfunctions, now bringing to the fore a support of $\mathbf{E}$ that is again fixed and independent of $\mathbf{F}$.

\appendices 

\section{Concepts Relating to the Design of the Schemes} \label{Basic-Concepts-and-Definitions}
Before describing our schemes and converses, we present in this section some basic properties and definitions that will be useful later on. 

We recall that our goal will be to design the communication matrix $\mathbf{D}$ and the computing matrix $\mathbf{E}$ that yield $\mathbf{DE} = \mathbf{F}$ (or  $\mathbf{DE} \approx \mathbf{F}$, depending on our case), under fixed $N,T$, and under a constraint of at most $\Delta$ non-zero elements in any column of $\mathbf{D}$ and at most $\Gamma$ non-zero elements in any row of $\mathbf{E}$.  
To do so, we will need some basic concepts and definitions relating to the approach\footnote{We quickly recall that for a matrix $\mathbf{A}$, then $\mathbf{A}(:,n)$ represents its $n$th column, $\mathbf{A}(n,:)$ its $n$th row, $\supp(\mathbf{A})$ the binary matrix indicating the support of $\mathbf{A}$, while for a vector $\mathbf{a}$, then $\text{supp}(\mathbf{a})$ represents the set of indices of $\mathbf{a}$ with non-zero elements.  Also, when we refer to a support constraint, this will be in the form of a binary matrix that indicates the support (the position of the allowed non-zero elements) of a matrix of interest.} in~\cite{le2023spurious}.
   \begin{Definition} \label{def-r1}
   Given two support constraints $\mathbf{I} \in \{0,1\}^{K \times NT}$ and $\mathbf{J} \in \{0,1\}^{NT \times L}$ of two matrices $\mathbf{D}\in\mathbb{R}^{K\times NT}$ and $\mathbf{E}\in\mathbb{R}^{NT\times L}$ respectively, then for any $n \in [NT]$, we refer to $\mathbf{S}_{n}(\mathbf{I}, \mathbf{J}) \triangleq \mathbf{I}(:,n) \mathbf{J}(n, :)^{} \in \mathbb{R}^{K\times L}$ as the $n$th \emph{rank-one contribution support\footnote{We have naturally assumed that $\mathbf{I}(:,n)$  and  $\mathbf{J}(n, :)$ each have at least one non-zero element.}} of $\mathbf{DE}$~\cite{le2023spurious}. 
   \end{Definition}
We note that when the supports are implied, we may shorten $\mathbf{S}_{n}(\mathbf{I}, \mathbf{J})$ to just $\mathbf{S}_n$. This will 
relate to the support of $\mathbf{DE}$. On this, we have the following lemma. 
\begin{Lemma}\label{Lemma-DE-Support}
    For $\mathbf{I} \triangleq \supp(\mathbf{D}) $ and $\mathbf{J} \triangleq \supp(\mathbf{E})$, then
    \begin{align}
         \cup^{N}_{n=1} \mathbf{S}_n(\mathbf{I}, \mathbf{J}) \supseteq \supp(\mathbf{D}\mathbf{E}).\label{Equation-nultiplicative-contribution}
    \end{align}
    \end{Lemma}
    \begin{proof}
The above follows from Definition~\ref{def-r1}, from having $\cup^{N}_{n=1} \mathbf{S}_n(\mathbf{I}, \mathbf{J}) = \cup^{N}_{n=1}\mathbf{I}(:,n) \mathbf{J}(n,:)$, and from the fact that $\mathbf{D} \mathbf{E} = \sum^{NT}_{n=1} \mathbf{D}(:,n) \mathbf{E}(n,:)$.
\end{proof}
 We also need the following definition. 
   \begin{Definition}\label{Def2}
       For $\mathbf{I} = \supp(\mathbf{D}) \in \{0,1\}^{K \times NT} $ and $\mathbf{J} = \supp(\mathbf{E}) \in \{0,1\}^{NT \times L}$, the \emph{equivalence classes of rank-one supports} are defined by the equivalence relation $i \sim j$ on $[NT]$ which holds if and only if $\mathbf{S}_i = \mathbf{S}_j$.  This relation allows us to define $\mathcal{C}$ to be the set of all equivalence classes~\cite{le2023spurious}.
   \end{Definition}
The above splits the columns of $\mathbf{D}$ (and correspondingly the rows of $\mathbf{E}$) such that the equivalence $i \sim j$ holds if and only if $\mathbf{I}(:,i) \mathbf{J}(i,:) = \mathbf{I}(:,j) \mathbf{J}(j,:)$.
   
\begin{Definition}\label{Def2b} 
    For two supports $\mathbf{I} \in \{0,1\}^{K \times NT}, \mathbf{J} \in \{0,1\}^{NT \times L}$ of $\mathbf{D}$ and $\mathbf{E}$ respectively, and for $\mathcal{C}$ being the collection of equivalence classes as in Definition~\ref{Def2}, then each class $\mathcal{P} \in \mathcal{C}$ will have a \emph{representative support} which we will denote as $\mathbf{S}_{\mathcal{P}}$.  
\end{Definition}    

\begin{Definition} \label{Def2bb} 
For a representative support $\mathbf{S}_{\mathcal{P}}$ of a class $\mathcal{P} \in \mathcal{C}$ (as in Definition~\ref{Def2b}), and for some $n \in \mathcal{P}$, we define $\mathbf{c}_{\mathcal{P}} \triangleq \mathbf{I}(:,n)$ (resp. $\mathbf{r}_{\mathcal{P}} \triangleq \mathbf{J}(n,:)$) to be the corresponding \emph{component column} (resp. \emph{component row}) of $\mathbf{S}_{\mathcal{P}}$, and we define $\mathcal{C}_{\mathcal{P}} \triangleq \text{supp}(\mathbf{c}_{\mathcal{P}} ) \subset [K] $ to be the set of indices of the non-zero elements in $\mathbf{c}_{\mathcal{P}}$, while we define $\mathcal{R}_{\mathcal{P}} \triangleq \text{supp}(\mathbf{r}_{\mathcal{P}}) \subset [L] $ to be the set of indices of the non-zero elements in $\mathbf{r}_{\mathcal{P}}$.  
\end{Definition}    

\begin{remark}
    As we will see later on, the non-zero part of $\mathbf{S}_{\mathcal{P}}$ will define the position and size of \emph{tile}\footnote{Note that $\mathbf{S}_n = \mathbf{S}_\mathcal{P}$ for any $n \in \mathcal{P}$.} $\mathcal{P}$ of $\mathbf{DE}$. 
    Furthermore, the non-zero part of $\mathbf{I}(:,\mathcal{P})$ and the non-zero part of $\mathbf{J}(\mathcal{P},:)$ will define the so-called tiles of $\mathbf{D}$ and $\mathbf{E}$ respectively\footnote{To help the reader with the notation, we remind here that $\mathbf{I}(:,\mathcal{P})$ simply refers to the columns of $\mathbf{I}$ labeled by the elements inside set $\mathcal{P}$. Similarly, $\mathbf{J}(\mathcal{P},:)$ corresponds to the rows of $\mathbf{J}$ indexed by $\mathcal{P}$.}, and they will naturally map onto tile $\mathcal{P}$ of $\mathbf{DE}$. Also note that a tile $\mathcal{P}$ (which corresponds to the representative contribution support $\mathbf{S}_{\mathcal{P}}$, i.e., which corresponds to the entire class $\mathcal{P}$) should not be confused with the rank-one contribution support $\mathbf{S}_{n}$. A tile may entail multiple (specifically $|\mathcal{P}|$ such) rank-one contribution supports. 
\end{remark}
We proceed with further definitions. 
\begin{Definition} \label{Def2c}
For every subset $\mathcal{C}' \subseteq \mathcal{C}$ of equivalence classes, we define the \emph{union of the representative supports} $\mathcal{S}_{\mathcal{C}'}\triangleq \cup_{\mathcal{P} \in \mathcal{C}'} \mathcal{S}_{\mathcal{P}}$ to be --- as defined in Section~\ref{NOTATIONS} --- the point-wise logical OR of the corresponding $\mathcal{S}_{\mathcal{P}}$. 
   \end{Definition}
In the above, $\mathcal{S}_{\mathcal{C}'}$ is simply the area of the product matrix $\mathbf{DE}$ covered by all the tiles $\mathcal{P}$ in $\mathcal{C}'$. Furthermore we have the following definition. 
\begin{Definition}[\!\!\cite{le2023spurious}] \label{maxrank}
The \emph{maximum rank of a representative support of class} $\mathcal{P} \in \mathcal{C}$ takes the form
   \begin{align}
       r_{\mathcal{P}} \triangleq \min (|\mathcal{C}_{\mathcal{P}}|, |\mathcal{R}_{\mathcal{P}}|). \label{max-rank}
   \end{align}
    \end{Definition}      
As one can readily see, when $\mathbf{I} = \supp(\mathbf{D})$ and $\mathbf{J} = \supp(\mathbf{E})$, then the part of matrix $\mathbf{D}\mathbf{E}$ covered by tile $\mathcal{S}_\mathcal{P}$, can have rank which is at most $r_{\mathcal{P}}$.
    \begin{figure}
       \centering
       \[\begin{array}{cc} \includegraphics[scale=0.45]{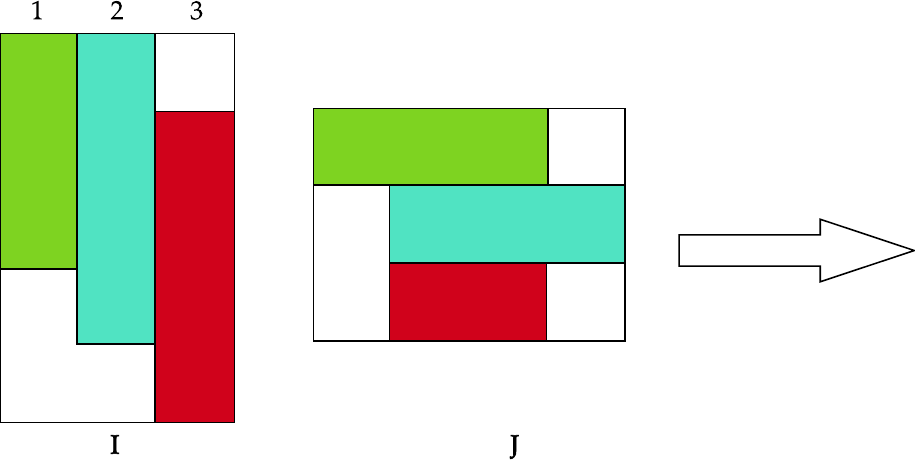} & \includegraphics[scale=0.45]{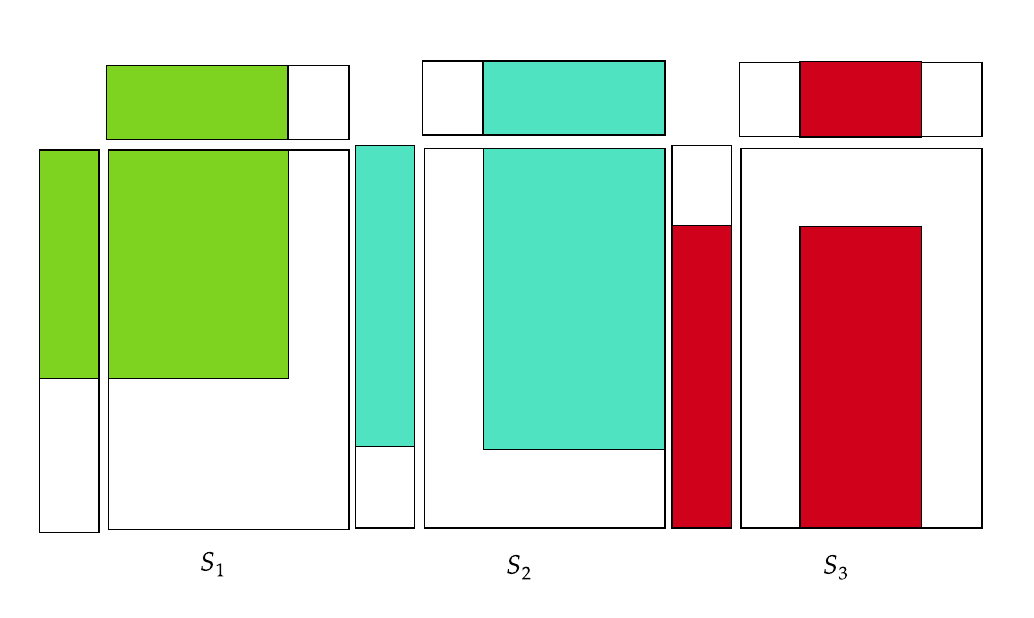} \end{array}\]
       \caption{The figure on the left illustrates the support constraints $\mathbf{I}$ and $\mathbf{J}$ on $\mathbf{D}$ and $\mathbf{E}$ respectively.  
       The constraints $\mathbf{I}(:,1)$ and $\mathbf{J}(1.:)$ on the columns and rows of $\mathbf{D}$ and $\mathbf{E}$ respectively are colored green, $\mathbf{I}(:,2)$ and $\mathbf{J}(2.:)$ are colored cyan and $\mathbf{I}(:,3)$ and $\mathbf{J}(3.:)$ are colored red.  The product of a column with a row of the same color, yields the corresponding rank-one contribution support $\mathbf{S}_{n}(\mathbf{I},\mathbf{J}),n=1,2,3,$ as described in Definition~\ref{def-r1}, and as illustrated on the right side of the figure.}
       \label{fig:my_label1}
   \end{figure}
\begin{figure}
       \centering
       \includegraphics[scale=0.6]{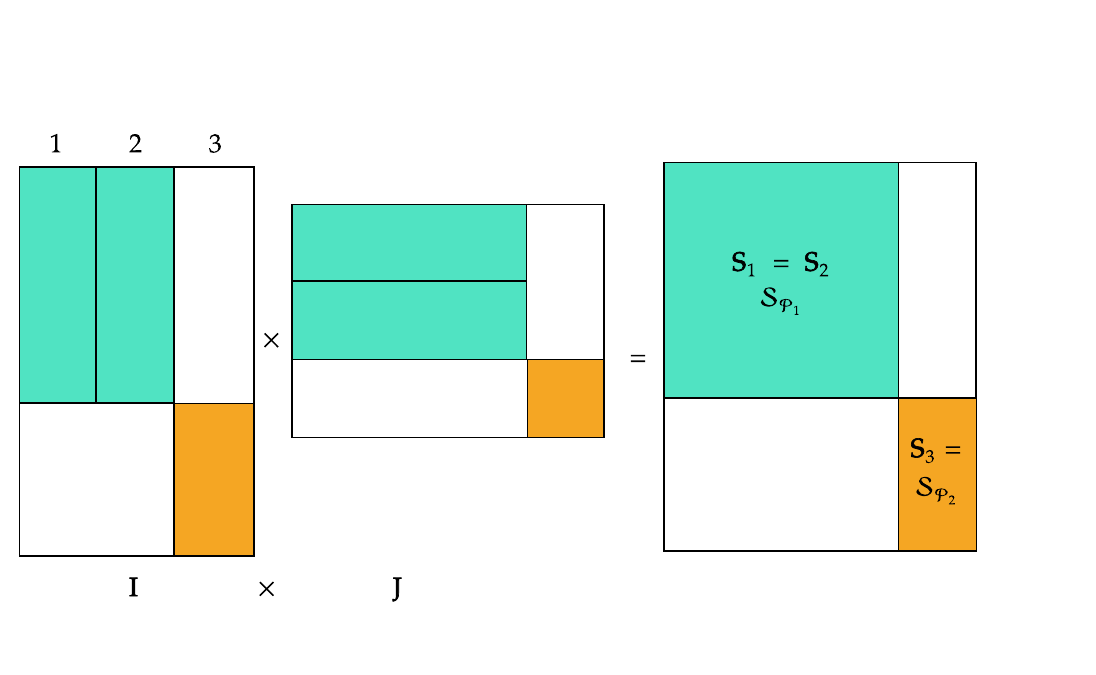}
       \caption{This figure illustrates three different rank-one contribution supports $\mathbf{S}_{1},\mathbf{S}_{2},\mathbf{S}_{3}$, where the first two fall into the same equivalence class $\mathcal{S}_{\mathcal{P}_1}  = \mathbf{S}_{1} = \mathbf{S}_{2}$, while $\mathcal{S}_{\mathcal{P}_2}  = \mathbf{S}_{3}$.  }
       \label{fig:difference}
   \end{figure}
\begin{remark}
    While in mathematics, tiling a matrix (or a surface area) corresponds to finding a way to `cover' this matrix with tiles of specific sizes, here our task is somewhat different. Here our goal is to cover a product matrix ($\mathbf{DE}$ in this case) with tiles that are the result of a product of properly placed tiles of $\mathbf{D}$ and of $\mathbf{E}$, where these last tiles must adhere to the sparsity constraints of $\mathbf{D}$ and $\mathbf{E}$.
    Furthermore, the size of the non-zero part of each $\mathbf{S}_{n}$, $n\in \mathcal{P}$, together with the number of such equivalent elements in $\mathcal{P}$, will jointly determine the degree of the approximation of the corresponding submatrix of $\mathbf{F}$.
\end{remark}

Before proceeding with the scheme, we here also give a very brief reminder on the basic concepts regarding SVD decompositions.

\subsection{Brief Primer on Matrix Approximation} \label{SVD}
For an $m \times n$ matrix $\mathbf{A}$, for some $k\leq \min\{m,n\}$, the rank-$k$ approximation of $\mathbf{A}$ takes the form\footnote{In the following, when needed we will be using matrix-subscripts to denote the dimensionality of matrices. For example, we will be using $\mathbf{B}_{m \times k}$ to emphasize that matrix $\mathbf{B}$ has dimensionality $m \times k$.} 
\begin{align}
    \mathbf{A}_{m \times n}  \simeq \mathbf{B}_{m \times k} \mathbf{C}_{k \times n}
\end{align}
where $\mathbf{B}_{m \times k}$ and $ \mathbf{C}_{k \times n}$ are two full rank matrices of dimension $m \times k$ and $k \times n$ respectively. 
Such decomposition allows us to represent $\mathbf{A}$ with at most $k(m+n)$ elements, which can be substantially fewer than the $mn$ elements required to represent $\mathbf{A}$.

As we will further recall below, matrices can be represented by employing SVD decomposition. 
For a rank-$r$ matrix $\mathbf{A}_{m \times n }$, this SVD takes the form
\begin{align}
    \mathbf{A} &= \mathbf{U}_{m \times r} \mathbf{S}_{r \times r} \mathbf{V}^{\intercal}_{r \times n} = \begin{bmatrix}
        \mathbf{u}_1 , \mathbf{u}_2 , \hdots , \mathbf{u}_{r}
    \end{bmatrix} 
    \begin{bmatrix}
        \sigma_1 & & & 0 \\
          & \sigma_2 & &  \\
           &  & \ddots & \\
             0 &  & & \sigma_r \\
    \end{bmatrix}
    \begin{bmatrix}
        \mathbf{v}^{\intercal}_1\\
        \mathbf{v}^{\intercal}_2\\
        \vdots\\
        \mathbf{v}^{\intercal}_r
    \end{bmatrix}
     = \sum^{r}_{i=1} \sigma_i u_i v_i^{\intercal}\label{Complete-SVD}
\end{align}
where $\mathbf{U}$ and $\mathbf{V}$ are orthogonal matrices, where $\mathbf{S}$ is diagonal with entries $\sigma_1 \geq \sigma_2 \geq \hdots \geq \sigma_{r} \geq 0$ being the singular values in descending order, where $\mathbf{u}_1,\mathbf{u}_2,\hdots, \mathbf{u}_r$ are the columns of $\mathbf{U}_{m \times r}$, and where $\mathbf{v}^{\intercal}_1,\mathbf{v}^{\intercal}_2,\hdots,\mathbf{v}^{\intercal}_r$ are the rows of $\mathbf{V}_{n \times r}$.

Subsequently, the optimal rank-$k$ approximation matrix $\mathbf{A}_k, k <r$ of $\mathbf{A}$, takes the form
 \begin{align}
     \mathbf{A}_k &= \mathbf{U}_{m \times k} (\mathbf{S}_k)_{k \times k} \mathbf{V}^{\intercal}_{k \times n} = \sum^{k}_{i=1} \sigma_i u_iv_i^{\intercal} = \mathbf{U}_{m \times k} \mathbf{U}^{\intercal}_{k \times m} \mathbf{A} \label{Truncated-SVD-1}\\&= (\sum^{k}_{i=1} u_iu_i^{\intercal}) \mathbf{A} = \mathbf{A} \mathbf{V}_{n \times k} \mathbf{V}^{\intercal}_{k \times n} = \mathbf{A}(\sum^{k}_{i=1} \mathbf{v}_i \mathbf{v}^{\intercal}_i)\label{Truncated-SVD}
 \end{align}
where $(\mathbf{S}_k)_{k \times k}= \diag(\sigma_1,\sigma_2, \hdots, \sigma_k) , \mathbf{U}_{m \times k} = [\mathbf{u}_1,\mathbf{u}_2, \hdots, \mathbf{u}_k]$ and $\mathbf{V}^{\intercal}_{k \times n} =[
     \mathbf{v}_1^{\intercal}, \mathbf{v}_2^{\intercal},
     \hdots, \mathbf{v}^{\intercal}_k]$. This approximation, also referred to as the \emph{truncated SVD approximation}, is simply 
the projection of $\mathbf{A}$ onto the space spanned by the strongest  $k$ singular vectors of $\mathbf{A}$. Directly from the well-known Eckart-Young Theorem~\cite{eckart1936approximation}, this approximation is optimal, as it guarantees $\| \mathbf{A} - \mathbf{B}\|_{F} \geq \|\mathbf{A} - \mathbf{A}_k\|_{F}  = \sqrt{\sigma^2_{k+1} + \hdots + \sigma^{2}_{r}}$ for any rank-$k$ matrix $\mathbf{B}$.

With the above in place, we proceed to describe the scheme for the lossless case.

\section{Scheme for Lossless Reconstruction (Achievability Proof of Theorem~\ref{Achievability-Converse})} \label{Achievability}
We recall that we wish to design the decomposition $\mathbf{DE} = \mathbf{F}$, for fixed $N, K, T, \Delta, \Gamma,$ that will guarantee the reconstruction of all requested function outputs, with each of the $N$ servers locally calculating up to $\Gamma$ subfunctions, and each engaging in communication with at most $\Delta$ users over the entirety of $T$ broadcast shots. 

In what follows, we will explain the tile design, elaborating on the positioning of these tiles in $\mathbf{D}$ and $\mathbf{E}$, as well as defining how these tiles are filled as a function of the SVD decomposition of carefully selected submatrices of $\mathbf{F}$.  This is done both for the single-shot as well as the multi-shot scenarios. Our design naturally abides by the number of servers available. In particular, the position and size of the tiles of $\mathbf{D}$ and $\mathbf{E}$ will be crucial, because they will determine the rank of corresponding submatrices of $\mathbf{DE}$, where in particular the sum of these ranks must be bounded by $NT$.   
Before proceeding with the scheme, we also note that we have Examples~\ref{Non-devidable-example} and \ref{multi-shot-example} that illustrate the design of our scheme for the more challenging case of $\Delta \nmid K, \Gamma \nmid L, T=1$ as well as of $\Delta | K, \Gamma | L, T>1$.

\subsection{Construction of $\mathbf{D,E}$}\label{Construction-DE}
 The construction will involve the following steps: a) Sizing and positioning the tiles of $\mathbf{D},\mathbf{E}$ and $\mathbf{DE}$, b) Filling the non-zero tiles in $\mathbf{D}\mathbf{E}$ as a function of $\mathbf{F}$, and finally c) Filling the tiles in $\mathbf{D}$ and $\mathbf{E}$. We elaborate on these steps below.

\paragraph{First step: Sizing and positioning the tiles of $\mathbf{D},\mathbf{E}$ and of $\mathbf{DE}$}
We first partition the set of equivalent classes $\mathcal{C}$ (cf. Definition~\ref{Def2}) into the following subsets of equivalence classes 
\begin{align}
    \mathcal{C}_{1} &\triangleq \{\mathcal{P}_{i,j}\: | \:\mathbf{c}_{\mathcal{P}_{i,j}}=  [\mathbf{0}_{(i-1)\Delta}^{\intercal}, \mathbf{1}_\Delta^{\intercal}, \mathbf{0}_{K- i \Delta}^{\intercal}]^{\intercal}, \mathbf{r}_{\mathcal{P}_{i,j}} =[\mathbf{0}_{(j-1)\Gamma}^{\intercal},  \mathbf{1}_{\Gamma}^{\intercal}, \mathbf{0}_{L- j \Gamma}^{\intercal}],(i,j) \in [\lfloor \frac{K}{\Delta}\rfloor] \times [\lfloor \frac{L}{\Gamma} \rfloor] \},\label{c-1}\\
    \mathcal{C}_{2} &\triangleq\{\mathcal{P}_{i,*} \:|\: \mathbf{c}_{\mathcal{P}_{i,*}}=  [\mathbf{0}_{(i-1)\Delta}^{\intercal}, \mathbf{1}_\Delta^{\intercal}, \mathbf{0}_{K- i \Delta}^{\intercal}]^{\intercal},
    \mathbf{r}_{\mathcal{P}_{i,*}} =[\mathbf{0}^{\intercal}_{L - \mod(L, \Gamma)},\mathbf{1}^{\intercal}_{\mod(L,\Gamma)}],i \in [\lfloor \frac{K}{\Delta}\rfloor] \}, \label{c-2} \\
  \mathcal{C}_{3} &\triangleq\{\mathcal{P}_{*,j} \:|\: \mathbf{c}_{\mathcal{P}_{*,j}} =  [\mathbf{0}^{\intercal}_{K - \mod(K,\Delta)} , \mathbf{1}^{\intercal}_{\mod(K, \Delta)}]^{\intercal}, \mathbf{r}_{\mathcal{P}_{*,j}} =[\mathbf{0}_{(j-1)\Gamma}^{\intercal},  \mathbf{1}_{\Gamma}^{\intercal}, \mathbf{0}_{L- j \Gamma}^{\intercal}],j \in [\lfloor \frac{L}{\Gamma} \rfloor] \},\label{c-3}\\
    \mathcal{C}_{4} &\triangleq  \{ \mathcal{P}\:|\: \mathbf{c}_{\mathcal{P}}=  [\mathbf{0}_{K - \mod(K,\Delta)}^{\intercal} , \mathbf{1}_{\mod(K, \Delta)}^{\intercal}]^{\intercal}, \ \mathbf{r}_{\mathcal{P}}^{\intercal} =[\mathbf{0}^{\intercal}_{L - \mod(L, \Gamma)},\mathbf{1}^{\intercal}_{\mod(L,\Gamma)}]\} \label{c-4} 
\end{align}
where $\mathbf{c}_{\mathcal{P}_{i,j}}, \mathbf{c}_{\mathcal{P}_{i,*}}, \mathbf{c}_{\mathcal{P}_{*,j}}, \mathbf{c}_{\mathcal{P}}$ are the corresponding component columns (as defined in Definition~\ref{Def2b}) of $\mathcal{C}_{1},\mathcal{C}_{2},\mathcal{C}_{3},\mathcal{C}_{4}$ respectively. Similarly $\mathbf{r}_{\mathcal{P}_{i,j}}, \mathbf{r}_{\mathcal{P}_{i,*}}, \mathbf{r}_{\mathcal{P}_{*,j}}, \mathbf{r}_{\mathcal{P}}$ are the corresponding component rows, from the same definition. These describe the exact position and size of each tile of $\mathbf{D}$, of $\mathbf{E}$, and thus automatically of each tile of $\mathbf{DE}$.

We now note that the number of classes in each subset $\mathcal{C}_i, \ i=1,2,3,4$, is equal to
\begin{align}
    |\mathcal{C}_1|=  \lfloor \frac{K}{\Delta} \rfloor \lfloor \frac{L}{\Gamma} \rfloor  , \: |\mathcal{C}_2| = \lfloor \frac{K}{\Delta} \rfloor , \: |\mathcal{C}_3| =\lfloor \frac{L}{\Gamma} \rfloor,\: 
    |\mathcal{C}_4|=1\label{rp-eq}
\end{align}
while we also note from~\eqref{max-rank} and Definition~\ref{maxrank}, that the maximum rank of each representative support (i.e. of each tile of $\mathbf{DE}$) takes the form
\begin{equation} \label{max-rank-formula}
    r_{\mathcal{P}}=
    \begin{cases}
      \min(\Delta, \Gamma), &  \mathcal{P} \in \mathcal{C}_1, \\
      \min(\mod(K,\Delta), \Gamma ), &  \mathcal{P} \in \mathcal{C}_2 ,\\
      \min(\mod(L, \Gamma) ,\Delta ), &  \mathcal{P} \in \mathcal{C}_3, \\
      \min(\mod(K,\Delta), \mod(L,\Gamma)), &  \mathcal{P} \in \mathcal{C}_4. \\
    \end{cases}
  \end{equation}
The above information will be essential in enumerating our equivalence classes and associating each such class to a collection of servers.

\paragraph{Second step: Filling the non-zero tiles in $\mathbf{D}\mathbf{E}$ as a function of $\mathbf{F}$}
Recall that we have a tile $\mathbf{S}_{\mathcal{P}}(\mathcal{R}_{\mathcal{P}}, \mathcal{C}_{\mathcal{P}})$ corresponding to the non-zero elements of $\mathbf{S}_{\mathcal{P}} $. This tile is now empty, in the sense that the non-zero entries are all equal to $1$. To fill this tile, we consider 
  \begin{align}
      \mathbf{F}_\mathcal{P} \triangleq (\mathbf{F} \odot  \mathbf{S}_{\mathcal{P}}) (\mathcal{R}_{\mathcal{P}}, \mathcal{C}_{\mathcal{P}}), \: \forall \mathcal{P} \in \cup^{4}_{i=1} \mathcal{C}_i \label{Formation-of-submatrices}
  \end{align}
where this filled tile, in our current lossless case, is simply the submatrix of $\mathbf{F}$ at the position defined by the non-zero elements of $\mathbf{S}_{\mathcal{P}}$. The schematic illustration in Figure~\ref{equivalence-classes} of Example~\ref{Non-devidable-example}, may help clarify the above.

\paragraph{Third step: Filling the tiles in $\mathbf{D}$ and $\mathbf{E}$} 
We now SVD-decompose each $\mathbf{F}_\mathcal{P}$, as 
\begin{align}
      \mathbf{F}_\mathcal{P} = \mathbf{D}_{\mathcal{P}} \mathbf{E}_{\mathcal{P}} \label{sub-SVD}
  \end{align}
  where $\mathbf{D}_{\mathcal{P}} \in  \mathbb{R}^{|\mathcal{R}_{\mathcal{P}}| \times r_{\mathcal{P}}}, \mathbf{E}_{\mathcal{P}} \in \mathbb{R}^{r_{\mathcal{P}} \times |\mathcal{C}_{\mathcal{P}}|} $.  
  Going back to \eqref{Complete-SVD}, our matrices $\mathbf{F}_{\mathcal{P}},\mathbf{D}_{\mathcal{P}},\mathbf{E}_{\mathcal{P}}$ are associated to $\mathbf{A}$, $ \mathbf{U} \cdot\mathbf{S}$ and $ \mathbf{V}$ respectively, and all correspond to complete SVD decompositions.   
 For a visual representation of this second step, we provide Figure~\ref{equivalence-classes} of Example \ref{Non-devidable-example}.

Let us now enumerate our classes (i.e., our tiles) as follows
\begin{align}\label{eq:tileEnumeration}
\cup^{4}_{i=1} \mathcal{C}_i=\{ \mathcal{P}_1, \mathcal{P}_2, \hdots, \mathcal{P}_{m}\}, \: m \in \mathbb{N}.\end{align}
Now let us position each tile of $\mathbf{D}$ as follows
\begin{align}\label{eq:tilePositionInD}
    \mathcal{R}_{\mathcal{P}_j},\:  [\sum^{j-1}_{i=1} T \lceil \frac{r_{\mathcal{P}_i}}{T} \rceil
+1: \sum^{j}_{i=1} T \lceil \frac{r_{\mathcal{P}_i}}{T} \rceil],\: \forall \mathcal{P}_j \in \cup^{4}_{i=1} \mathcal{C}_i 
\end{align} 
and each tile of $\mathbf{E}$ as follows
\begin{align} \label{eq:tilePositionInE}
[\sum^{j-1}_{i=1} T \lceil \frac{r_{\mathcal{P}_i}}{T} \rceil
+1: \sum^{j}_{i=1} T \lceil \frac{r_{\mathcal{P}_i}}{T} \rceil],\: \mathcal{C}_{\mathcal{P}_j}, \: \forall \mathcal{P}_j \in \cup^{4}_{i=1} \mathcal{C}_i
\end{align}
where the above describes the indices of columns and rows where each tile resides. 

At this point, we note that for the single shot case of $T=1$, the above yields
\begin{align}
    \mathbf{D}(\mathcal{R}_{\mathcal{P}_j}, [\sum^{j-1}_{i=1} r_{\mathcal{P}_i} +1,\sum^{j}_{i=1} r_{\mathcal{P}_i} ]) = \mathbf{D}_{\mathcal{P}_j}
    \label{Decoding-Encoding1}
\end{align}
and 
\begin{align}
    \mathbf{E}([\sum^{j-1}_{i=1} r_{\mathcal{P}_i} +1,\sum^{j}_{i=1} r_{\mathcal{P}_i} ], \mathcal{C}_{\mathcal{P}_j})=\mathbf{E}_{\mathcal{P}_j}\label{Decoding-Encoding2}
\end{align}
while naturally the remaining non-assigned elements of $\mathbf{D}$ and $\mathbf{E}$ are zero.  This step can also be visualized in Figure~\ref{equivalence-classes} which illustrates this step as it applies to Example~\ref{Non-devidable-example}.

We recall from Section \ref{Formulating} (cf.~\eqref{EncodingMatrix},~\eqref{DecodingMatrix}), that each server $n \in [N]$ corresponds to a column and row index of $\mathbf{D}$ and $\mathbf{E}$ respectively. To make the connection between $N$ and the parameters employed in our scheme,
let us recall our tiles $\mathcal{P} \in \cup_{i \in [4]} \mathcal{C}_i$ and the corresponding $\mathbf{D}_\mathcal{P}$ as seen in~\eqref{sub-SVD}, and let us recall, as described in~\eqref{Decoding-Encoding1}, that all such $\mathbf{D}_\mathcal{P}$  occupy $\sum^{m}_{i=1} r_{\mathcal{P}_{i}}$ columns in total. Combining this information with the value of $r_{\mathcal{P}}$ in~\eqref{max-rank-formula}, yields that
\begin{align}
    N_{} = \sum_{i \in [4]} \sum_{\mathcal{P} \in \mathcal{C}_i} r_{\mathcal{P}} |\mathcal{C}_i| &=
    \min(\Delta,\Gamma) \lfloor \frac{K}{\Delta} \rfloor \lfloor \frac{L}{\Gamma} \rfloor + \min(\mod(K,\Delta), {\Gamma})  \lfloor \frac{L}{\Gamma} \rfloor \nonumber \\&+\min(\mod(L,\Gamma), {\Delta}) \lfloor \frac{K}{\Delta}\rfloor + \min(\mod(K,\Delta), \mod(L, \Gamma)) \label{result-2-f}
\end{align}
which proves Theorem~\ref{Achievability-Converse} for the case of $T=1$.  An additional clarifying illustration for the single shot case can be found in Example~\ref{Non-devidable-example}.

For the case of $T>1$, the difference is in the third step, where we now replace~\eqref{Decoding-Encoding1} and \eqref{Decoding-Encoding2}, where by accounting for~\eqref{eq:tilePositionInE}, we see that the tiles of $\mathbf{D}$ and $\mathbf{E}$ respectively take the form
    \begin{align}
\mathbf{D}_{\mathcal{P}_j} =     \mathbf{D}(\mathcal{R}_{\mathcal{P}_j}, [\sum^{j-1}_{i=1} T \lceil \frac{r_{\mathcal{P}_i}}{T} \rceil
+1: \sum^{j}_{i=1} T \lceil \frac{r_{\mathcal{P}_i}}{T} \rceil]) 
    \label{Decoding-EncodingT1}
\end{align}
and 
\begin{align}
\mathbf{E}_{\mathcal{P}_j} =     \mathbf{E}([\sum^{j-1}_{i=1} T \lceil \frac{r_{\mathcal{P}_i}}{T} \rceil
+1: \sum^{j}_{i=1} T \lceil \frac{r_{\mathcal{P}_i}}{T} \rceil], \mathcal{C}_{\mathcal{P}_j}).\label{Decoding-EncodingT2}
\end{align}
One aspect in our design that distinguishes the multi-shot from single-shot case, regards the association of tiles to servers. An additional level of complexity here has to do with what one might describe as an ``accumulation of rank" at the different servers. As we have realized, there is an association between tiles and servers. In the case of $T=1$, a single tile $\mathcal{P}_i$ of rank $r_{\mathcal{P}_i}$, could be associated with $r_{\mathcal{P}_i}$ servers, each contributing to a single rank. If the number of servers associated with a tile reached the rank of that tile, then that tile could be decomposed in a lossless manner. Now though, as one can imagine, having multiple shots (corresponding to $T>1$) can allow a single server to span more than one dimension, i.e., to contribute to more than one rank inside that tile. For arbitrary $T$ and $r_{\mathcal{P}_i}$ though, one can imagine the possibility of having underutilized servers. Imagine for example a tile with rank $r_{{\mathcal{P}}_{i}}$, which could fully utilize $\lfloor \frac{r_{\mathcal{P}_i}}{T} \rfloor$ servers, leaving the tile with an accumulated rank that is $\mod({r_{\mathcal{P}_i},T})$ smaller than its desired $r_{\mathcal{P}_i}$, and leaving us also with a server that is underutilized (in terms of accumulating rank) in the decomposition of the particular tile. One might consider as a remedy, the possibility of associating the underutilized server with an additional tile. This though, runs the risk of violating the communication and computation constraints, simply because this additional (second) tile of $\mathbf{DE}$ may correspond to a tile of $\mathbf{D}$ and $\mathbf{E}$, whose union with the aforementioned (first) tile of $\mathbf{D}$ or $\mathbf{E}$ associated to this same underutilized server, may have a number of non-zero elements that exceeds the number allowed by the communication and computation constraints. The scheme that we present accounts for this, and provably maintains a small and bounded under-utilization of our servers compared to the optimal case. 

With this in mind, we associate $\lceil \frac{r_{\mathcal{P}}}{T}\rceil$ servers to each equivalence class where, continuing as before, we can evaluate ${r}_{\mathcal{P}}$ using~\eqref{max-rank-formula} for any $\mathcal{P} \in \mathcal{C}_j, j \in [4]$. This yields automatically 
\begin{align}
    N &= \lceil\frac{\min(\Delta, \Gamma) }{T} \rceil \lfloor \frac{K}{\Delta}\rfloor  \lfloor \frac{L}{\Gamma}\rfloor + 
    \lceil\frac{\min(\textrm{{\normalfont mod}}(K,\Delta), {\Gamma} )}{T}\rceil  \lfloor \frac{L}{\Gamma} \rfloor \nonumber\\ &+ 
    \lceil\frac{\min(\mod(L,\Gamma) ,  {\Delta})}{T}\rceil\lfloor \frac{K}{\Delta} \rfloor + \lceil\frac{ \min(\mod(K,\Delta), \mod(L, \Gamma))}{T}\rceil.\label{end-formula-theorem1}
\end{align}
To complete the achievability proof of Theorem~\ref{Achievability-Converse}, we proceed to evaluate \eqref{Zero-error-capacity}, which gives our capacity to be
      \begin{align*}
            C_{0} & \overset{(a)}{=}\frac{K}{N_{}} \overset{(b)}{=} \frac{ \Delta \Gamma }{\lceil\frac{\min(\Delta, \Gamma) }{T} \rceil L }
            \overset{(c)}=   \begin{cases}
     (T/L) ({\Delta \Gamma}/{\min(\Delta,\Gamma)}),\:\:& \text{if } T\: |\: \min(\Delta, \Gamma), \\
     (\Delta \Gamma)/ L,\:\: & \text{if } T > \min(\Delta, \Gamma)\:\:\:
    \end{cases}
        \end{align*}
        where $(a)$ follows by definition, $(b)$ follows from Theorem~\ref{Achievability-Converse}, and $(c)$ follows after basic algebraic manipulations where we see that if $T | \min(\Delta, \Gamma)$ then $\lceil\frac{\min(\Delta, \Gamma) }{T} \rceil = \frac{\min(\Delta, \Gamma) }{T} $, while if $T> \min(\Delta,\Gamma)$ then $\lceil\frac{\min(\Delta, \Gamma) }{T} \rceil =1$. The proof is then completed directly after applying \eqref{Nonasymptotic-constatnts} and \eqref{eq:zetaKappa} to evaluate the RHS of the last equality.

\subsection{Examples}
We now present two small examples, first for the single-shot and then for the multi-shot case, which can help us better understand the concepts of representative  supports (tiles), and the tessellation pattern that yields $\mathbf{D}$ and $\mathbf{E}$. 
\begin{example}\label{Non-devidable-example}
 Consider a single-shot scenario with $K=7$ users, $L=11$ subfunctions (thus corresponding to a demand matrix $\mathbf{F} \in \mathbb{R}^{7 \times 11}$), under the constraint $\Delta =3$ and $\Gamma =5$.
 Let us go through the design steps described above.
    \begin{figure}
       \centering
      \[ \begin{array}{cc} \hspace{-10pt}\includegraphics[scale=0.5]{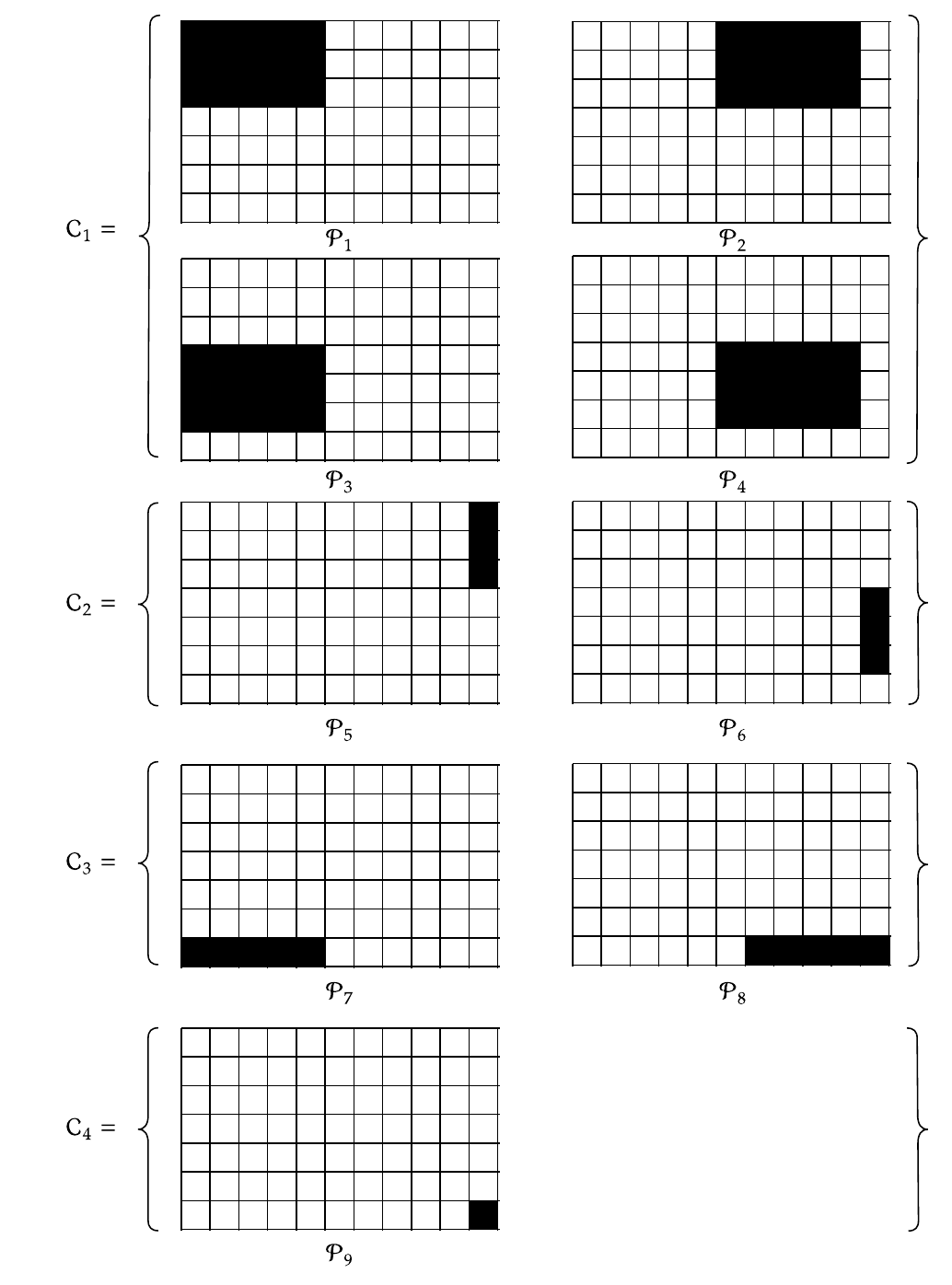} & \hspace{-120pt} \includegraphics{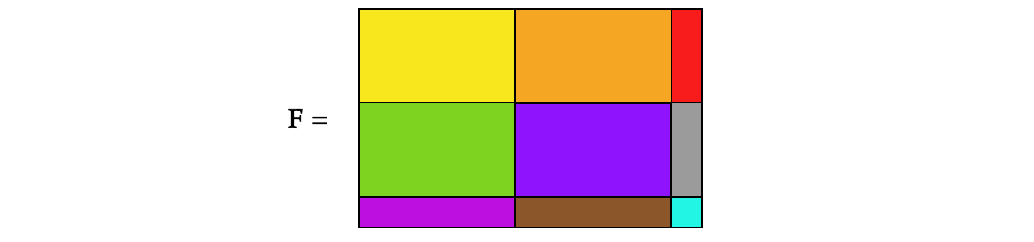}[scale=0.2] \end{array} \]
       \caption{Corresponding to Example \ref{Non-devidable-example}, the figure on the left represents in black the families of the equivalent classes (cf.~\eqref{c-1}--\eqref{c-4}). The $9$ equivalent classes (right) cover the entire $\mathbf{F}$.} \label{equivalence-classes}
   \end{figure}
\begin{itemize}
\item \emph{First step --- Sizing and positioning the tiles of $\mathbf{D},\mathbf{E}$ and of $\mathbf{DE}$:}  
The positions of the tiles are derived according to \eqref{c-1}--\eqref{c-4}, yielding the corresponding tessellation pattern illustrated in Figure~\ref{equivalence-classes}. As we can see, the pattern entails four tile families $\mathcal{C}_1,\mathcal{C}_2,\mathcal{C}_3,\mathcal{C}_4$, of respective sizes $3\times 5, 3\times 1, 1\times 5 $ and $1\times 1$. Each family has the following number of tiles
$
    |\mathcal{C}_1|=  \lfloor \frac{K}{\Delta} \rfloor \lfloor \frac{L}{\Gamma} \rfloor = 2 \times 2 =4 , \: |\mathcal{C}_2| = \lfloor \frac{K}{\Delta} \rfloor =2 , \: |\mathcal{C}_3| =\lfloor \frac{L}{\Gamma} \rfloor =2,\: 
    |\mathcal{C}_4|=1$, and the tiles have a maximum rank (cf. Definition~\ref{maxrank}) equal to $r_{\mathcal{P}}= 3$ for $\mathcal{P} \in \mathcal{C}_1$, and $r_{\mathcal{P}}= 1$ for the rest. 
Figure~\ref{equivalence-classes} also illustrates how the designed tessellation pattern successfully covers $\mathbf{F}$, which --- in the lossless case --- is a necessary condition. 

\item \emph{Second step --- Filling the non-zero tiles in $\mathbf{D}\mathbf{E}$:}  The master node extracts the submatrices corresponding to each of the tiles as described in~\eqref{Formation-of-submatrices}, and we have now matrices $\mathbf{F}_\mathcal{P} = (\mathbf{F} \odot  \mathbf{S}_{\mathcal{P}}) (\mathcal{R}_{\mathcal{P}}, \mathcal{C}_{\mathcal{P}})$, which tell us how the tiles of $\mathbf{F}$ are filled.

\item \emph{Third step --- Filling the tiles in $\mathbf{D}$ and $\mathbf{E}$:} In this final step, the master node proceeds to perform complete SVD decompositions for all matrices $\mathbf{F}_\mathcal{P}$ above, where each SVD decomposition takes the form $\mathbf{F}_\mathcal{P} = \mathbf{D}_\mathcal{P}  \mathbf{E}_\mathcal{P} $, thus yielding all $\mathbf{D}_{\mathcal{P}}$ and $\mathbf{E}_{\mathcal{P}},\: \forall \mathcal{P} \in  \mathcal{C} $, as described in \eqref{sub-SVD}. Note that there are four different types of SVD decompositions, depending on whether $\mathcal{P}$ comes from $\mathcal{C}_1 , \mathcal{C}_2, \mathcal{C}_3 $ or $\mathcal{C}_4$.
These filled tiles are placed inside $\mathbf{D}$ and $\mathbf{E}$ respectively (as illustrated in Figure~\ref{Blockformation5}), in accordance to the positioning steps in~\eqref{Decoding-EncodingT1} and \eqref{Decoding-EncodingT2}.  

\begin{figure} 
    \centering
    \includegraphics[scale=0.65]{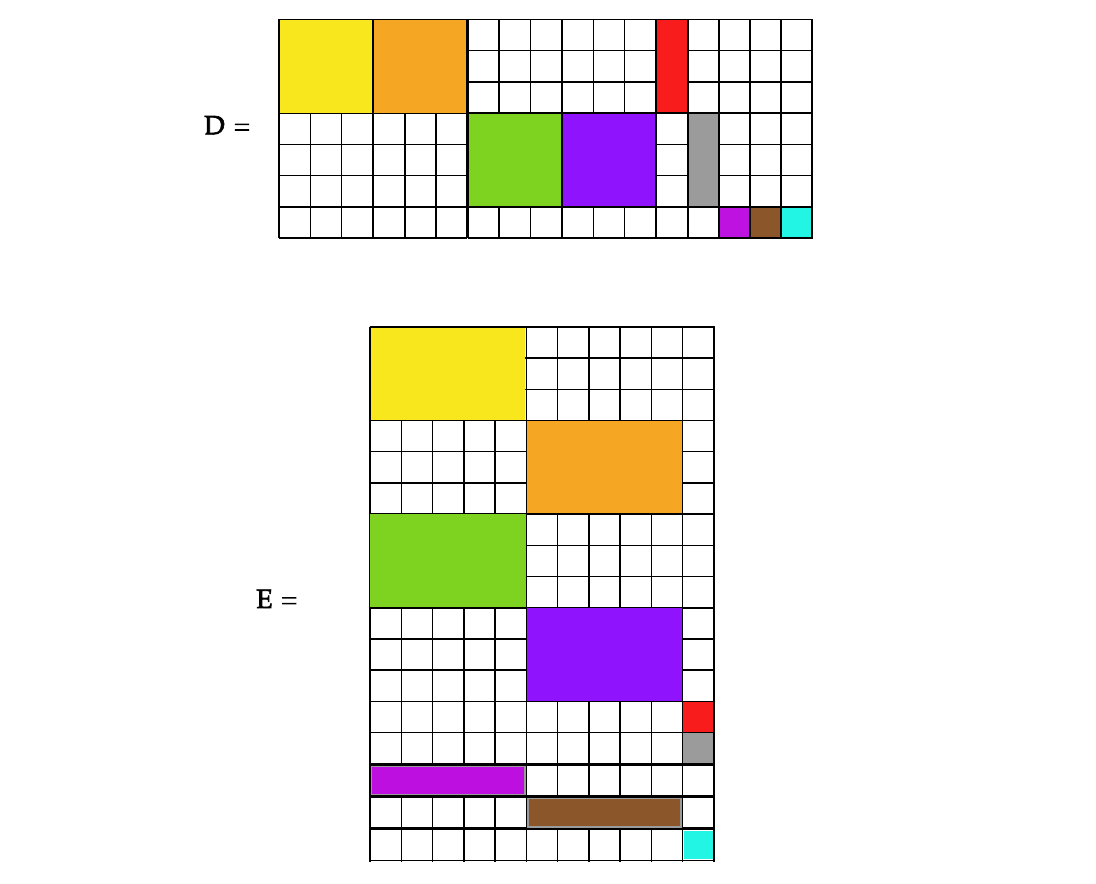}
    \caption{Creating our communication and computing matrices $\mathbf{D}, \mathbf{E}$ and  applying the coordinates given in~\eqref{eq:tilePositionInD}--\eqref{Decoding-Encoding2}.}\label{Blockformation5}
\end{figure}
At this point, we calculate the number of used servers (cf.~\eqref{Non-devidable-example}) to be $N = 3 \times 4 + 1 \time 1 + 1\times 2 + 1 \times 2  = 17$.
\end{itemize}
\end{example}

We proceed with an additional example, now for the multi-shot case. 
\begin{example}\label{multi-shot-example}
{ We consider the same setting as in Example~\ref{single-shot-example-simple}, with $K=6$ users, $L=10$ subfunctions, a communication cost $\Delta =3$ and a computation cost of $\Gamma =5$, but where now we consider each server to be able to communicate to $T=2$ different sets of users, in two respective shots (one shot to one group of users, and another one to a potentially different group of users). Let us go through the steps described in our section above.
\begin{itemize}
\item \emph{First step --- Sizing and positioning the tiles of $\mathbf{D},\mathbf{E}$ and of $\mathbf{DE}$:}  
The sizes and locations of the tiles of $\mathbf{D},\mathbf{E}$ and $\mathbf{DE}$, are described by the tessellation pattern given directly from \eqref{c-1}--\eqref{c-4}, as also illustrated on the left side of Figure~\ref{ex1}. This pattern, in our simpler case here, entails tiles only from $\mathcal{C}_1$, where we find a total of $ |\mathcal{C}_1|=  \lfloor \frac{K}{\Delta} \rfloor \lfloor \frac{L}{\Gamma} \rfloor = 2 \times 2 =4$ tiles, each of size $3\times 5$, and each of maximum rank $r_{\mathcal{P}}= 3$ (cf.~Definition~\ref{maxrank}).
   We again note that the pattern successfully covers $\mathbf{F}$.
\item \emph{Second step --- Filling the non-zero tiles in $\mathbf{D}\mathbf{E}$:}  Exactly as in the single-shot case, again here the master node extracts the submatrices corresponding to each of the tiles as described in~\eqref{Formation-of-submatrices}. This yields all the $\mathbf{F}_\mathcal{P}$ which are simply the  filled tiles of $\mathbf{F}$.
\item \emph{Third step --- Placing the filled cropped tiles $\mathbf{D}_{\mathcal{P}}$ and $\mathbf{E}_{\mathcal{P}}$ in $\mathbf{D}$ and $\mathbf{E}$:} Finally, the master SVD decomposes each $\mathbf{F}_\mathcal{P}$ as $\mathbf{F}_\mathcal{P} = \mathbf{D}_\mathcal{P}  \mathbf{E}_\mathcal{P} $, which provides the required $\mathbf{D}_{\mathcal{P}}$ and $\mathbf{E}_{\mathcal{P}},\: \forall \mathcal{P} \in  \mathcal{C} $, as described in \eqref{sub-SVD}. Finally, these tiles are placed in $\mathbf{D}$ and $\mathbf{E}$ respectively, in direct accordance to the coordinates described in~\eqref{eq:tilePositionInD}--\eqref{Decoding-Encoding2}. This last part is illustrated in Figure~\ref{ex22}.  
\end{itemize}
For this setting, applying directly \eqref{end-formula-theorem1} tells us that we need $N = 8$ servers. We also observe that while the tiles of $\mathbf{F}$ remain the same as in the corresponding single shot case of Example\ref{single-shot-example-simple} (as also illustrated in Fig~\ref{ex1}), indeed our tiles of $\mathbf{D}$ and $\mathbf{E}$ change\footnote{We note here that if we had forced $N$ down to $N=6$ as in Example~\ref{single-shot-example-simple} (corresponding to the tiling in Figure~\ref{ex1}), then some servers would violate the communication and computation cost constraints. For example, server $2$ would have been forced to support a communication cost of $6$ ($6$ links to different users), since the union of supports of the third and fourth column of $\mathbf{D}$ would have size $6$. Similarly the corresponding $\mathbf{E}$ would entail a computation cost of $10$ ($10$ subfunctions locally calculated), since the size of the union of the supports of the third and the fourth columns of $\mathbf{E}$ would have been $10$.}, as illustrated in Figure~\ref{ex22}. The same figure also illustrates how each two consecutive columns of $\mathbf{D}$ and two consecutive rows of $\mathbf{E}$, correspond to a server.  Note that although the rank of each tile is at most $r_{\mathcal{P}}=3$, the number of servers associated with each tile is $\lceil\frac{r_{\mathcal{P}}}{T}\rceil=\lceil\frac{3}{2}\rceil = 2$. This reflects the fact that the second shot of the even-numbered servers remains unused, to avoid having a server be associated to two tiles, and thus to avoid violating the computation and communication constraints. 
\begin{figure}
    \centering
    \includegraphics[scale=0.65]{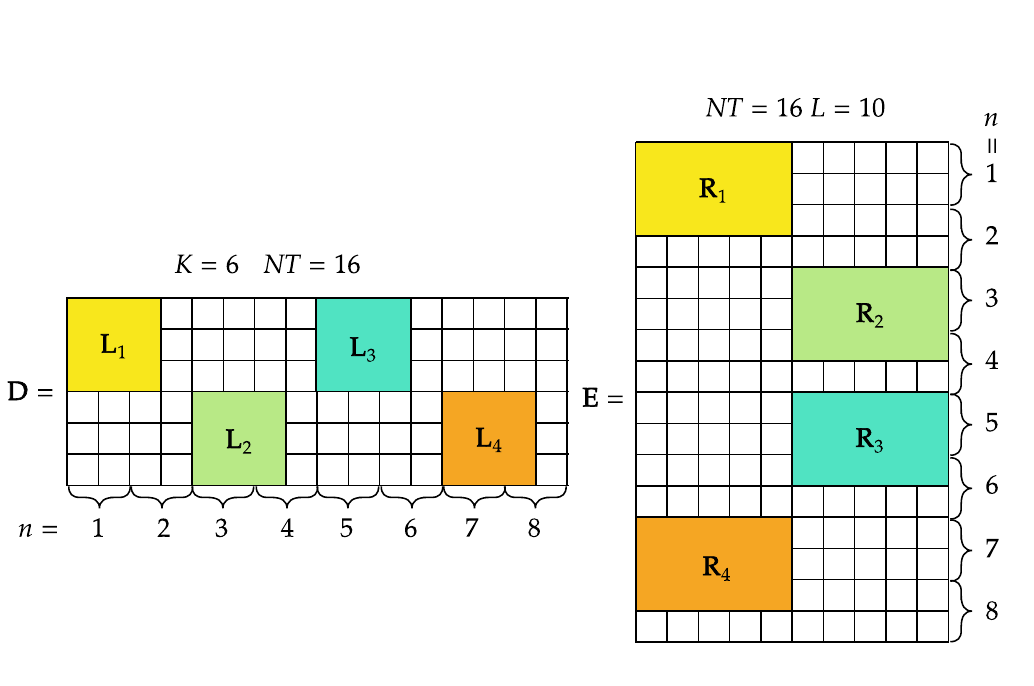}
    \caption{Corresponding to Example~\ref{multi-shot-example}, this figure illustrates the tiling  of $\mathbf{D}$ and $\mathbf{E}$ respectively with $\mathbf{D}_{\mathcal{P}} \in \mathbb{R}^{3 \times 3}$ and $\mathbf{E}_{\mathcal{P}} \in \mathbb{R}^{3 \times 5}$, for $\mathcal{P} \in \{P_1,P_2,P_3,P_4\}$. Guaranteeing $\mathbf{D}_{\mathcal{P}}  \mathbf{E}_{\mathcal{P}} = \mathbf{F}_{\mathcal{P}} \in \mathbb{R}^{3 \times 5}, \forall \mathcal{P} \in \{P_1,P_2,P_3,P_4\}$, in turn guarantees lossless function reconstruction.  We can see that the union of the supports of $T=2$ consecutive columns of $\mathbf{D}$ includes at most $\Delta=3$ non-zero elements thus guaranteeing the communication constraint. We also see that the union of the supports of $T=2$ consecutive rows of $\mathbf{E}$ includes at most $\Gamma=5$ non-zero elements thus guaranteeing the computation constraint.
    }
    \label{ex22}
\end{figure}}
\end{example}

\section{Appendix:Proof of The Converse for Theorem \ref{Achievability-Converse}}\label{Converse}
The converse that we provide here will prove that the scheme for the single shot case is exactly optimal when $\Gamma \geq  \Delta, \Gamma | L, T | \Delta$ or $\Delta \geq \Gamma, \Delta |K, T|\Gamma$ and also the scheme for the multi-shot case is optimal when  $T\geq \min(\Delta,\Gamma)$.
We begin with three lemmas that will be useful later on. 
First, Lemma~\ref{Server-Tile} will state the necessity of having a one-to-one correspondence between each rank-one contribution support\footnote{Recall that the $n$th rank-one contribution support takes the form $\mathbf{S}_{n}(\mathbf{I}, \mathbf{J}) = \mathbf{I}(:,n) \mathbf{J}(n, :)^{}$.} and each server, then Lemma~\ref{Coverage-nessecity} will state the necessity of having a tessellation pattern that covers the whole area of $\mathbf{F}$, and then Lemma~\ref{size-of-tiles} will elaborate more on size limits of each tile as a consequence of the communication and computation constraints. Lemma~\ref{Disjoint-Support-Assumption: Equivalence} then  establishes the conceptual equivalence between Definition~\ref{disjointSupportAssumption} and disjoint tiles. Lemma~\ref{Relation-rank-sub-tile}, lower bounds the number of tiles is each equivalence class by the number of subtiles (cf. Definition~\ref{Relation-rank-sub-tile}) corresponding to a tile. Subsequently, Lemma~\ref{minim-r}, gives the necessary number of subtiles, so that a covering scheme can be constructed, then via combining the last two of the aforementioned lemmas, we give a lower bound on the number of rank-one contribution supports for any covering scheme with proper sizes, and then using Lemma~\ref{Server-Tile}, we finalize our converse by giving a lower-bound on the number of servers. 

When focusing on the general case of $T>1$ in Section~\ref{multi-shot-convers}, we will substitute Lemma~\ref{Server-Tile} with Lemma~\ref{Server-Tile2} to eventually show how each server's transmission (to one set of users) must correspond to a rank-one contribution support. Then using the same argument as in the single-shot case, we lower bound the necessary number of servers by the number $\frac{KL}{T\max(\Gamma ,\Delta)}$. Then the lower bound is further tightened for the cases where $T \geq  \min(\Delta, \Gamma)$, since then each tile corresponds to one server which will enable us, using Lemma~\ref{minim-tiles}, to obtain a new lower bound of  $\lceil \frac{K}{\Delta}\rceil \lceil \frac{L}{\Gamma} \rceil$, which matches the value obtained by the achievable scheme.
\subsection{Converse for The Single-Shot Case of $T=1$}\label{single-shot-convers}

 Before presenting the lemmas, let us recall that rank-one contribution supports were defined in Definition~\ref{def-r1}, that representative supports (i.e. tiles) were defined in Definition~\ref{Def2b}, as well as let us recall from the same definition that the collection of all classes is represented by $\mathcal{C}$.

The following lemma, while stating the obvious, will be useful in associating the number of servers to the number of rank-one contribution supports.
\begin{Lemma}\label{Server-Tile}
    For any $\mathbf{D,E}$ with respective supports $\mathbf{I} = \supp(\mathbf{D}) \in \{0,1\}^{K \times N} $ and $\mathbf{J} = \supp(\mathbf{E}) \in \{0,1\}^{N \times L}$, there exists a one-to-one mapping between the server indices $n\in[N]$ and  rank-one contribution supports $\mathbf{S}_{n}(\mathbf{I}, \mathbf{J})$.  
\end{Lemma}
\begin{proof}
    The proof is direct by first recalling Definition~\ref{def-r1} which, for any $n  \in [N]$, says that $\mathbf{S}_{n}(\mathbf{I}, \mathbf{J}) = \mathbf{I}(:,n) \mathbf{J}(n, :)$, and then by recalling from~\eqref{encoding-vectors-per-shot},\eqref{decoding-vectors-per-shot-1},\eqref{EncodingMatrix} and \eqref{DecodingMatrix} that, in the single-shot setting, each server $n \in \mathbb{N}$ corresponds to the $n$th column of $\mathbf{D}$ (itself corresponding to $\mathbf{I}(:,n)$) and the $n$th row of $\mathbf{E}$ (corresponding to $\mathbf{J}(n, :)$).
\end{proof}
We proceed with the next lemma, which simply says that every element of $\mathbf{F}$ must belong to at least one representative support (tile). 

\begin{Lemma}\label{Coverage-nessecity} In any lossless function reconstruction scheme corresponding to $\mathbf{DE}=\mathbf{F}$, for each $(i,j) \in [K]\times [L]$, there exists a class $\exists \mathcal{P} \in \mathcal{C}$ such that $(i,j) \in \mathcal{R}_{\mathcal{P}} \times \mathcal{C}_{\mathcal{P}}$.
\end{Lemma}
\begin{proof}
    The lemma aims to prove that there exists no element $\mathbf{F}(i,j)$ of $\mathbf{F}$ that has not been mapped to a tile $\mathbf{F}_\mathcal{P}$.  We will prove that for each $(i,j) \in [K]\times [L]$ then $\exists \mathcal{P} \in \mathcal{C} \ : \ (i,j) \in \mathcal{R}_{\mathcal{P}} \times \mathcal{C}_{\mathcal{P}}  $ and we will do so by contradiction. Let us thus assume that there exists $(i,j) \in [K]\times [L]$ such that $\nexists \mathcal{P} \in \mathcal{C} \ : \ (i,j) \in \mathcal{R}_{\mathcal{P}} \times \mathcal{C}_{\mathcal{P}}  $, which would in turn imply --- directly from \eqref{Decoding-Encoding1}, \eqref{Decoding-Encoding2} ---  that $\mathbf{DE}(i,j) = 0$ as well as would imply the aforementioned fact that the non-assigned (by the process in \eqref{Decoding-Encoding1}, \eqref{Decoding-Encoding2}) elements of $\mathbf{D}$ and $\mathbf{E}$, are zero.
    Now we see that 
    \begin{align}
       \|(\mathbf{D} \mathbf{E} - \mathbf{F})\mathbf{w}\|^{2}_2  &=\|  \sum^{K}_{k=1}(\mathbf{D} \mathbf{E} - \mathbf{F})(k,:) \mathbf{w}\|^{2}_2 \nonumber \\ &= \sum^{K}_{k=1,k\neq i}\sum^{L}_{\ell=1,\ell\neq j}[(\mathbf{D}\mathbf{E} -\mathbf{F})(k,\ell) \mathbf{w}(\ell)]^2 +(\mathbf{D}\mathbf{E} - \mathbf{F})^2(i,j) \mathbf{w}^2(j) \nonumber \\ &+ 2 (\mathbf{D}\mathbf{E}-\mathbf{F})(i,j) \mathbf{w}(j)  \sum^{K}_{k=1,k\neq i}\sum^{L}_{\ell=1,\ell\neq j}(\mathbf{D}\mathbf{E} -\mathbf{F})(k,\ell) \mathbf{w}(\ell)\nonumber \\ &= \sum^{K}_{k=1,k\neq i}\sum^{L}_{\ell=1,\ell\neq j}[(\mathbf{D}\mathbf{E} -\mathbf{F})(\ell,k) \mathbf{w}(\ell)]^2  \nonumber \\ &+ \mathbf{F}^2(i,j)\mathbf{w}^{2}(j)-2 \mathbf{F}(i,j)   \mathbf{w}(j)  \sum^{K}_{k=1,k\neq i}\sum^{L}_{\ell=1,\ell\neq j}(\mathbf{D}\mathbf{E} -\mathbf{F})(\ell,k) \mathbf{w}(\ell). \label{NonCovering}
    \end{align}
    Let us now recall that lossless function reconstruction implies that $\|(\mathbf{D} \mathbf{E} - \mathbf{F})\mathbf{w}\|^{2}_2 = 0 $ for all $\mathbf{F}\in \mathbb{R}^{K\times L}$ and all $\mathbf{w} \in \mathbb{R}^L$. Under the special case of $\mathbf{w}(\ell)=0, \forall \ell \in [L] \backslash \{j\}$ and $\mathbf{w}(j)\mathbf{F}(i,j) \neq 0$, we see --- directly from \eqref{NonCovering} --- that $\|(\mathbf{D} \mathbf{E} - \mathbf{F})\mathbf{w}\|^{2}_2 = \mathbf{F}^2(i,j)\mathbf{w}^{2}(j) \neq 0$, which contradicts the lossless assumption, thus concluding the proof of the lemma.
\end{proof}
The next lemma now limits the sizes of each tile.  
\begin{Lemma}\label{size-of-tiles}
    For any feasible scheme yielding $\mathbf{DE}=\mathbf{F}$, then  each representative support $\mathcal{P} \in \mathcal{C}$ satisfies 
    \begin{align}
        0<\|\mathbf{S}_{\mathcal{P}}(k,:)\|_{0} \leq \Gamma, \forall k \in\mathcal{R}_{\mathcal{P}}\:, \:\:\:
        0<\|\mathbf{S}_{\mathcal{P}}(:,l)\|_{0} \leq \Delta, \forall l \in \mathcal{C}_{\mathcal{P}}\:
    \end{align}
which means that each $\mathbf{S}_{\mathcal{P}}$ can have at most $\Gamma$ non-zero elements in each row and $\Delta$ non-zero elements in each column. 
\end{Lemma}
\begin{proof}
We first recall from Definition~\ref{def-r1} that $\supp(\mathbf{D}) =\mathbf{I}, \supp(\mathbf{E}) = \mathbf{J}$. We also recall that  $
\max_{n \in [N]} |\cup^{T}_{t=1}\supp(\mathbf{D}(:,(n-1)T +t)) | \leq \Delta$ and  $\max_{n \in [N]} |\cup^{T}_{t=1}\text{supp}(\mathbf{E}((n-1)T +t,:))| \leq \Gamma $  (cf. \eqref{Communication-cost-condition},\eqref{Computation-cost-condition}) must hold for any feasible scheme. Hence for all $n \in [NT]$, we have that $\|\mathbf{I}(:,n )\|_{0} \leq \Delta, \|\mathbf{J}(n,: )\|_{0} \leq \Gamma $, and consequently since $\mathbf{S}_{n}=\mathbf{I}(:,n )\mathbf{J}(n,: )$ (cf.~Definition~\ref{def-r1}), we must have that $\|\mathbf{S}_{n}(k,:)\|_{0} \leq \Gamma, \forall k \in [K]\:,
        \|\mathbf{S}_{n}(:,l)\|_{0} \leq \Delta, \forall l \in [L]\: \forall n \in [N]$. Then from Definition~\ref{Def2b}, we see that for all $\mathcal{P} \in \mathcal{C}$, there exists  an $n \in [N]$ such that $\mathbf{S}_{n}= \mathcal{S}_{\mathcal{P}}$. Note that the lower bound follows from Definition~\ref{Def2bb} where $\mathcal{R}_{\mathcal{P}}, \mathcal{C}_{\mathcal{P}}$ are defined.
\end{proof}

Continuing with the main proof,
in order to relate the notion of tiles to the rank-one contribution supports, we need first to define the notion of sub-tiles, where each entry of a sub-tile is a matrix coordinate. We also recall that $\mathcal{R}_{\mathcal{P}}$ and $\mathcal{C}_{\mathcal{P}}$ are respectively the row and column indices of tile $\mathcal{P}$, as given in Definition~\ref{Def2bb}, as well as note that we here regard $\mathcal{R}_{\mathcal{P}}$ and $\mathcal{C}_{\mathcal{P}}$ as arbitrarily ordered sets.
    
\begin{Definition}\label{sub-tiles}
For each tile $\mathcal{P}$, the set of (at most) $\Delta$ horizontal sub-tiles takes the form
    \begin{align}
     \mathcal{H}_{\mathcal{P},h_{\mathcal{P}}} &\triangleq\{(\mathcal{R}_{\mathcal{P}}(h_{\mathcal{P}}),j) | j \in \mathcal{C}_{\mathcal{P}}, \ h_{\mathcal{P}} \in [\Delta]\}\label{horizontal2}
    \end{align}
while the set of (at most) $\Gamma$ vertical sub-tiles takes the form
    \begin{align}
     \mathcal{V}_{\mathcal{P},{v_{\mathcal{P}}}} &\triangleq \{(i,\mathcal{C}_{\mathcal{P}}(v_{\mathcal{P}})) | i \in \mathcal{R}_{\mathcal{P}}, \ v_{\mathcal{P}} \in [\Gamma]\}.\label{vertical2}
    \end{align}
\end{Definition}
In the above, $\mathcal{R}_{\mathcal{P}}(h_{\mathcal{P}})$ represents the $h_{\mathcal{P}}$-th element of $\mathcal{R}_{\mathcal{P}}$, and similarly  $\mathcal{C}_{\mathcal{P}}(v_{\mathcal{P}})$ represents the $v_{\mathcal{P}}$-th element of $\mathcal{C}_{\mathcal{P}}$. We also note (cf. Lemma \ref{size-of-tiles}) that each horizontal (resp. vertical) sub-tile can have at most $\Gamma$ (resp. $\Delta$) elements. We also need the following function definition. 
\begin{Definition}\label{minimum-rank-sub-tiles}
For $\mathcal{G}$ being the power set of all horizontal and vertical sub-tiles $\{\mathcal{H}_{\mathcal{P},h_{\mathcal{P}}}, \mathcal{V}_{\mathcal{P},v_{\mathcal{P}}}\}_{\mathcal{P}\in\mathcal{C}, h_{\mathcal{P}} \in [\Delta],v_{\mathcal{P}} \in [\Gamma]} $, we define the function $\Phi(.): \{0,1\}^{K \times L} \rightarrow  \mathcal{G}$ as 
    \begin{align}
        \Phi(\mathbf{S}_{\mathcal{P}}) \triangleq \begin{cases}
      \{\mathcal{H}_{\mathcal{P},h_{\mathcal{P}}} |h_{\mathcal{P}} \in [\Delta] \}, &  \text{if } \: |\mathcal{R}_{\mathcal{P}}| \leq |\mathcal{C}_{\mathcal{P}}|, \\
       \{\mathcal{V}_{\mathcal{P},v_{\mathcal{P}}} |v_{\mathcal{P}} \in [\Gamma] \}, &  \text{if }\:  |\mathcal{R}_{\mathcal{P}}| > |\mathcal{C}_{\mathcal{P}}|.\label{Phi-function}
    \end{cases}
    \end{align}
\end{Definition}
We now proceed to bound the number of rank-one contribution supports, and we do so under our previously stated assumption of disjoint supports (cf. Definition~\ref{disjointSupportAssumption}), which is equivalent to disjoint tiles assumptions via the following Lemma,
 \begin{Lemma} \label{Disjoint-Support-Assumption: Equivalence}
For two matrices $\mathbf{D},\mathbf{E}$, the representative supports $\{\mathbf{S}_{\mathcal{P}_i}\}_{i=1}^m$ of $\mathbf{DE}$ are disjoint (i.e., $\mathbf{S}_{\mathcal{P}_i} \cap \mathbf{S}_{\mathcal{P}_j} = \mathbf{0}, \ j\neq i $) if and only if $\mathbf{D}$ and $\mathbf{E}$ accept the disjoint   support assumption of Definition~\ref{disjointSupportAssumption}.
 \end{Lemma}
 \begin{proof}
     Assuming that $\mathbf{D} \in \mathbb{R}^{K \times NT},\mathbf{E} \in \mathbb{R}^{NT \times L}$ abide by the disjoint   support assumption from Definition~\ref{disjointSupportAssumption}, then for all $i,i'\in [NT]$, we have that either
$\text{Supp}(\mathbf{D}(:,i)\mathbf{E}(i,:)) = \text{Supp}(\mathbf{D}(:,i')\mathbf{E}(i',:) )$ or that $\text{Supp}(\mathbf{D}(:,{i})\mathbf{E}(i,:)) \cap \text{Supp}(\mathbf{D}(:,{i'})\mathbf{E}({i'},:)) = \emptyset$. This in turn implies that for $\mathbf{I}=\Supp(\mathbf{D}) \in \{0,1\}^{K \times NT}, \mathbf{J} = \Supp(\mathbf{E}) \in \{0,1\}^{NT \times L}$, then either  $\mathbf{I}(:,i)\mathbf{J}(i,:) = \mathbf{I}(:,j)\mathbf{J}(j,:)$ or $\mathbf{I}(:,i)\mathbf{J}(i,:) \cap \mathbf{I}(:,j)\mathbf{J}(j,:) = \mathbf{0}_{K \times L}$, which in turn yields the assumption in Definition~\ref{Def2} of disjoint representative support equivalence classes. 

In reverse, if  $\mathbf{D}\mathbf{E}$  accepts the disjoint representative support assumption, and if $\mathcal{C}=\{\mathcal{P}_1, \hdots, \mathcal{P}_m\}$ is the collection of the equivalence classes, then $\forall \mathcal{P},\mathcal{P}' \in \mathcal{C}, \mathcal{C}_{\mathcal{P}} = \mathcal{C}_{\mathcal{P'}}$ or $\mathcal{C}_{\mathcal{P}} \cap \mathcal{C}_{\mathcal{P'}} = \emptyset$ and similarly $\forall \mathcal{P},\mathcal{P}' \in \mathcal{C}, \mathcal{R}_{\mathcal{P}} = \mathcal{R}_{\mathcal{P'}}$ or $\mathcal{R}_{\mathcal{P}} \cap \mathcal{R}_{\mathcal{P'}} = \emptyset$, which in turn implies that $\forall i,i' \in [NT]$ then $\supp(\mathbf{D}(:,i)) = \supp(\mathbf{D}(:,i'))$ or $\supp(\mathbf{D}(:,i)) \cap \supp(\mathbf{D}(:,i'))= \emptyset$, as well implies that $\forall i,i' \in [NT]$ then $\supp(\mathbf{E}(i,:)) = \supp(\mathbf{E}(i',:))$ or $\supp(\mathbf{E}(i,:)) \cap \supp(\mathbf{E}(i',:))= \emptyset$. Consequently, if $\supp(\mathbf{D}(:,i)) = \supp(\mathbf{D}(:,i')) $ and $\supp(\mathbf{E}(i,:)) = \supp(\mathbf{E}(i,:))$ both hold, then $\text{supp}(\mathbf{D}(:,i)\mathbf{E}(i,:)) = \text{supp}(\mathbf{D}(:,i')\mathbf{E}(i',:) )$  or  other wise $\text{supp}(\mathbf{D}(:,i)\mathbf{E}(i,:)) \cap \text{supp}(\mathbf{D}(:,i')\mathbf{E}(i',:) ) = \emptyset$, which in turn yields the assumption in Definition~\ref{disjointSupportAssumption} that $\mathbf{D} $ and $\mathbf{E}$ satisfy the disjoint   support assumption. 
 \end{proof}
 
We proceed with the next lemma. 
\begin{Lemma}\label{Relation-rank-sub-tile}
In any lossless function reconstruction scheme corresponding to $\mathbf{DE}=\mathbf{F}$, the number of rank-one contribution supports $|\mathcal{P}|$ in each class $\mathcal{P}$, satisfies 
$|\mathcal{P}| \geq |\Phi(\mathbf{S}_{\mathcal{P}})| $.
\end{Lemma}
\begin{proof}
    Recall from Definition~\ref{disjointSupportAssumption} and Lemma~\ref{Disjoint-Support-Assumption: Equivalence} that in the context of lossless schemes, each representative support is disjoint. Then we can see that 
    \begin{align}
        |\mathcal{P}| &\overset{}{\geq} \min ( \min(|\mathcal{R}_{\mathcal{P}}| ,|\mathcal{P}|) , \min(|\mathcal{C}_{\mathcal{P}}|, |\mathcal{P}|)) \\ & \overset{(a)}{=} \min(\rank(\mathbf{D}_{\mathcal{P}}), \rank(\mathbf{E}_{\mathcal{P}})) \overset{(b)}{\geq} \rank(\mathbf{F}_{\mathcal{P}})  \overset{(c)}{=}\min(|\mathcal{R}_{\mathcal{P}}|, |\mathcal{C}_{\mathcal{P}}|)  = r_{\mathcal{P}}
    \end{align}
    where (a) follows from Definition~\ref{disjointSupportAssumption} and Lemma~\ref{Disjoint-Support-Assumption: Equivalence} which tells us that in the context of lossless schemes then each representative support is disjoint which in turn tells us that \eqref{sub-SVD} applies in which case we have $\mathbf{D}(\mathcal{R}_{\mathcal{P}}, \mathcal{P})= \mathbf{D}_{\mathcal{P}}$ and $\mathbf{E}( \mathcal{P}, \mathcal{C}_{\mathcal{P}})= \mathbf{E}_{\mathcal{P}}$. Subsequently $(b)$ follows from the fact that $\mathbf{D}_{\mathcal{P}}\mathbf{E}_{\mathcal{P}} = \mathbf{F}_{\mathcal{P}}$, $(c)$ follows from the dimensionality of $\mathbf{F}_{\mathcal{P}}$, and the last equality holds from the definition of $r_{\mathcal{P}}$.

\end{proof}

We now proceed with a lemma that lower bounds the minimum number of horizontal or vertical sub-tiles needed\footnote{To ``cover" in this context means that every element of $\mathbf{F}$ has to be in at least one representative support.} to cover $\mathbf{F} \in \mathbb{R}^{K \times L}$.
\begin{Lemma}\label{minim-r}
For any single-shot lossless function reconstruction scheme, and for the corresponding $\mathbf{DE} = \mathbf{F}$ decomposition, the minimum number of sub-tiles needed to cover $\mathbf{F}$ is at least $\frac{KL}{\max(\Delta,\Gamma)}$.

\end{Lemma}
\begin{proof}
    Suppose first that $\Gamma \geq \Delta$ and consider a scheme that covers the entire $\mathbf{F}$, with $m_1$ horizontal sub-tiles and $m_2$ vertical sub-tiles. We wish to show that $m_1 + m_2 \geq \frac{KL}{\max(\Delta,\Gamma)}$.     
    To see this, we first note that since there is no intersection between each of the tiles (cf. Definition~\ref{disjointSupportAssumption} and Lemma~\ref{Disjoint-Support-Assumption: Equivalence}), and since there is no intersection between each sub-tile inside a tile (Definition~\ref{minimum-rank-sub-tiles}), then we can conclude that for any $\mathcal{P},\mathcal{P}' \in \mathcal{C} $ and any $\mathcal{S} \in \Phi(\mathbf{S}_{\mathcal{P}}), \mathcal{S}' \in \Phi(\mathbf{S}_{\mathcal{P}'})$, we must have $\mathcal{S} \cap \mathcal{S}' = \emptyset$. This in turn implies that the sub-tiles can now cover at most $m_1 \Gamma + m_2 \Delta$  elements of $\mathbf{F}$, which in turn means that $m_1 \Gamma + m_2 \Delta \geq KL$, which means that $\frac{KL}{\Gamma} - m_2 \frac{\Delta}{\Gamma} \leq  m_1$, which means that $\frac{KL}{\Gamma}  + (1- \frac{\Delta}{\Gamma})m_2 \leq m_1 + m_2$. Since $\frac{\Delta}{\Gamma}  \leq  1$, we have that $ \frac{KL}{\Gamma} \leq \frac{KL}{\Gamma}  + (1- \frac{\Delta}{\Gamma})m_2$, which directly tells us that $m_1 + m_2\geq \frac{KL}{\Gamma}$. This concludes the proof for the case of $\Gamma \geq \Delta$. The same process follows directly also for the case of $\Gamma \leq \Delta$, thus concluding the proof.
\end{proof}
At this point we can combine our results. We know from Definitions~\ref{sub-tiles},\ref{minimum-rank-sub-tiles} and from Lemma~\ref{minim-r} that $\sum_{\mathcal{P} \in \mathcal{C}}|\Phi(\mathbf{S}_{\mathcal{P}})|> \frac{KL}{\max(\Gamma,\Delta)
)}$, while we know from Lemma~\ref{minimum-rank-sub-tiles} that $\sum_{\mathcal{P} \in \mathcal{C}}|\mathcal{P}| \geq \sum_{\mathcal{P} \in \mathcal{C}}|\Phi(\mathbf{S}_{\mathcal{P}})|$.
Now by recalling that $\sum_{\mathcal{P} \in \mathcal{C}} |\mathcal{P}|$ is the number of rank-one contribution supports, and by recalling from Lemma~\ref{Server-Tile} that each rank-one contribution support corresponds to a server in any lossless scheme, we can use the lower bound on the number of sub-tiles in Lemma~\ref{minim-r} as a lower bound on the number of servers, allowing us to thus conclude that $N_{opt} \geq \frac{KL}{\max(\Delta,\Gamma)}$, thus concluding the proof of our converse for the single-shot case.   
\subsection{The General Multi-Shot Case of $T>1$}\label{multi-shot-convers}
We begin with our first lemma for this case. 
\begin{Lemma}\label{Server-Tile2}
To guarantee lossless function reconstruction with constraints $\Delta, \Gamma$, any $T$-shot scheme with computation and communication matrices $\mathbf{D,E}$, must associate each server transmission (shot) to a unique rank-one contribution support $\mathbf{S}_{n}(\mathbf{I}, \mathbf{J}), n \in [NT]$, where $\mathbf{I} \in \{0,1\}^{K \times NT}$ and $\mathbf{J} \in \{0,1\}^{NT \times L}$ are the support constraints of $\mathbf{D}$ and $\mathbf{E}$ respectively as in Definition \ref{def-r1}.  
\end{Lemma}
\begin{proof}
    The proof follows directly the proof steps of Lemma~\ref{Server-Tile}, after noting that in our current multi-shot setting, each column of $\mathbf{D}$ and row of $\mathbf{E}$ correspond to a single transmission by a unique server.
\end{proof}
We use Lemma~\ref{Server-Tile2} to recall that  $NT$ is equal to the number of rank-one contribution supports. Then we apply the same proof steps found at the last paragraph of  Appendix~\ref{single-shot-convers} to see that the number of rank-one contribution supports is no less than $\frac{KL}{\max(\Gamma,\Delta)}$, which in turn implies that $N \geq \frac{KL}{T \max(\Delta,\Gamma)}$. This holds for all $T$. 
The following lemma will allow us to provide a tighter bound, for the case of $T\geq \min(\Delta,\Gamma)$.  After proving the following lemma, we will complete the proof of the converse for the multi-shot case. In the following, we recall that the term ``cover" in our context means that every element of $\mathbf{F}$ has to be in at least one representative support.

\begin{Lemma} \label{minim-tiles} For lossless function reconstruction, the corresponding $\mathbf{DE} = \mathbf{F}$ decomposition needs at least $\ceil{\frac{K}{\Delta}}  \ceil{\frac{L}{\Gamma}} $ representative supports (tiles) to cover the entire matrix $\mathbf{F} \in \mathbb{R}^{K \times L}$.
\end{Lemma}
\begin{proof}
Let us first recall from Lemma~\ref{Coverage-nessecity} that (the tiles corresponding to) any lossless optimal scheme must cover $\mathbf{F}$. Let $\mathcal{C}$ be the collection of classes of an optimal scheme, and let $\mathbf{I}=\Supp(\mathbf{D}),\mathbf{J} = \Supp(\mathbf{E})$ and $\Supp(\mathbf{F}) = \Supp(\mathbf{DE})= \mathbf{I}\mathbf{J}$.
  
Let us first consider the simplest instance where $\Delta \vert  K,  \Gamma | L$, in which case we first note that $\mathbf{F}$ has a total of $KL$ elements, and that each tile can cover at most $\Delta  \Gamma$ elements of $\mathbf{F}$ (cf. Lemma~\ref{size-of-tiles}), which in turn means that the minimum number of disjoint covering tiles is simply $\frac{KL}{\Delta \Gamma}$ (cf. Lemma~\ref{disjointSupportAssumption}), which then completes the proof of the lemma for this instance. Subsequently, for the case of $\Delta \nmid  K,  \Gamma | L$, we first split the rows of $\mathbf{F}$ into an upper part $\mathcal{R}_1$ with $K_1=\lfloor \frac{K}{\Delta}\rfloor \Delta$ rows, and a lower part $\mathcal{R}_2$ with $K_2 =\mod(K,\Delta) = K-K_1$ rows. Here $\mathcal{R}_1$ and $\mathcal{R}_2$ are the corresponding row indices. Using the argument from the previous case of $\Delta \vert  K,  \Gamma | L$, we can see that the upper part $\mathbf{F}(\mathcal{R}_1, :)$ needs at least $\lfloor
 \frac{K}{\Delta}\rfloor \frac{L}{\Gamma}$ tiles to complete covering.  For the lower submatrix $\mathbf{F}(\mathcal{R}_2, :)$, we know that having $\lfloor
 \frac{K}{\Delta}\rfloor \frac{L}{\Gamma}$ tiles is not enough since we would only be able to cover $K_1  L = \lfloor \frac{K}{\Delta}\rfloor \Delta L$ elements, thus leaving $L$ uncovered elements in each row in $\mathcal{R}_2$. Since each tile can cover at most $\Gamma$ elements in each column (cf. Lemma~\ref{size-of-tiles}), there has to be at least $\frac{L}{\Gamma}$ additional tiles to cover all of $\mathbf{F}$, which implies a total of $\lfloor \frac{K}{\Delta} \rfloor \frac{L}{\Gamma} + \frac{L}{\Gamma} = \lceil \frac{K}{\Delta} \rceil \frac{L}{\Gamma}$ tiles, which proves our claim for the case of $\Delta \nmid  K,  \Gamma | L$.  The third case of $\Delta |  K,  \Gamma \nmid L$ is similar.
     
We now consider the more involved, general case of $K = q_1 \Delta + r_1, L=q_2 \Gamma + r_2$, where $q_1,q_2, r_1,r_2 \in \mathbb{N} \cup \{0\},0 < r_1 \leq  \Delta, 0 < r_2 \leq  \Gamma$.   We first recall from Lemma~\ref{size-of-tiles} that each tile $\mathcal{P} \in \mathcal{C}$ can cover at most $\Gamma$ elements of a row of $\mathbf{F}$, which in turn implies that $\|\mathbf{S}_{\mathcal{P}}(k,:) \cap \Supp(\mathbf{F})(k',:)\|_{0} \leq \Gamma,\: \forall k,k' \in [N], \forall \mathcal{P} \in \mathcal{C}$. Similarly, we also know that each tile can cover up to $\Delta$ elements in one column of $\mathbf{F}$, which in turn implies that $\|\mathbf{S}_{\mathcal{P}}(:,\ell) \cap \Supp(\mathbf{F})(:,\ell')\|_{0} \leq \Delta,\: \forall \ell,\ell'  \in [L], \forall \mathcal{P} \in \mathcal{C}$.

We will employ a two-dimensional version of the pigeon-hole principle. In reference to this principle, our ``pigeon" here will correspond to an element of $\mathbf{F}$, a ``hole" will correspond to a tile, while now also each row and column of a tile (which corresponds to a horizontal and vertical sub-tile (cf.Definition~\ref{sub-tiles})) will correspond to a `sub-hole'. Let us recall from Definition~\ref{sub-tiles} that each tile consists of $\Delta$ horizontal sub-tiles, and then recall from Lemma \ref{size-of-tiles} that each such horizontal sub-tile (sub-hole) has up to $\Gamma$ elements\footnote{Recall that a tile, and by extension, a sub-tile, is a set of indices.}. Similarly we recall that each tile consists of $\Gamma$ vertical sub-tiles, each having at most $\Delta$ elements. For any scheme whose corresponding $\mathbf{D,E}$ satisfy the disjoint support assumption (Definition~\ref{disjointSupportAssumption} and Lemma~\ref{Disjoint-Support-Assumption: Equivalence}), we now define a mapping function $\Xi_{\mathbf{D},\mathbf{E}}: [K] \times [L] \rightarrow \mathcal{C} \times \mathcal{C}_{h} \times \mathcal{C}_{v}$, where $\mathcal{C}, \mathcal{C}_{h}, \mathcal{C}_{v}$ are respectively the set of all tiles, horizontal sub-tiles, and vertical sub-tiles involved in the  scheme. This function is here defined to take the form
\begin{align}
    \Xi_{\mathbf{D},\mathbf{E}}((i,j)) \triangleq \{(\mathcal{P},h_{\mathcal{P}},v_{\mathcal{P}})\}, (i,j)  \in [K] \times [L] 
\end{align}
where the input $(i,j)$ is the location of the element in $\mathbf{F}$, and where the output consists of the tile $\mathcal{P}$ that covers $(i,j)$, and the corresponding sub-tile index $h_{\mathcal{P}} \in [\Delta]$ of the horizontal sub-tile that covers $(i,j)$, as well as sub-tile index $v_{\mathcal{P}} \in [\Gamma]$ of the vertical sub-tile again covering $(i,j)$. For ease of reading, we recall here from Definition~\ref{sub-tiles} that horizontal and vertical sub-tiles are defined as
 \begin{align}
     \mathcal{H}_{\mathcal{P},h_{\mathcal{P}}} &= \{(\mathcal{R}_{\mathcal{P}}(h_{\mathcal{P}}),j) | j \in \mathcal{C}_{\mathcal{P}}\}\label{horizontal}\\
     \mathcal{V}_{\mathcal{P},{v_{\mathcal{P}}}} & = \{(i,\mathcal{C}_{\mathcal{P}}(v_{\mathcal{P}})) | i \in \mathcal{R}_{\mathcal{P}}\}\label{vertical}
 \end{align}
  where $\mathcal{R}_{\mathcal{P}}$ and $\mathcal{C}_{\mathcal{P}}$, are defined in Definition~\ref{Def2bb}. We also recall that we regard 
$\mathcal{R}_{\mathcal{P}}$ and $\mathcal{C}_{\mathcal{P}}$ as ordered sets, with an arbitrary order, where $\mathcal{R}_{\mathcal{P}}(h_{\mathcal{P}})$ is the $h_{\mathcal{P}}$-th element of $\mathcal{R}_{\mathcal{P}}$ and where similarly $\mathcal{C}_{\mathcal{P}}(v_{\mathcal{P}})$ is the $v_{\mathcal{P}}$-th element of $\mathcal{C}_{\mathcal{P}}$. 
 Now we claim that 
 \begin{align}
     \forall i \in [K]: |\{\mathcal{P} | \Xi_{\mathbf{D},\mathbf{E}}(i,j) = \{(\mathcal{P},h_{\mathcal{P}},v_{\mathcal{P}}) , j \in [L]\}|\geq q_2 +1 \label{lower-bound-h-sub-hole}
 \end{align}
which says that the intersection of each row of $\mathbf{F}$ with any specific tile, is at least $q_2 +1$. This is obvious because if this intersection was less than $q_2 +1$ then --- as a consequence of the pigeon-hole principle, where again each element of $\mathbf{F}(i,:)$ is our ``pigeon" and each horizontal sub-tile as a sub-hole --- there would exist at least one sub-hole covering at least $P_1$ elements, where 
\begin{align}
P_1 = \lceil \frac{L}{q_2} \rceil =  \lceil \frac{q_2 \Gamma + r_2}{q_2} \rceil  =   \Gamma +1
\end{align}
which would though contradict Lemma~\ref{size-of-tiles} which guarantees that there can exist no tile covering more than $\Gamma$ elements in a row, which simply translates to having no horizontal sub-tile with more than $\Gamma$ elements.  

With~\eqref{lower-bound-h-sub-hole} in place, we now conclude that\footnote{ For the case where $q_2 =0$,  Lemma~\ref{Coverage-nessecity} simply guarantees that each row has to be at least in one tile. } in each row, at least $q_2 +1$ horizontal sub-tiles  reside. 
Recall that by Lemma~\ref{size-of-tiles}, each horizontal sub-tile in a tile can cover at most $\Gamma$ elements. Note also that by Definition~\ref{sub-tiles}, for any given row of $\mathbf{F}$, we cannot encounter two or more sub-tiles from the same tile (they must be from different tiles). In this context, for any given row $i$ of $\mathbf{F}$, let $(i, 1),(i,2),\cdots,(i, u_i)$ be the indices of the corresponding horizontal sub-tiles that reside in that row, from left to right. Recall from~\eqref{lower-bound-h-sub-hole} that $u_i \geq q_2+1$.

We proceed to provide a similar enumeration, now though for vertical sub-tiles.

We first note from Lemma~\ref{size-of-tiles} that each column of $\mathbf{F}$ can intersect at most $\Delta$ \emph{horizontal} sub-tiles.
Similar to before, we can also see that 
 \begin{align}
     \forall j \in [L]: |\{\mathcal{P} | \Xi_{\mathbf{D},\mathbf{E}}(i,j) = \{(\mathcal{P},h_{\mathcal{P}},v_{\mathcal{P}})\}, i \in [K]\}|\geq q_1 +1 \label{lower-bound-v-sub-hole}
 \end{align}
which says that the intersection of each column of $\mathbf{F}$ with any specific tile, entails at least $q_1 +1$ tiles, because if this intersection was less than $q_1 +1$ then --- as a consequence of the pigeon-hole principle, where now each horizontal sub-tile is our ``pigeon" and each tile is a hole --- there would exist at least one sub-hole covering at least $P_2$ elements, where
\begin{align}
P_2 = \lceil \frac{K}{q_1} \rceil =  \lceil \frac{q_1 \Delta + r_1}{q_1} \rceil  =   \Delta+1
\end{align}
which though would contradict Lemma~\ref{size-of-tiles} which now guarantees that there can exist no tile covering more than $\Delta$ elements in a column, which now simply translates to having no vertical sub-tile intersecting with more than $\Gamma$ horizontal sub-tiles. We can now conclude that\footnote{For the case where $q_1 =0$. Lemma~\ref{Coverage-nessecity} simply guarantees that each column has to be at least in one tile.} each row of $\mathbf{F}$ intersects with at least $q_2+1$ tiles. After recalling from Definition~\ref{sub-tiles} that two horizontal sub-tiles of the same tile have the same beginning and end\footnote{More rigorously, we can rephrase the above by saying `after recalling from Definition~\ref{sub-tiles} that $\forall h_{\mathcal{P}}, h_{\mathcal{P}}' \in [\Delta]:\: \{j|(i,j) \in \mathcal{H}_{\mathcal{P}, h_{\mathcal{P}}}\} = \{j|(i,j) \in \mathcal{H}_{\mathcal{P}, h_{\mathcal{P}}'}\}$'.}, we can conclude that two horizontal sub-tiles $(i',j')$ and $(i'',j'')$  (where $j' \neq j''$), cannot be found in the same tile, because their respective column indices differ by at least one element. We will now only consider the parts of $\mathbf{F}$ that correspond to horizontal sub-tiles with indices restricted to the set $[K]\times [q_2+1]$. Note that this ``pruning" is in line with our effort to lower bound the number of required tiles for covering $\mathbf{F}$. We proceed by combining the fact that each column of $\mathbf{F}$ intersects with at least $q_1+1$ tiles (cf.~\eqref{lower-bound-v-sub-hole}), together with the aforementioned fact that the column coordinates of two horizontal sub-tiles of the same tile are identical. Thus we can finally adopt the tile indexing 
$(1,\hat{j}),(2,\hat{j}),\hdots, (v_{\hat{j}}, {\hat{j}})$ for all $\hat{j} \in [q_2+1]$. The fact that this indexing applies to any conceivable covering tessellation scheme, allows us to conclude that any covering scheme must entail at least $\sum^{q_2+1}_{\hat{j}=1}v_{\hat{j}}$ tiles. Noting now from~\eqref{lower-bound-h-sub-hole}, we have that $v_{\hat{j}} \geq q_1 +1, \forall \hat{j} \in [q_2+1]$. This fact allows us to conclude that there must exist at least $(q_1+1)(q_2+1)$ tiles, which completes the proof after recalling that $K = q_1 \Delta + r_1, L=q_2 \Gamma + r_2$ which in turn says that $q_1 =\ceil{\frac{K}{\Delta}}, q_2 =  \ceil{\frac{L}{\Gamma}}$.
\end{proof}

With the above lemma in place, focusing again on the particular case of $T\geq \min(\Delta,\Gamma)$, we can now tighten the bound on $N$, from $N \geq \frac{KL}{T \max(\Delta,\Gamma)}$ to $N\geq \lceil \frac{K}{\Delta}\rceil \lceil \frac{L}{\Gamma}\rceil $. We see this by first noting that in our case of $T\geq \min(\Delta,\Gamma)$ each tile corresponds to a server, and then by combining this with Lemma~\ref{Server-Tile2} which additionally tells us that each server corresponds to $T$ distinct rank-one contribution supports. Furthermore, we also know from Definition~\ref{minim-r} and Lemma~\ref{Relation-rank-sub-tile} that $\min(\Delta,\Gamma) = r_{\mathcal{P}} = |\Phi(\mathbf{S}_{\mathcal{P}})|$. Finally, from the fact that no server can be associated with two different tiles\footnote{This was shown in appendix Section~\ref{Construction-DE}, to be a necessary condition for guaranteeing the $\Gamma$ and $\Delta$ constraints.}, and directly from Definition~\ref{minimum-rank-sub-tiles}, we can conclude that the optimal number of servers will be the minimum number of covering tiles, which was shown Lemma~\ref{minim-tiles} to be equal to $\lceil \frac{K}{\Delta}\rceil \lceil \frac{L}{\Gamma}\rceil $. This concludes the proof that $N\geq \lceil \frac{K}{\Delta}\rceil \lceil \frac{L}{\Gamma}\rceil $ for $T\geq \min(\Delta,\Gamma)$. This concludes the converse for the multi-shot case. 

\subsection{Proof of Corollary~\ref{Gap}} \label{proof-gap}
Our aim here is to show that for $T < \min(\Delta,\Gamma)$, then the achievable $N$ (and thus the achievable rate) in \eqref{eq-upper-N} is at most a multiplicative factor of $8$ from the corresponding converse expression in \eqref{eq-upper-lower-N}. The proof can be derived from the following sequence of expressions 
    \begin{align*}
        \frac{N_{\text{upper}}}{N_{\text{Lower}}} &\leq \frac{(\min(\Delta,\Gamma)/T +1) (KL/(\Delta\Gamma) + L / \Gamma + K / \Delta ) + \min(\Delta,\Gamma)/T ) }{ KL / (T \max(\Gamma,\Delta))} \\&= 1 + \frac{T \max(\Delta,\Gamma)}{\Delta \Gamma} + \frac{T \max(\Delta,\Gamma) \min(\Delta,\Gamma)}{ T K \Gamma} \\&+ \frac{T \max(\Gamma,\Delta) L}{KL\Gamma} + \frac{T \max(\Delta,\Gamma) \min(\Delta,\Gamma)}{T \Delta  L} + \frac{T \max(\Delta,\Gamma)}{L \Delta} \\&+ \frac{T \max(\Delta,\Gamma) \min(\Delta,\Gamma)}{TKL} + \frac{T \max(\Delta,\Gamma)}{KL}
        \\& \leq 1 + 4 \frac{T}{\min(\Delta,\Gamma)} + \delta + \gamma + \delta \gamma < 8
    \end{align*}
    where the final answer results by noting  that $T <\min(\Delta,\Gamma), \Delta \leq K, \Gamma \leq L$.
\section{Proof of The achievability and converse of Theorem \ref{asymptotic-capacity}}\label{ProofOfTheorem2}
The proof incorporates an achievable and a converse argument.
We first prove the following lemma used in~\eqref{eq:Err1}, corresponding to the average normalized error\footnote{We recall from \eqref{Decoding-Criteria} our instantaneous error $\mathcal{E} = \sum^{K}_{k=1}{|F'_k - F_k|}^2, \: \mathcal{E} \in \mathbb{R},\:\: \: \forall k \in [K]$.} $\epsilon = \frac{\mathbb{E}_{\:{\mathbf{F},\mathbf{w}}}\{\mathcal{E} \}}{{KL}}$ of any scheme with communication matrix $\mathbf{D}$ and computing matrix $\mathbf{E}$. Recall that in our asymptotic setting, the parameter $N$ scales to infinity, while the calibrating ratios $\delta, \gamma,\kappa,R$ remain constant.  We also note that $T$ here is a non-scaling constant, while we also recall that the elements of $\mathbf{w}$ are i.i.d, independent of $\mathbf{D,E,F}$, and have unit variance. We proceed with the first lemma:
\begin{Lemma}\label{Lemma-Epsilon-normalization}
    In the limit of large $N$, the average normalized error $\epsilon = \frac{\mathbb{E}_{\:{\mathbf{F},\mathbf{w}}}\{\mathcal{E} \}}{{KL}}$ of any lossy function reconstruction scheme corresponding to $\mathbf{DE} = \mathbf{F}$, under the assumptions of Theorem~\ref{asymptotic-capacity}, takes the form 
    \begin{align}
        \epsilon &\overset{}{=} \frac{\mathbb{E}_{\mathbf{F}}\{\| \mathbf{D}\mathbf{E} -\mathbf{F}\|^{2}_{F}\}}{KL}.
        \label{distortion-equality}
    \end{align}
\end{Lemma}
\begin{proof}
The proof starts with the definition of the average error from \eqref{eq:Err1} where we defined this error to be $ \epsilon \ \frac{\mathbb{E}_{\mathbf{F}}\{\| \mathbf{D}\mathbf{E} -\mathbf{F}\|^{2}_{F}\}}{KL}$, then the following sequence of steps hold
\begin{align}
\epsilon 
 &\overset{}{=}  \frac{\mathbb{E}_{\:{\mathbf{F},\mathbf{w}}}\{  [(\mathbf{D}\mathbf{E} - \mathbf{F})\mathbf{w}]^{\intercal} 
 [(\mathbf{D}\mathbf{E} - \mathbf{F})\mathbf{w}]^{}
 \}}{K L}
 \overset{}{=}  \frac{\mathbb{E}_{\:{\mathbf{F},\mathbf{w}}}\{  \mathbf{w}^{\intercal} (\mathbf{D}\mathbf{E} - \mathbf{F})^{\intercal}
 (\mathbf{D}\mathbf{E} - \mathbf{F})\mathbf{w}
 \}}{K L}
\nonumber\\
 &\overset{(a)}{=}  \frac{\mathbb{E}_{\:{\mathbf{F},\mathbf{w}}}\{\tr(  \mathbf{w}^{\intercal} (\mathbf{D}\mathbf{E} - \mathbf{F})^{\intercal}
 (\mathbf{D}\mathbf{E} - \mathbf{F})\mathbf{w})
 \}}{K L}
\overset{(b)}{=}  \frac{\mathbb{E}_{\:{\mathbf{F},\mathbf{w}}}\{\tr(   (\mathbf{D}\mathbf{E} - \mathbf{F})^{\intercal}
 (\mathbf{D}\mathbf{E} - \mathbf{F})\mathbf{w}\mathbf{w}^{\intercal})
 \}}{K L}
\nonumber\\
 &\overset{(c)}{=}  \frac{\tr(  \mathbb{E}_{\:{\mathbf{F},\mathbf{w}}}\{ (\mathbf{D}\mathbf{E} - \mathbf{F})^{\intercal}
 (\mathbf{D}\mathbf{E} - \mathbf{F})\mathbf{w}\mathbf{w}^{\intercal}\})
 }{KL }
\overset{(d)}{=}  \frac{\tr(  \mathbb{E}_{\:{\mathbf{F}}}\{ (\mathbf{D}\mathbf{E} - \mathbf{F})^{\intercal}
 (\mathbf{D}\mathbf{E} - \mathbf{F})\}\mathbb{E}_{\:{\mathbf{w}}}\{\mathbf{w}\mathbf{w}^{\intercal}\})
 }{K L }
\nonumber \\
&\overset{(e)}{=}  \frac{\tr(  \mathbb{E}_{\:{\mathbf{F}}}\{ (\mathbf{D}\mathbf{E} - \mathbf{F})^{\intercal}
 (\mathbf{D}\mathbf{E} - \mathbf{F})\})
 }{K L}
 \overset{(f)}{=}  \frac{ \mathbb{E}_{\:{\mathbf{F}}} \{\tr( (\mathbf{D}\mathbf{E} - \mathbf{F})^{\intercal}
 (\mathbf{D}\mathbf{E} - \mathbf{F}))\}
 }{KL }
 \overset{(g)}{=}\frac{\mathbb{E}_{\mathbf{F}}\{\| \mathbf{D}\mathbf{E} -\mathbf{F}\|^{2}_{F}\}}{KL}
 \label{eq-11}
\end{align}
where $(a)$ holds since the trace argument is a scalar, $(b)$ and $(c)$ follow from the cyclic property of the trace operation and its linearity, $(d)$ holds since $\mathbf{w}$ and $\mathbf{F}$ are independent, $(e)$ holds since $\mathbb{E}_{\mathbf{w}}\{\mathbf{w}\mathbf{w}^{\intercal}\}= \mathbf{I}_L$, $(f)$ follows from the interchangeability of the trace and expectation operations, and $(g)$ holds since $\tr(\mathbf{A}^{\intercal} \mathbf{A}) = \sum^{I}_{i=1}\sum^{J}_{j=1} \mathbf{A}(i,j)^{2}= \|\mathbf{A}\|^{2}_{F},\: \forall \mathbf{A} \in \mathbb{R}^{I \times J}$. 
\end{proof}
In the following, we provide the achievability part of the proof, where we present in this appendix,  the lossy variant of the scheme of Theorem \ref{Achievability-Converse}, while in this appendix we also derive the average of the normalized error. For the error analysis, in Lemma~\ref{M-P distribution} (see also Remark~\ref{M-P-CDF}) we adapt the well known Marchenko–Pastur Theorem to our setting, to then yield the proof of Lemma~\ref{MP-relation-Lemma} that tells us how the approximation error corresponding to each tile is defined by the truncated first moment of the Marchenko–Pastur distribution. Subsequently, in Lemma~\ref{Sum-tile}, we describe the total error $\|\mathbf{D}\mathbf{E}- \mathbf{F}\|^2_{F}$ as the sum of errors for each tile, reflecting the nature of our tile assignment from Theorem~\ref{Achievability-Converse}. The last step combines Lemmas~\ref{MP-relation-Lemma} and \ref{Sum-tile} to derive the corresponding upper bound on the normalized error.  
Note that all the definitions in appendix~\ref{Basic-Concepts-and-Definitions} apply to our setting. 

\subsection{Scheme Design}\label{Construction-DE-Asymptotic}
Similar to Section~\ref{Construction-DE} of this appendix, the construction of $\mathbf{D},\mathbf{E}$ will involve the steps of: a) sizing and positioning the tiles of $\mathbf{D}$, of $\mathbf{E}$, and of $\mathbf{DE}$, b) filling the non-zero tiles in $\mathbf{D}\mathbf{E}$ as a function of $\mathbf{F}$, and c) filling the tiles in $\mathbf{D}$ and $\mathbf{E}$. Crucial to our lossy-variant of our scheme, will be the rank $\alpha r_{\mathcal{P}}  \in \mathbb{N}$ of each tile $\mathcal{P}$, where this rank will be defined by the desired error performance as we will see in this appendix.
We proceed with the description of the steps. 
\paragraph{Sizing and positioning the tiles of $\mathbf{D}$, of $\mathbf{E}$, and of $\mathbf{DE}$} Our first step applies the corresponding step in the achievable scheme of the appendix Section~\ref{Construction-DE}, to the current setting of $\mathcal{C}_2= \mathcal{C}_3= \mathcal{C}_4 = \emptyset$, where this latter equality follows after noting that --- in terms of the derived performance --- $\Delta | K, \:\Gamma | L$ and $T | \min(\Delta,\Gamma)$ hold directly in our current setting of constant $T$ and scaling $\Delta$ and $\Gamma$.

\paragraph{Filling the non-zero tiles in $\mathbf{D}\mathbf{E}$ as a function of $\mathbf{F}$}
We  here first approximate each $ \mathbf{F}_{\mathcal{P}}$ (cf.~\eqref{Formation-of-submatrices}) by a low-rank matrix $\mathbf{F}^{\alpha}_{\mathcal{P}}$, whose rank does not exceed  $\alpha r_{\mathcal{P}}$, for some auxiliary variable $\alpha \in \mathbb{R} , 0 < \alpha \leq1, \alpha r_{\mathcal{P}}  \in \mathbb{N}$ \footnote{As we will see soon, the choice of $\alpha$ here will determine the degree of the approximation of $\mathbf{F}_{\mathcal{P}}$. As we will elaborate later on, this parameter will take the form $\alpha  =T \frac{\max(\Delta , \Gamma)}{R}$, while guaranteeing that the rank $\alpha r_{\mathcal{P}}$ is an integer. Note that since $\Delta | K, \Gamma | L$, then $r_{\mathcal{P}} = \min(\Delta,\Gamma)$ (cf.~\eqref{max-rank-formula}) which naturally scales with $\Delta$ and $\Gamma$.}. To obtain the desired    
\begin{align}
\mathbf{F}^{\alpha}_{\mathcal{P}} & \triangleq \underset{\mathbf{A}}{\text{argmin}} \{\|\mathbf{A}
 - \mathbf{F}_{\mathcal{P}}\|_{F}, \ \ : \ \  \text{rank}(\mathbf{A}) \leq \alpha r_{\mathcal{P}}\} \label{low-F-approximation}
 \end{align}
we employ the well-known Echart-Young Theorem \cite{kishore2017literature} to get 
\begin{align}\label{Filling-2}
\mathbf{F}^{\alpha}_{\mathcal{P}} & =  \mathbf{D}^{\alpha}_{\mathcal{P}} \mathbf{E}^{\alpha}_{\mathcal{P}}
\end{align}
where $\mathbf{D}_{\mathcal{P}} \in \mathbb{R}^{|\mathcal{R}_{\mathcal{P}}| \times \alpha r_\mathcal{P} }, {\mathbf{E}_{\mathcal{P}} \in \mathbb{R}^{\alpha r_\mathcal{P} \times |\mathcal{C}_{\mathcal{P}}| }}$ are the results of the truncated SVD algorithm\footnote{In particular, $\mathbf{F}^{\alpha}_{\mathcal{P}}$,  $\mathbf{D}^{\alpha}_{\mathcal{P}}$ and $\mathbf{E}^{\alpha}$ are respectively associated to $\mathbf{A}_k$, $\mathbf{U}\mathbf{S}$ and $\mathbf{V}$ in \eqref{Truncated-SVD-1} of appendix Section~\ref{Basic-Concepts-and-Definitions}.}. 
Naturally, since $\mathcal{C}_2, \mathcal{C}_3, \mathcal{C}_4 = \emptyset$, we have that $|\mathcal{R}_{\mathcal{P}}| = \Delta, |\mathcal{C}_{\mathcal{P}}| =\Gamma$ (cf.\eqref{max-rank-formula}).  

\paragraph{Filling the tiles in $\mathbf{D}$ and $\mathbf{E}$} In this last step, after considering $\mathcal{C}_1 = \{\mathcal{P}_1,\mathcal{P}_2,\hdots, \mathcal{P}_m\}$, then for each $j\in[m]$, we set  
\begin{align}
    \mathbf{D}(\mathcal{R}_{\mathcal{P}_j}, [\sum^{j-1}_{i=1} \alpha r_{\mathcal{P}_i} +1,\sum^{j}_{i=1} \alpha r_{\mathcal{P}_i} ]) = \mathbf{D}_{\mathcal{P}_j}
    \label{Decoding-Encoding-A1}
\end{align}
and 
\begin{align}
    \mathbf{E}([\sum^{j-1}_{i=1} \alpha r_{\mathcal{P}_i} +1,\sum^{j}_{i=1} \alpha r_{\mathcal{P}_i} ], \mathcal{C}_{\mathcal{P}_j})=\mathbf{E}_{\mathcal{P}_j}\label{Decoding-Encoding-A2}
\end{align}
while  the remaining non-assigned elements of $\mathbf{D}$ and $\mathbf{E}$ remain equal to zero.

\subsection{Normalized Error Analysis of the Designed Scheme}\label{Normalized-Error-Analysis}
We proceed to evaluate in Lemma~\ref{MP-relation-Lemma} the approximation error for each tile as a function of the truncated first moment of the Marchenko–Pastur distribution, and then to show in Lemma~\ref{Sum-tile} that for our scheme, the total approximation error $\|\mathbf{D}\mathbf{E}- \mathbf{F}\|^2_{F}$, for each instance of the problem, is equal to the sum of the approximation errors, where the sum is over all tiles.  The proof for the achievability part of the error analysis is completed by properly combining the two aforementioned lemmas.

Directly from the Echart-Young Theorem, the truncated SVD solution yields the optimal approximation error for each tile, which takes the form
\begin{align}
\|\mathbf{F}^{\alpha}_{\mathcal{P}} 
 - \mathbf{F}_{\mathcal{P}}\|_{F} =  \| \mathbf{D}^{\alpha}_{\mathcal{P}} \mathbf{E}^{\alpha}_{\mathcal{P}} - \mathbf{F}_{\mathcal{P}} 
 \|_{F}= \sqrt{\sum^{ r_{\mathcal{P}}}_{i=\alpha r_{\mathcal{P}} +1} \sigma^{2}_{i}(\mathbf{F}_{\mathcal{P}})}\label{Error-Sigma}
\end{align}
where $\sigma_{i}(\mathbf{F}_{\mathcal{P}}),i \in [r_{\mathcal{P}}]$ are the singular values of $\mathbf{F}_{\mathcal{P}}$ (cf.~\eqref{Formation-of-submatrices}), in decreasing order. 

Let us now recall that for each $\mathcal{P} \in \mathcal{C}_1$, then $\mathbf{F}_{\mathcal{P}} \in \mathbb{R}^{\Delta \times \Gamma}$ is a random matrix of \emph{i.i.d} elements having zero mean and $\bar{\sigma}$ variance. Let $\lambda_1(\mathbf{Y}_{\mathcal{P}}) \leq \lambda_2(\mathbf{Y}_{\mathcal{P}}) \leq \hdots \leq \lambda_{\Delta}(\mathbf{Y}_{\mathcal{P}})$ be the ordered eigenvalues of the symmetric matrix  
\begin{align}
\mathbf{Y}_{\mathcal{P}} \triangleq  \frac{1}{\Gamma} \mathbf{F}_{\mathcal{P}} \mathbf{F}_{\mathcal{P}}^{\intercal}
\end{align}
and consider 
\begin{align} \label{mu1}
     \mu_{\Delta}(x) = \frac{1}{\Delta} \#\{\lambda_j(\mathbf{Y}_{\mathcal{P}})  \leq  x \}
 \end{align}
which will be associated below to the CDF $F_{MP,\lambda} = \mu_{\Delta}(x)$ of the Marchenko-Pastur distribution. 

 The following describes the well known Marchenko-Pastur law (\!\!\cite{marchenko1967distribution}) as applied to our setting.

\begin{Lemma}\label{M-P distribution}[Marchenko-Pastur Law]
Consider a random matrix $\mathbf{F}_{\mathcal{P}} \in \mathbb{R}^{\Delta \times \Gamma}$ having \emph{i.i.d} elements with  zero mean and variance $\bar{\sigma}$. Let $\lambda_1(\mathbf{Y}_{\mathcal{P}}) \leq \lambda_2(\mathbf{Y}_{\mathcal{P}}) \leq  \hdots \leq \lambda_{\Delta}(\mathbf{Y}_{\mathcal{P}})$ be the ordered eigenvalues of $\mathbf{Y}_{\mathcal{P}} =  \frac{1}{\Gamma} \mathbf{F}_{\mathcal{P}} \mathbf{F}_{\mathcal{P}}^{\intercal}$. Let $\Delta, \Gamma \rightarrow \infty$, and let $\frac{\Delta}{\Gamma}\rightarrow \lambda$ for some $\lambda>0$. Then $\mu_{\Delta}$ (cf.~\eqref{mu1}) converges in distribution to
\begin{align}
        \mu_{\Delta}(x)=
    \begin{cases}
     (1- \frac{1}{\lambda}){\mathlarger{\mathbf{1}}}_{(0 \leq x)} + v(x), & \text{if}\:\:\lambda > 1,\\
     v(x),  & \text{if} \:\: 0<\lambda \leq 1
    \end{cases}
    \end{align}
    where $v(x)$ is such that 
    \begin{align}
     d(v(x))=\frac{1}{2 \pi \bar{\sigma}^2} \frac{\sqrt{(\lambda_{+} -x)(x -\lambda_{-})}}{\lambda x} \mathbf{1}_{x \in [\lambda_{-}, \lambda_{+}]} dx
    \end{align} for any point $x \in \mathbb{R}$. This in turn means that the PDF of the Marchenko-Pastur distribution takes the form
    \begin{align}
     f_{MP,\lambda}(x) &= \frac{1}{2 \pi \bar{\sigma}^2} \frac{\sqrt{(\lambda_{+} -x)(x -\lambda_{-})}}{\lambda x} \mathbf{1}_{x \in [\lambda_{-}, \lambda_{+}]} +  {\mathlarger{\mathbf{1}}}_{(1 < \lambda)} (1 - \frac{1}{\lambda})\delta(x)
    \end{align}
    where $\lambda_{\pm}\triangleq \sigma^2(1 \pm \sqrt{\lambda})^2 $. 
 \end{Lemma}

 \begin{remark}\label{M-P-CDF}
     With the same notation as in  Lemma~\ref{M-P distribution}, the CDF's explicit form is
     \begin{align}
        F_{\text{MP}, \lambda}(x)&=
    \begin{cases}
     \frac{\lambda-1}{\lambda} {\mathlarger{\mathlarger{\mathbf{1}}}}_{x \in [0, \lambda_{-})} + \left(\frac{\lambda -1 }{2\lambda}+ F_{\text{MP}}(x)\right) {\mathlarger{\mathlarger{\mathbf{1}}}}_{x \in [\lambda_{-}, \lambda_{+})} + {\mathlarger{\mathlarger{\mathbf{1}}}}_{x \in [ \lambda_{+}, +\infty)},\:& \text{if $\lambda > 1$,}\\
     F_{\text{MP}}(x) {\mathlarger{\mathlarger{\mathbf{1}}}}_{x \in [\lambda_{-}, \lambda_{+})} + {\mathlarger{\mathlarger{\mathbf{1}}}}_{[\lambda_{+}, \infty)},\: &\text{if}{\: 0 \leq\lambda \leq 1,}
     \end{cases}\\
     F_{\text{MP}}(x) &= \frac{1}{2 \pi \lambda} \Big(\pi \lambda + \sigma^{-2} \sqrt{(\lambda_{+} - x)(x- \lambda_{-})} \nonumber \\&- (1 + \lambda) \text{arctan}(\frac{r(x)^2 -1}{2r(x)}) + (1 - \lambda) \text{arctan}(\frac{\lambda_{-} r(x)^{2} - \lambda_{+}}{2 \sigma^2(1 - \lambda)r(x)}) \Big) \\
     \text{where } r(x) &= \sqrt{\frac{\lambda_{+} - x}{ x - \lambda_{-}}}.
     \end{align}
 \end{remark}
 Now utilizing Lemma~\ref{M-P distribution}, we provide an expression for the error contributed by any tile $\mathcal{P} \in \mathcal{C}_1$. The following lemma holds under the same assumptions as in  Theorem \ref{asymptotic-capacity} and Lemma \ref{M-P distribution}.
 \begin{Lemma} \label{MP-relation-Lemma}
  The average error attributed to each tile $\mathcal{P}$, described in \eqref{Error-Sigma},  takes the form
\begin{align}
   \mathbb{E}_{\:{\mathbf{F}}}\{ \|\mathbf{F}^{\alpha}_{\mathcal{P}} - \mathbf{F}_{\mathcal{P}} 
 \|^{2}_{F} \} = \Gamma \Delta  \int^{F^{-1}_{MP,\lambda}(1- (\min(\Delta,\Gamma)/ \Delta) \alpha)}_{ (1 - \sqrt{\lambda})^2} x f_{MP,\lambda}(x) \:dx.\label{Class_Error}
\end{align}
 \end{Lemma}
 \begin{proof}
Addressing first the case where $\Delta \leq \Gamma$, we see that
 \begin{align*}
        \mathbb{E}_{\:{\mathbf{F}}}\{ \|\mathbf{F}^{\alpha}_{\mathcal{P}} 
 - \mathbf{F}_{\mathcal{P}}\|^{2}_{F} \} &\overset{(a)}{=}  \mathbb{E}_{\:{\mathbf{F}}}\{  {\sum^{ (1 - \alpha)r_{\mathcal{P}}}_{i=1}  \sigma^{2}_{i + \alpha r_{\mathcal{P}}}(\mathbf{F}_{\mathcal{P}})}\}\\
 &\overset{}{=} \Gamma \mathbb{E}_{\:{\mathbf{F}}}\{  {\sum^{ (1 - \alpha)r_{\mathcal{P}}}_{i=1} \frac{ \sigma^{2}_{i + \alpha r_{\mathcal{P}}}(\mathbf{F}_{\mathcal{P}})}{\Gamma}}\}
 \\&\overset{(b)}{=} \Gamma \mathbb{E}_{\:{\mathbf{F}}}\{
 {\int^{+ \infty}_{ - \infty} \sum^{ (1 - \alpha)r_{\mathcal{P}}}_{i=1} \delta(x -\frac{ \sigma^{2}_{i + \alpha r_{\mathcal{P}}}(\mathbf{F}_{\mathcal{P}} )}{\Gamma})\: x dx}\}
 \\&\overset{(c)}{=} \Gamma \mathbb{E}_{\:{\mathbf{F}}}\{
 { \sum^{ (1 - \alpha)r_{\mathcal{P}}}_{i=1} \int^{+ \infty}_{ - \infty} \delta(x - \frac{\sigma^{2}_{i + \alpha r_{\mathcal{P}}}(\mathbf{F}_{\mathcal{P}} )}{\Gamma})\: x dx}\}
  \\&\overset{(d)}{=} \Gamma
  \mathbb{E}_{\:{\mathbf{F}}}\{{ \sum^{ (1 - \alpha)r_{\mathcal{P}}}_{i=1} \int^{+ \infty}_{ - \infty} \lim_{\delta x \rightarrow 0} \frac{\mathbf{1}( x  <  {\sigma^{2}_{i + \alpha r_{\mathcal{P}}}(\mathbf{F}_{\mathcal{P}} )} / {\Gamma}\leq x + \delta x)}{\delta x}
 \: x dx} \}
 \\&\overset{(e)}{=} \Gamma
 { \sum^{ (1 - \alpha)r_{\mathcal{P}}}_{i=1} \int^{+ \infty}_{ - \infty} \lim_{\delta x \rightarrow 0} \mathbb{E}_{\:{\mathbf{F}}}\{\frac{\mathbf{1}(  \sigma^{2}_{i + \alpha r_{\mathcal{P}}}(\mathbf{F}_{\mathcal{P}} ) / \Gamma \leq x + \delta x)- \mathbf{1}(  \sigma^{2}_{i + \alpha r_{\mathcal{P}}}(\mathbf{F}_{\mathcal{P}} ) / \Gamma \leq x )}{\delta x}\}
 \: x dx}
 \\&\overset{(f)}{=}
 { \Gamma \Delta  \int^{F^{-1}_{MP,\lambda}(1 - \alpha)}_{ - \infty} \lim_{\delta x \rightarrow 0} \frac{ F_{MP}(x + \delta x) - F_{MP,\lambda}(x)}{\delta x}
 \: x dx}
 \\&\overset{(g)}{=}
 { \Gamma \Delta  \int^{F^{-1}_{MP,\lambda}(1 - \alpha)}_{ (1 - \sqrt{\lambda})^2} x f_{MP,\lambda}(x) \:dx}
  \end{align*}
where  $(a)$ holds because of \eqref{Error-Sigma},   $(b)$ is true because of the continuous delta Dirac function properties,  $(c)$ follows from the linearity of both the summation and integral operators, while $(d)$ and $(e)$  follow from the fact that $d(\mathbf{1}(x-\sigma^{2}_{i + \alpha r_{\mathcal{P}}} / \Gamma))/dx =\lim_{\delta x \rightarrow 0} \frac{\mathbf{1}(\sigma^{2}_{i + \alpha r_{\mathcal{P}}}(\mathbf{F}_{\mathcal{P}}) / \Gamma \leq x + \delta x) - \mathbf{1}(\sigma^{2}_{i + \alpha r_{\mathcal{P}}}(\mathbf{F}_{\mathcal{P}}) / \Gamma \leq x)}{\delta x}$. Then, to show $(f)$,  we first note that $\frac{1}{\Delta} \# \{ \lambda_{j}(\mathbf{Y}_{\mathcal{P}}) = \frac{\sigma^2_{j}(\mathbf{F}_{\mathcal{P}})}{\Gamma} \leq x, j \in [r_{\mathcal{P}}]\} \rightarrow F_{MP,\lambda}(x)$ as $\Delta, \Gamma$ scale to infinity. 
Now, again in $(f)$, to understand the change of limits in the integral, we note that given that $r_{\mathcal{P}} = \Delta \leq \Gamma$, we seek to find an  $x'$  such that $(1 - \alpha) r_{\mathcal{P}}= (1 - \alpha) \Delta= \Delta  F_{MP}(x')$, which will allow $x'$ to be the upper integration limit\footnote{In essence, this step simply says that the first $(1 - \alpha) r_{\mathcal{P}} $ eigenvalues (corresponding to the summation) correspond to the eigenvalues accounted for by $F_{MP}^{-1}(1 - \alpha)$.} of the integral in the RHS of equality $(f)$, yielding $\sum^{(1- \alpha)r_{\mathcal{P}}}_{i=1}\mathbf{1}({\sigma^{2}_{i + \alpha r _{\mathcal{P}}}(\mathbf{F}_{\mathcal{P}})}/ \Gamma \leq x) =  \sum^{r_{\mathcal{P}}}_{i=1}\mathbf{1}({\sigma^{2}_{i}}(\mathbf{F}_{\mathcal{P}}) / \Gamma \leq x) = \Delta F_{MP,\lambda}(x) $ since  $x \leq x_1'= F_{MP}^{-1}(1 - \alpha)$, thus yielding $(f)$. Finally $(g)$ is simply the result of the relationship between CDF and PDF, after considering the support of the Marchenko-Pastur distribution.  

For the case where $\Delta > \Gamma$, we have that 
\begin{align*}
        \mathbb{E}_{\:{\mathbf{F}}}\{ \|\mathbf{F}^{\alpha}_{\mathcal{P}} 
 - \mathbf{F}_{\mathcal{P}}\|^{2}_{F} \} &\overset{(a)}{=}  \mathbb{E}_{\:{\mathbf{F}}}\{  {\sum^{ (1 - \alpha)r_{\mathcal{P}}}_{i=1}  \sigma^{2}_{i + \alpha r_{\mathcal{P}}}(\mathbf{F}_{\mathcal{P}})}\}\\
 &\overset{}{=} \Gamma \mathbb{E}_{\:{\mathbf{F}}}\{  {\sum^{ (1 - \alpha)r_{\mathcal{P}}}_{i=1} \frac{ \sigma^{2}_{i + \alpha r_{\mathcal{P}}}(\mathbf{F}_{\mathcal{P}})}{\Gamma}}\}
 \\&\overset{(b)}{=} \Gamma \mathbb{E}_{\:{\mathbf{F}}}\{
 {\int^{+ \infty}_{ - \infty} \sum^{ (1 - \alpha)r_{\mathcal{P}}}_{i=1} \delta(x -\frac{ \sigma^{2}_{i + \alpha r_{\mathcal{P}}}(\mathbf{F}_{\mathcal{P}} )}{\Gamma})\: x dx}\}
 \\&\overset{(c)}{=} \Gamma \mathbb{E}_{\:{\mathbf{F}}}\{
 { \sum^{ (1 - \alpha)r_{\mathcal{P}}}_{i=1} \int^{+ \infty}_{ - \infty} \delta(x - \frac{\sigma^{2}_{i + \alpha r_{\mathcal{P}}}(\mathbf{F}_{\mathcal{P}} )}{\Gamma})\: x dx}\}
  \\&\overset{(d)}{=} \Gamma
  \mathbb{E}_{\:{\mathbf{F}}}\{{ \sum^{ (1 - \alpha)r_{\mathcal{P}}}_{i=1} \int^{+ \infty}_{ - \infty} \lim_{\delta x \rightarrow 0} \frac{\mathbf{1}( x  <  {\sigma^{2}_{i + \alpha r_{\mathcal{P}}}(\mathbf{F}_{\mathcal{P}} )} / {\Gamma}\leq x + \delta x)}{\delta x}
 \: x dx} \}
 \\&\overset{(e)}{=} \Gamma
 { \sum^{ (1 - \alpha)r_{\mathcal{P}}}_{i=1} \int^{+ \infty}_{ - \infty} \lim_{\delta x \rightarrow 0} \mathbb{E}_{\:{\mathbf{F}}}\{\frac{\mathbf{1}(  \sigma^{2}_{i + \alpha r_{\mathcal{P}}}(\mathbf{F}_{\mathcal{P}} ) / \Gamma \leq x + \delta x)- \mathbf{1}(  \sigma^{2}_{i + \alpha r_{\mathcal{P}}}(\mathbf{F}_{\mathcal{P}} ) / \Gamma \leq x )}{\delta x}\}
 \: x dx}
 \\&\overset{(f)}{=}
 { \Gamma \Delta \int^{F^{-1}_{MP,\lambda}((\Gamma / \Delta )(1 - \alpha) +(1 - \frac{\Gamma}{\Delta}))}_{ - \infty} \lim_{\delta x \rightarrow 0} \frac{ F_{MP}(x + \delta x) - F_{MP,\lambda}(x)}{\delta x}
 \: x dx}
 \\&\overset{(g)}{=}
 { \Gamma \Delta  \int^{F^{-1}_{MP,\lambda}(1 - \alpha \frac{\Gamma}{\Delta})}_{ (1 - \sqrt{\lambda})^2} x f_{MP,\lambda}(x) \:dx}
  \end{align*}
where firstly $(a)$ through $(e)$ and $(g)$ follow as in the first case of $\Delta \leq \Gamma$. Then, regarding  $(f)$, we first note that $\frac{1}{\Delta} \# \{ \lambda_{j}(\mathbf{Y}_{\mathcal{P}}) = \frac{\sigma^2_{j}(\mathbf{F}_{\mathcal{P}})}{\Gamma} \leq x, j \in [r_{\mathcal{P}}]\} \rightarrow F_{MP,\lambda}(x)$ as $\Delta, \Gamma$ scale to infinity. We also note that $r_{\mathcal{P}} = \Gamma <\Delta$. Thus, if we find an $x'$ such that $(1 - \alpha) r_{\mathcal{P}}= (1 - \alpha) \Gamma+ (1- \frac{\Gamma}{\Delta}) \Delta=   \Delta F_{MP}(x')$, then $x'$ can act as the upper limit of integration, thus yielding $\sum^{(1- \alpha)r_{\mathcal{P}}}_{i=1}\mathbf{1}({\sigma^{2}_{i + \alpha r _{\mathcal{P}}}(\mathbf{F}_{\mathcal{P}})}/ \Gamma \leq x) =  \sum^{r_{\mathcal{P}}}_{i=1}\mathbf{1}({\sigma^{2}_{i}}(\mathbf{F}_{\mathcal{P}}) / \Gamma \leq x) = F_{MP,\lambda}(x) $ because, as before, in the integration we have that $x \leq x_1'= F_{MP}^{-1}(\frac{\Gamma}{\Delta}(1 - \alpha)+(1 - \frac{\Gamma}{\Delta}))$.  In essence, the change in the upper integration limit in $(f)$ here (as compared to the same integration limit found  the previous case of $\Delta \leq \Gamma$) accounts for the fact that the first $\Delta-\Gamma$ eigenvalues are now zero. This proves $(f)$ and the entire lemma. 
 \end{proof}
 
We proceed with an additional lemma that bounds the approximation error of the design described in \eqref{Decoding-Encoding-A1} and \eqref{Decoding-Encoding-A2}.

\begin{Lemma}\label{Sum-tile}
    The $\mathbf{D}$ and $\mathbf{E}$ design in
    \eqref{Decoding-Encoding-A1} and \eqref{Decoding-Encoding-A2}, guarantees that  
    \begin{align}
       \| \mathbf{D}\mathbf{E} - \mathbf{F} \|_{F}=\sqrt{\sum^{}_{\mathcal{P} \in 
    \mathcal{C}_1} \| \mathbf{F}^{\alpha}_{\mathcal{P}} -\mathbf{F}_{\mathcal{P}} \|^{2}_{F} } .\label{Partition}
    \end{align}
\end{Lemma} 
\begin{proof}   
We first note that for  
\begin{align}
    \mathbf{U} \triangleq \mathbf{D} \mathbf{E} - \mathbf{F}\label{impo3}
\end{align}
then following~\eqref{Decoding-Encoding-A1} and \eqref{Decoding-Encoding-A2}, we see that 
\begin{align}
    \mathbf{U}(\mathcal{R}_{\mathcal{P}}, \mathcal{C}_{\mathcal{P}}) = \mathbf{D}^{\alpha}_{\mathcal{P}} \mathbf{E}^{\alpha}_{\mathcal{P}} - \mathbf{F}_{\mathcal{P}} =\mathbf{F}^{\alpha}_{\mathcal{P}} - \mathbf{F}_{\mathcal{P}} \label{Important-bridge}
\end{align}
where $\mathbf{F}_{\mathcal{P}}$ is defined in \eqref{Formation-of-submatrices}. 
Now 
we note that
\begin{align}
  \|\mathbf{U}\|_{F} = \sqrt{\sum^{(K,L)}_{(k,l)=(1,1)} \mathbf{U}^{2}({k,l})} & \overset{(a)}{ =} 
\sqrt{\sum_{\mathcal{P} \in \mathcal{C}_1} \sum_{(i,j) \in \mathcal{R}_{\mathcal{P}} \times \mathcal{C}_{\mathcal{P}}} \mathbf{U}^{2}(i,j)} \label{impo1}\\  
& \overset{(b)}{ =} 
\sqrt{\sum_{\mathcal{P} \in \mathcal{C}_1} \|\mathbf{U}(\mathcal{R}_{\mathcal{P}}, \mathcal{C}_{\mathcal{P}})\|^{2}_{F}}\\
& \overset{(c)}{ =} 
\sqrt{\sum_{\mathcal{P} \in \mathcal{C}_1} \|\mathbf{F}^{\alpha}_{\mathcal{P}} - \mathbf{F}_{\mathcal{P}} \|^{2}_{{F}}}
\end{align}
where $(a)$ follows from the fact that the representative supports $\mathcal{P} \in \mathcal{C}_1$ partition matrix $\mathbf{F} \in \mathbb{R}^{K \times L }$\footnote{It means that the tiles are disjoint and cover the whole matrix $\mathbf{F}$.}, $(b)$ follows from the definition of the Frobenius norm, while $(c)$ follows as a direct consequence of \eqref{Important-bridge}. Combining \eqref{impo3} and   \eqref{impo1}  proves the lemma. 
\end{proof}
We can now prove the theorem. For the case where $\Delta \leq  \Gamma$, we have
\begin{align}
       \epsilon & \overset{(a)}{=}  \mathbb{E}_{\:{\mathbf{F}}}\{\| \mathbf{D}\mathbf{E} - \mathbf{F} \|^2_F\}\frac{1}{K L } \\
       &\overset{(b)}{=}\mathbb{E}_{\:{\mathbf{F}}}\{\sum^{}_{\mathcal{P} \in 
    \mathcal{C}_1} \| \mathbf{F}^{\alpha}_{\mathcal{P}} -\mathbf{F}_{\mathcal{P}} \|^{2}_{F}  \}\frac{1}{K L }
    \\
       &\overset{(c)}{=}\sum^{}_{\mathcal{P} \in 
    \mathcal{C}_1}  \mathbb{E}_{\:{\mathbf{F}}}\{\| \mathbf{F}^{\alpha}_{\mathcal{P}} -\mathbf{F}_{\mathcal{P}} \|^{2}_{F}  \}\frac{1}{K L }\\
     &\overset{(d)}{=}\frac{K}{\Delta} \frac{L}{\Gamma} \mathbb{E}_{\:{\mathbf{F}}}\{\| \mathbf{F}^{\alpha}_{\mathcal{P}} -\mathbf{F}_{\mathcal{P}} \|^{2}_{F}  \}\frac{1}{K L}
     \\ &\overset{(e)}{=}
 {   \int^{F^{-1}_{MP,\lambda}(1 - \alpha)}_{ (1 - \sqrt{\lambda})^2} x f_{MP,\lambda}(x) \:dx} 
  \\ &\overset{(f)}{=}
 {\int_{(1- \sqrt{\frac{\delta \kappa}{\gamma}})^2}^{F^{-1}_{MP,\lambda}(1-T\frac{\gamma}{R})}}{x f_{MP,\lambda}(x) \:dx}
\end{align}
where $(a)$ follows from~\eqref{distortion-equality}, $(b)$ follows from \eqref{Partition}, $(c)$ results from the interchangeability of the statistical mean and the summation, $(d)$ is true since the number of representative supports in $\mathcal{C}_1$ is $\frac{K}{\Delta} \frac{L}{\Gamma}$, $(e)$ results from \eqref{Class_Error}, and $(f)$ results from the fact that $\alpha = T\frac{\gamma}{R}$, which follows from the fact that $ r_{\mathcal{P}} = \min(\Delta , \Gamma) = \Delta $ and from the fact that $NT=\alpha \min(\Delta,\Gamma) \frac{K}{\Delta}  \frac{L}{\Gamma} = \alpha  \frac{K L }{\Gamma}$ (cf.~\eqref{Filling-2},\eqref{Decoding-Encoding-A1},\eqref{Decoding-Encoding-A2}), which in turn says that $R = \frac{K}{N} = T \frac{\Gamma}{\alpha  L} = T \frac{ \gamma}{\alpha}$. The proof of $(f)$ is concluded by noting that $\lambda = \frac{\Delta}{\Gamma} = \frac{\delta \kappa}{ \gamma}$ (cf.~\eqref{eq:zetaKappa}).

Similarly, for the case of $\Delta >\Gamma$, we have 
\begin{align}
    \epsilon 
     &\overset{(a')}{=}\frac{K}{\Delta} \frac{L}{\Gamma} \mathbb{E}_{\:{\mathbf{F}}}\{\| \mathbf{F}^{\alpha}_{\mathcal{P}} -\mathbf{F}_{\mathcal{P}} \|^{2}_{F}  \}\frac{1}{KL}
\\ &\overset{(b')}{=}
 {   \int^{F^{-1}_{MP,\lambda}(1 - \Gamma / \Delta \alpha)}_{ (1 - \sqrt{\lambda})^2} x f_{MP,\lambda}(x) \:dx}   
 \\ &\overset{(c')}{=}
 {\int_{(1- \sqrt{\frac{\delta \kappa}{\gamma}})^2}^{F^{-1}_{MP,\lambda}(1-T \frac{\gamma}{R})}}{x f_{MP,\lambda}(x) \:dx}
\end{align}
where $(a')$ follows from equalities $(a),(b),(c),(d)$ corresponding to the above case of $\Delta \leq \Gamma$, where $(b')$ results from \eqref{Class_Error}, and where $(c')$ results after substituting  $\gamma / \kappa \delta = \Gamma / \Delta, \alpha = T \frac{\kappa \delta}{R}$, which follows after noting that $r_{\mathcal{P}}=\min(\Delta,\Gamma) = \Gamma$ and that $NT = \alpha \min(\Delta,\Gamma) \frac{KL}{\Delta \Gamma} =\alpha \frac{KL}{\Delta}$ (cf.~\eqref{Filling-2},\eqref{Decoding-Encoding-A1},\eqref{Decoding-Encoding-A2}), which in turn says that $R = \frac{T \Delta}{ \alpha L}=T \frac{\kappa \delta}{\alpha}$. The proof of $(f)$ and of the entire lemma
is concluded after recalling our asymptotic setting and also that $T |
\min(\Delta,\Gamma)$ and that $\lambda =\frac{\delta \kappa}{\gamma}$
(cf.~\eqref{eq:zetaKappa})..

\subsection{Converse and Proof of Optimality}
We will here provide a converse on the average normalized error. The converse will then allow us to show that under the disjoint balanced support assumption of Definition~\ref{disjointBalancedSupportAssumption}, the scheme proposed in this appendix is asymptotically optimal. First, Lemma~\ref{lowerbound-on-approximation} will offer a lower bound on the approximation error attributed to each tile, and then  Lemma~\ref{Disjoint-Balanced-Support-Assumption: Equivalence} proves that any scheme satisfying Definition~\ref{disjointBalancedSupportAssumption}, has to have disjoint and tiles with the same $|\mathcal{R}_{\mathcal{P}}|$ and $|\mathcal{C}_{\mathcal{P}}|$. 
Subsequently, Lemma~\ref{Asymptotic-Coverage} will establish the covering requirement for any optimal scheme.  Combining the above will then yield the proof.

Our goal is to prove that
in the limit of large $N$ and constant $\delta, \gamma,\kappa, R$, and under the disjoint balanced  support assumption, the average optimal reconstruction error is bounded as 
\begin{align} \label{lowerboundAverageAsymptotic}
\hat{\epsilon} \geq \Phi_{\text{MP},\lambda}(t,r)=\int_{r}^{t} x f_{\text{MP},\lambda}(x) dx 
\end{align}
where $\lambda = \frac{\delta K}{\gamma L} = \frac{\Delta}{\Gamma}, r=(1 - \sqrt{\lambda})^2$, and where $t$ is the solution to $F_{MP,\lambda}(t)= 1- T \frac{\gamma N}{K}$. To achieve this, we begin with the following lemmas.

\begin{Lemma}\label{lowerbound-on-approximation}
Consider approximating any $\mathbf{F}$ by a product $\mathbf{DE}$ with limited $\Delta, \Gamma$, and consider a disjoint set of $m$ tiles $\mathcal{C}=\{\mathcal{P}_i\}_{i=1}^m$ (corresponding to an arbitrary set of disjoint submatrices $\{\mathbf{F}_{\mathcal{P}_i}\}_{i=1}^m$ of $\mathbf{F}$) each allocated $\alpha_i r_{\mathcal{P}_i}$ contribution supports where $\alpha_i r_{\mathcal{P}_i}\in \mathbb{N}, \alpha_i \in \mathbb{R}, \alpha_i \leq 1$. Then 
\begin{align}
    \sum^{m}_{i=1} \sum^{r_{\mathcal{P}_i}}_{j= (1-\alpha_i) r_{\mathcal{P}_i}+1} \sigma^{2}_j(\mathbf{F}_{\mathcal{P}_i}) \leq \|\mathbf{D}\mathbf{E} -\mathbf{F}\|^{2}_{F}
\end{align}
where $\sigma_j(\mathbf{F}_{\mathcal{P}_i}), j \in [r_{\mathcal{P}_i}], i \in [m]$ are the singular values of each $ \mathbf{F}_{\mathcal{P}_i}$, in descending order.
\end{Lemma}
\begin{proof}
The above can be seen by noting that
    \begin{align}
         \sum^{m}_{i=1} \sum^{r_{\mathcal{P}_i}}_{j= (1-\alpha_i) r_{\mathcal{P}_i}+1} \sigma^{2}_j(\mathbf{F}_{\mathcal{P}_i}) 
         &\overset{(a)}{\leq}\sum^{m}_{i=1} \| \mathbf{D}_{\mathcal{P}_i} \mathbf{E}_{\mathcal{P}_i} - \mathbf{F}_{\mathcal{P}_i}\|^{2}_{F} \\
         &\overset{(b)}{\leq} \| \mathbf{D}\mathbf{E} - \mathbf{F}\|^{2}_{F}
    \end{align}
where, for $\mathbf{D}_{\mathcal{P}_i} \mathbf{E}_{\mathcal{P}_i}$ being a rank-$\alpha r_{\mathcal{P}_i}$ approximation of $\mathbf{F}_{\mathcal{P}_i}$, then $(a)$ follows directly from the Eckart-Young Theorem that guarantees that $\sum^{r_{\mathcal{P}_i}}_{j= (1-\alpha_i) r_{\mathcal{P}_i}+1} \sigma^{2}_j(\mathbf{F}_{\mathcal{P}_i})  \leq  \| \mathbf{D}_{\mathcal{P}_i} \mathbf{E}_{\mathcal{P}_i} - \mathbf{F}_{\mathcal{P}_i}\|^{2}_{F}$, while $(b)$ follows after considering \eqref{Formation-of-submatrices}, after considering that the representative supports are disjoint, that the rank of $\mathbf{D}_{\mathcal{P}_i}\mathbf{E}_{\mathcal{P}_i}$ is $\alpha_i r_{\mathcal{P}_i}$, and after considering that parts of $\mathbf{F}$ may remain uncovered by the tiles. 
\end{proof}

We proceed with the next lemma. 
 \begin{Lemma} \label{Disjoint-Balanced-Support-Assumption: Equivalence}
For two matrices $\mathbf{D},\mathbf{E}$, the representative supports $\{\mathbf{S}_{\mathcal{P}_i}\}_{i=1}^m$ of $\mathbf{DE}$ are disjoint (i.e., $\mathbf{S}_{\mathcal{P}_i} \cap \mathbf{S}_{\mathcal{P}_j} = \mathbf{0}, \ j\neq i $)  and $|\mathcal{R}_{\mathcal{P}_i}| = |\mathcal{R}_{\mathcal{P}_j}|,\mathcal{C}_{\mathcal{P}_i}| = |\mathcal{C}_{\mathcal{P}_j}| \: \forall i,j \in [m]$ if and only if $\mathbf{D}$ and $\mathbf{E}$ accept the disjoint balanced  support assumption of Definition~\ref{disjointBalancedSupportAssumption}.
 \end{Lemma}
 \begin{proof}
  
   The proof follows from the same proof of Lemma~\ref{disjointSupportAssumption}, after additionally now noting that for any class $\mathcal{P}_{i} \in \mathcal{C}$ then $|\mathcal{R}_{\mathcal{P}_{i}} | = \| \mathbf{I}(:,i)\|_0$ for some  $i \in [NT]$, and after noting that due to the disjoint balanced support assumption of Definition~\ref{disjointBalancedSupportAssumption} we have that $\|\mathbf{D}(:,i)\|_0 = \|\Supp(\mathbf{D}(:,i))\|_0 = \|\mathbf{D}(:,j)\|_0 = \|\Supp(\mathbf{D}(:,j))\|_0, \forall i \neq j, i ,j \in [NT] $, which in turn implies that $|\mathcal{R}_{\mathcal{P}_i} | = |\mathcal{R}_{\mathcal{P}_j} |, \forall i, j \in [m]$. Similar considerations to $\mathcal{R}_{\mathcal{P}}$, also apply when focusing on $\mathcal{C}_{\mathcal{P}}$.  
 \end{proof}
We also have the following lemma. 
 \begin{Lemma} \label{Asymptotic-Coverage}
  In the limit of large $N$ and fixed $R, \frac{K}{L}$, and under the assumption of demand matrices  $\mathbf{F} \in \mathbb{R}^{K \times L}$ drawn with \emph{i.i.d} entries having zero mean and unit variance, then any optimal scheme that satisfies the disjoint balanced support assumption of Definition~\ref{disjointBalancedSupportAssumption} having fixed $\frac{|\mathcal{R}_{\mathcal{P}_i}|}{K}, \frac{|\mathcal{C}_{\mathcal{P}_j}|}{L}, \forall \mathcal{P}_i \in \mathcal{C} =\{\mathcal{P}_1,\mathcal{P}_2,\hdots,\mathcal{P}_m\}$, must guarantee that the union of all representative supports covers $\mathbf{F}$.  
    \end{Lemma}

\begin{proof}
We prove this by contradiction. Suppose that there exists an optimal scheme whose union of representative supports (union of tiles) does not cover $\mathbf{F}$. Since the  scheme satisfies  Definition~\ref{disjointBalancedSupportAssumption},  by Lemma~\ref{Disjoint-Support-Assumption: Equivalence}, we can let   $|\mathcal{R}_{\mathcal{P}_i}| = |\mathcal{R}_{\mathcal{P}_j}| = X, \forall i,j \in [m] $ and  $|\mathcal{C}_{\mathcal{P}_i}| = |\mathcal{C}_{\mathcal{P}_j}| = Y, \forall i,j \in [m] $. We also see due to the fact that the scheme does not cover $\mathbf{F}$, then 
\begin{align}
[K] \times [L] \backslash \cup_{\mathcal{P} \in \mathcal{C}} \mathcal{R}_{\mathcal{P}} \times \mathcal{C}_{\mathcal{P}} \neq \emptyset.\end{align}Furthermore we see that 
\begin{align}
    [K] \times [L] \backslash \cup_{\mathcal{P} \in \mathcal{C}} \mathcal{R}_{\mathcal{P}} \times \mathcal{C}_{\mathcal{P}} &= \cap_{\mathcal{P} \in \mathcal{C}}([K] \times [L] \backslash \mathcal{R}_{\mathcal{P}} \times \mathcal{C}_{\mathcal{P}})\\ &= \cap_{\mathcal{P} \in \mathcal{C}} ((([K]\backslash \mathcal{R}_{\mathcal{P}}) \times [L] )\cup ([K] \times [L] \backslash \mathcal{C}_{\mathcal{P}} ))
    \\ &= \cup_{\mathcal{A} \subseteq \mathcal{C}} ((([K]\backslash \cup_{i \in \mathcal{A}}\mathcal{R}_{\mathcal{P}_i}) \times [L] )\cap ([K] \times [L] \backslash \cup_{i \in (\mathcal{C}{\backslash}\mathcal{A})}\mathcal{C}_{\mathcal{P}_I} ))
     \\ &= \cup_{\mathcal{A} \subseteq \mathcal{C}} ([K]\backslash \cup_{i \in \mathcal{A}}\mathcal{R}_{\mathcal{P}_i} \times[L] \backslash \cup_{i \in (\mathcal{C}{\backslash}\mathcal{A})}\mathcal{C}_{\mathcal{P}_i} ) \neq \emptyset\label{hole-indices}
\end{align} 
and thus there exists a subset $\mathcal{A} \subseteq \mathcal{C}$ such that $[K]\backslash \cup_{i \in \mathcal{A}}\mathcal{R}_{\mathcal{P}_i} \times[L] \backslash \cup_{i \in (\mathcal{C}{\backslash}\mathcal{A})}\mathcal{C}_{\mathcal{P}_i} \neq \emptyset$, which in turn says that $[K]\backslash \cup_{i \in \mathcal{A}}\mathcal{R}_{\mathcal{P}_i}\neq \emptyset$ and $[L] \backslash \cup_{i \in (\mathcal{C}{\backslash}\mathcal{A})}\mathcal{C}_{\mathcal{P}_i}\neq \emptyset$. Since $|[K]\backslash \cup_{i \in \mathcal{A}}\mathcal{R}_{\mathcal{P}_i}| = K - |\cup_{i \in \mathcal{A}}\mathcal{R}_{\mathcal{P}_i}|$, then from Definition~\ref{disjointBalancedSupportAssumption} and Lemma~\ref{disjointSupportAssumption}, we see that either $\mathcal{R}_{\mathcal{P}_i} = \mathcal{R}_{\mathcal{P}_j},\: \forall i,j \in \mathcal{A}$ or that $\mathcal{R}_{\mathcal{P}_i} \cap \mathcal{R}_{\mathcal{P}_j} = \emptyset,\: \forall i\neq j, i,j \in \mathcal{A}$. We also recall that $|\mathcal{R}_{\mathcal{P}_i}| = X, \forall i \in [m]$, and thus that $X$ divides $|\cup_{i \in \mathcal{A}}\mathcal{R}_{\mathcal{P}_i}|$. Since also $X$ divides $K$, we can see that $X$ divides $|[K]\backslash \cup_{i \in \mathcal{A}}\mathcal{R}_{\mathcal{P}_i}|$. A similar argument can be made to show that $Y$ divides $\: |[L] \backslash \cup_{i \in (\mathcal{C}{\backslash}\mathcal{A})}\mathcal{C}_{\mathcal{P}_i}|$. 

Now, to show that this assumption results in a contradiction, we construct another scheme for the same set of parameters and analyse its performance.  We first recall that $NT = \sum^{m}_{i=1} \text{rank}(\mathbf{F}^{\alpha}_{\mathcal{P}}) = \sum^{m}_{i=1} \alpha_i r_{\mathcal{P}_i}$, i.e., that the sum of the ranks of the matrices $\mathbf{F}^{\alpha}_{\mathcal{P}}$ (cf. \eqref{Error-Sigma}) across $\mathcal{P} \in \mathcal{C}$ is equal to $NT$. Assuming, without loss of generality, that $X \leq Y$, yields $r_{\mathcal{P}} = X, \forall \mathcal{P} \in \mathcal{C}$ which in turn yields
\begin{align}
    NT = \sum^{m}_{i=1} \alpha_i r_{\mathcal{P}_i} = \sum^{m}_{i=1} \alpha_i X. \label{Server-tiles-c-optimal}
\end{align}

Now, consider another scheme with the same positioning of tiles, though now an additional tile $\mathcal{P}_{m+1}$, with a set of row indices $\mathcal{R}_{\mathcal{P}_{m+1}} \subseteq \cup_{i \in \mathcal{A}}\mathcal{R}_{\mathcal{P}_i}, |\mathcal{R}_{\mathcal{P}}| = X$ and column indices $\mathcal{C}_{\mathcal{P}_{m+1}} \subseteq [L] \backslash \cup_{i \in (\mathcal{C}{\backslash}\mathcal{A})}\mathcal{C}_{\mathcal{P}_i}, |\mathcal{C}_{\mathcal{P}_{m+1}}| = Y$. Let the now additionally introduced tiles be placed inside the previously ``uncovered" area of the previous scheme, which means that $ \mathcal{R}_{\mathcal{P}_{m+1}} \times \mathcal{C}_{\mathcal{P}_{m+1}}\subseteq [K] \times [L] \backslash \cup_{\mathcal{P} \in \mathcal{C}} \mathcal{R}_{\mathcal{P}} \times \mathcal{C}_{\mathcal{P}}$. Let us now choose $\beta < \alpha_1, \beta r_{\mathcal{P}_1} \in \mathbb{N}$ and then let $ \mathbf{F}_{\mathcal{P}_1}$ be now estimated by a matrix $\mathbf{F}^{\alpha_1 - \beta}_{\mathcal{P}_1}$ of  rank $(\alpha_1 - \beta)X$ (rather than by matrix    $\mathbf{F}^{\alpha_1}_{\mathcal{P}_1}$ of rank $\alpha_1 X$.). We now recall that the error of the first tile of the optimal scheme is
\begin{align}
    \|\mathbf{F}^{\alpha_1}_{\mathcal{P}_1}  - \mathbf{F}_{\mathcal{P}_1} \|_{F} =  {\sum^{ (1 - \alpha_1)r_{\mathcal{P}}}_{i=1}  \sigma^{2}_{i + \alpha_1 r_{\mathcal{P}}}(\mathbf{F}_{\mathcal{P}})}\label{The-error-of-the-first-tile}
\end{align} 
which we know to be optimal from the Eckart-Young Theorem. 
Now we see that the error corresponding to the first tile, for the new scheme --- which again employs the optimal truncated SVD method --- takes the form 
\begin{align}
    \|\mathbf{F}^{\alpha_1- \beta}_{\mathcal{P}_1}  - \mathbf{F}_{\mathcal{P}_1} \|_{F} =  {\sum^{ (1 - \alpha_1 + \beta )r_{\mathcal{P}}}_{i=1}  \sigma^{2}_{i + (\alpha_1 -\beta )r_{\mathcal{P}}}(\mathbf{F}_{\mathcal{P}})}.\label{The-error-of-the-first-tile-c}
\end{align}
For the same new scheme, the last  used $\mathbf{F}_{\mathcal{P}_{m+1}}$ is being approximated by matrix $\mathbf{F}^{\beta}_{\mathcal{P}_{m+1}}$, which  has rank $\beta r_{\mathcal{P}_{m+1}} = \beta X $, yielding a last-tile error of the form
\begin{align}
    \|\mathbf{F}^{\beta}_{\mathcal{P}_{m+1}}  - \mathbf{F}_{\mathcal{P}_{m+1}} \|_{F} =  {\sum^{ (1 - \beta )r_{\mathcal{P}}}_{i=1}  \sigma^{2}_{i + \beta r_{\mathcal{P}}}(\mathbf{F}_{\mathcal{P}})}. \label{The-error-of-the-first-tile-m+1}
\end{align}
We also note that both the original and the new scheme incur $NT = (\alpha_1 - \beta) r_{\mathcal{P}_1}  +   \sum^{m}_{i=2} \alpha_i r_{\mathcal{P}_i} + \beta r_{\mathcal{P}_{m+1}} = ((\alpha_1 - \beta)  +  \sum^{m}_{i=2} \alpha_i +\beta  ) X =  \sum^{m}_{i=1} \alpha_i X$ which comes from~\eqref{Server-tiles-c-optimal}. 

Now we can see that the average normalized error of the scheme can be thus expressed as
\begin{align}
    \epsilon_{opt} =  \frac{\mathbb{E}_{F}\{ \sum_{i=1}^{(1-\alpha_1) X} \sigma_{i + \alpha_1  X }(\mathbf{F}_{\mathcal{P}_i})\}}{KL} + \epsilon_{\{2,\hdots,m\}}+ \frac{\mathbb{E}_{\mathbf{F}}\{\sum_{(i,j) \in   [K] \times [L] \backslash \cup_{\mathcal{P} \in \mathcal{C}} \mathcal{R}_{\mathcal{P}} \times \mathcal{C}_{\mathcal{P}} } \mathbf{F}(i,j)^2 \}}{KL}\label{normalized-error-optimal-scheme}
\end{align}
where the first fraction describes the error due to the first tile (cf. \eqref{The-error-of-the-first-tile}), the second term $\epsilon_{\{2,\hdots,m\}}$ is the average normalized error of the second to  last tiles, while the last term represents the Frobenius norm for the part of the matrix that represents the ``hole" left uncovered. Then the average normalized error of the constructed  scheme can be expressed as
\begin{align}
    \epsilon_{c} &= \frac{\mathbb{E}_{F}\{ \sum_{i=1}^{(1-\alpha_1 + \beta) X} \sigma_{i + (\alpha_1 -\beta)X }(\mathbf{F}_{\mathcal{P}_i})\}}{KL} + \epsilon_{\{2,\hdots,m\}}\notag\\ &+ \frac{ \mathbb{E}_{\mathbf{F}}\{\sum_{(i,j) \in   [K] \times [L] \backslash \cup_{\mathcal{P} \in \mathcal{C}} \mathcal{R}_{\mathcal{P}} \times \mathcal{C}_{\mathcal{P}}} \mathbf{F}(i,j)^2 - \sum_{i=1}^{\beta X} \sigma_{i }(\mathbf{F}_{\mathcal{P}_{m+1}}) \}}{KL} \label{normalized-error-constructed-scheme}
\end{align}
where the first term corresponds to the first tile (cf.~\eqref{The-error-of-the-first-tile-c}), the second term $\epsilon_{\{2,\hdots,m\}}$ is again the average normalized error corresponding to the second to $m$th tiles, while the last error term is attributed to tile $m+1$ and to the still uncovered area. This last expression is simplified after considering the summation of the average Frobenious norm of the uncovered area that yields error  $ \mathbb{E}_{\mathbf{F}} \{\sum_{(i,j) \in   [K] \times [L] \backslash \cup_{\mathcal{P} \in \mathcal{C}} \mathcal{R}_{\mathcal{P}} \times \mathcal{C}_{\mathcal{P}}} \mathbf{F}(i,j)^2  - \|\mathbf{F}_{\mathcal{P}_{m+1}}\|_{\mathbf{F}}\}$, as well as after considering that the error $\mathbb{E}_{\mathbf{F}} \{\|\mathbf{F}^{\beta}_{\mathcal{P}_{m+1}}  - \mathbf{F}_{\mathcal{P}_{m+1}} \|_{F}\}$ of the $(m+1)$th tile takes the form in~\eqref{The-error-of-the-first-tile-m+1}, and finally after noting that $ \|\mathbf{F}^{\beta}_{\mathcal{P}_{m+1}}  - \mathbf{F}_{\mathcal{P}_{m+1}} \|_{F} - \|\mathbf{F}_{\mathcal{P}_{m+1}}\|_{\mathbf{F}} = - \sum_{i=1}^{\beta X} \sigma_{i }(\mathbf{F}_{\mathcal{P}_{m+1}})$. Thus we can conclude that the normalized error attributed to tile $m+1$ and to the still uncovered area,  takes the form of the numerator of \eqref{normalized-error-constructed-scheme}. 

Thus we can now see that 
\begin{align}
    \epsilon_{opt} - \epsilon_{c} = \frac{XY }{KL}( -  \int^{F^{-1}_{MP,\lambda}(1 - \alpha_1 + \beta)}_{F^{-1}_{MP,\lambda}(1 - \alpha_1)} x f_{MP,\lambda}(x) \:dx +   \int^{F^{-1}_{MP,\lambda}(1)}_{F^{-1}_{MP,\lambda}(1 - \beta)} x f_{MP,\lambda}(x) \:dx ),\: \lambda = \frac{X}{Y} \label{differnece}
\end{align}
which follows after considering~\eqref{normalized-error-optimal-scheme}, \eqref{normalized-error-constructed-scheme}, as well as after following the corresponding steps found in the proof of Lemma~\ref{MP-relation-Lemma}. We can now also note that 
\begin{align*}
    \frac{d}{dt} \int^{F^{-1}_{MP,\lambda}(t)}_{F^{-1}_{MP,\lambda}(t-  \beta)} x f_{MP,\lambda}(x) \:dx &= \frac{1}{f_{MP,\lambda}(F^{-1}_{MP,\lambda}(t))}  F^{-1}_{MP,\lambda}(t) f_{MP,\lambda}(F^{-1}_{MP,\lambda}(t)) \\&- \frac{1}{f_{MP,\lambda}(F^{-1}_{MP,\lambda}(t-\beta))}  F^{-1}_{MP,\lambda}(t-\beta) f_{MP,\lambda}(F^{-1}_{MP,\lambda}(t-\beta))\\& =  F^{-1}_{MP,\lambda}(t) -  F^{-1}_{MP,\lambda}(t-\beta) >0 
\end{align*}
where the last inequality is due to the fact that $F^{-1}_{MP,\lambda}(t)$ is a strictly monotonically increasing function of $t$, which means that the function $\int^{F^{-1}_{MP,\lambda}(t)}_{F^{-1}_{MP,\lambda}(t-  \beta)} x f_{MP,\lambda}(x) \:dx$  for all $\beta \leq t\leq 1$ is a strictly monotonically increasing function of $t$, which now yields $\int^{F^{-1}_{MP,\lambda}(1)}_{F^{-1}_{MP,\lambda}(1 - \beta)} x f_{MP,\lambda}(x) \:dx  > \int^{F^{-1}_{MP,\lambda}(1 - \alpha_1 + \beta)}_{F^{-1}_{MP,\lambda}(1 - \alpha_1)} x f_{MP,\lambda}(x) \:dx$, and thus --- after considering~\eqref{differnece} --- yields $ \epsilon_{opt} - \epsilon_{c}  >0$, which contradicts our initial optimality assumption. By showing that there can always be a scheme that offers strictly lower error than the best scheme that fails to fully cover $\mathbf{F}$, we prove that optimal schemes must cover $\mathbf{F}$.   
\end{proof}

Continuing with our main proof, we recall that $\mathbf{F}$ is a $K \times L$ real random matrix whose entries are \emph{i.i.d} with zero mean and unit variance. In this context,
and in the limit of large $N$, 
we recall Lemma~\ref{Disjoint-Support-Assumption: Equivalence} and Definition~\ref{disjointBalancedSupportAssumption}, and after seeing that $|\mathcal{R}_{\mathcal{P}}| = X, |\mathcal{C}_{\mathcal{P}} |= Y, \forall \mathcal{P} \in \mathcal{C}$, we first consider the case of $ X \leq Y$, where we have $r=\frac{X}{Y}$ and $\sum^{m}_{i=1} \alpha_i X = NT$, and proceed to see that 
\begin{align*}
    \mathbb{E}_{\mathbf{F}}\{\sum^{m}_{i=1} \sum^{r_{\mathcal{P}_i}}_{j= (1-\alpha_i) r_{\mathcal{P}_i}+1} \sigma^{2}_j(\mathbf{F}_{\mathcal{P}_i})\} &\overset{(a)}{=}Y  \mathbb{E}_{\mathbf{F}}\Big \{\frac{  \sum^{m}_{i=1} \sum^{\alpha_i X}_{j=1} \sigma^{2}_{j+ (1-\alpha_i) X} (\mathbf{F}_{\mathcal{P}_i})}{Y}\Big \}\\
     &\overset{(b)}{=}Y  \mathbb{E}_{\mathbf{F}}\Big \{ \sum^{m}_{i=1} \sum^{\alpha_i X}_{j=1} \int^{\infty}_{-\infty}\delta(x - \frac{ \sigma^{2}_{j+ (1-\alpha_i) X} (\mathbf{F}_{\mathcal{P}_i}) }{Y}) x dx\Big \}\\
&\overset{(c)}{=}Y  \mathbb{E}_{\mathbf{F}}\Big \{ \sum^{m}_{i=1} \sum^{\alpha_i X}_{j=1} \int^{\infty}_{-\infty}(1 / \delta x)\mathbf{1}(  \frac{ \sigma^{2}_{j+ (1-\alpha_i) X} (\mathbf{F}_{\mathcal{P}_i}) }{Y} \leq x + \delta x)\\ &-\mathbf{1}(  \frac{ \sigma^{2}_{j+ (1-\alpha_i) X} (\mathbf{F}_{\mathcal{P}_i}) }{Y} \leq x ) x dx\Big \} 
\\
&\overset{(d)}{=}Y  \sum^{m}_{i=1} \mathbb{E}_{\mathbf{F}}\Big \{\sum^{\alpha_i X}_{j=1} \int^{\infty}_{-\infty}(1 / \delta x)\mathbf{1}(  \frac{ \sigma^{2}_{j+ (1-\alpha_i) X} (\mathbf{F}_{\mathcal{P}_i}) }{Y} \leq x + \delta x)\\ &-\mathbf{1}(  \frac{ \sigma^{2}_{j+ (1-\alpha_i) X} (\mathbf{F}_{\mathcal{P}_i}) }{Y} \leq x ) x dx\Big \}
\\
&\overset{(e)}{=}Y X  \sum^{m}_{i=1}  \int^{F^{-1}_{MP,\lambda}(1 -\alpha_i)}_{(1 - \sqrt{r})^{2}}(1 / \delta x)[F_{MP}(x + \delta x) - F_{MP,\lambda}(x)] x dx
\\
&\overset{(f)}{=}YX \sum^{m}_{i=1}  \int^{F^{-1}_{MP,\lambda}(1 -\alpha_i)}_{(1 - \sqrt{r})^{2}}f_{MP}(x ) x dx
\\
&\overset{(g)}{\geq }Y X m \int^{F^{-1}_{MP,\lambda}(1 -\frac{NT}{mX})}_{(1 - \sqrt{r})^{2}}f_{MP}(x ) x dx
\\
&\overset{(h)}{= } KL \int^{F^{-1}_{MP,\lambda}(1 -\frac{NTY}{KL})}_{(1 - \sqrt{r})^{2}}f_{MP}(x ) x dx
\\
&\overset{(i)}{\geq } KL \int^{F^{-1}_{MP,\lambda}(1 -\frac{NT\Gamma}{KL})}_{(1 - \sqrt{\frac{\Delta}{\Gamma}})^{2}}f_{MP}(x ) x dx
\end{align*}
where $(a)$ follows since $Y$ is a constant, $(b)$ follows from basic properties of the delta Dirac function, $(c)$ follows from the fact that the delta Dirac function is a derivative of the step function, $(d)$ follows from the fact that expectation and sum are interchangeable functions, $(e)$ follows from the same reasoning that allowed equality $(f)$ for the case $\Delta \leq \Gamma$ of the proof of Lemma  \ref{MP-relation-Lemma}, and $(f)$ here follows from the fact that $f_{MP}$ is the derivative of $F_{MP}$. Furthermore, $(g)$ follows by first setting  $g(\alpha_i)\triangleq \int^{F^{-1}_{MP,\lambda}(1 -\alpha_i)}_{(1 - \sqrt{r})^{2}}f_{MP}(x )$, and then by recalling that $\sum^{m}_{i=1} \alpha_i = \frac{NT}{X}$, and subsequently by noting that $g(\alpha_i)$ is a monotonic decreasing convex function of $\alpha_i, \forall 0\leq \alpha_i \leq 1$, which in turn implies that its Lagrange optimizer function will be $\mathcal{L}(\alpha_1, \alpha_2,\hdots, \alpha_m) = Y \sum^{m}_{i=1} g(\alpha_i) - \lambda(\sum^{m}_{i=1} \alpha_i - \frac{NT}{X})$, where to ensure  $\frac{\partial \mathcal{L}}{\partial \alpha_i} = 0, \forall i \in [m]$ we have $\frac{\partial g(\alpha_i)}{\partial \alpha_i} = F^{-1}_{MP,\lambda}(1 - \frac{NT}{mX})= \lambda, \forall \alpha_i$, which in turn means that $\alpha_1 =\alpha_2= \hdots = \alpha_m =\alpha$ which, combined with the optimization constraints, yields that $\alpha= \frac{NT}{mX}$. Then $(h)$ results by noting that $m= \frac{KL}{XY}$, which itself holds since, by Lemma~\ref{Asymptotic-Coverage}, we have that for any asymptotic scheme with bounded normalized error, the representative supports or tiles must cover the whole matrix $\mathbf{F}$. Finally $(i)$ follows from the fact that $X\leq \Delta, Y \leq \Gamma$ and after seeing that the derivative of the RHS side of $(h)$ with respect $X$ is $\frac{KL}{X}((-2(1 -  \sqrt{\frac{X}{Y}}))(\frac{1}{Y})(-\frac{1}{2})\frac{X}{Y}^{-1 / 2} f_{MP}((1 - \sqrt{X/Y})^2)(1 - \sqrt{\frac{X}{Y}})^2) -\frac{KL}{X^2} \int_{(1-\sqrt{\frac{X}{Y}})^{2}}^{F_{MP}^{-1}(1 - \frac{NT}{KL} Y)}x f_{MP,\lambda}(x) dx $, as well as by noting that since $ f_{MP}((1 - \sqrt{X/Y})^2) =0$, the derivative of the RHS side of $(h)$  is always negative, which in turn means that the RHS side of $(h)$ is a monotonically decreasing function of $X$ which in turn yields that $X = \Delta$ gives the optimal solution. This is then combined with the fact that the derivative of the RHS side of $(h)$ with respect regarding $Y$ is $-\frac{NT}{KL} \frac{1}{f_{MP}(F^{-1}(1 - \frac{NTY}{KL}))}{F^{-1}_{MP,\lambda}(1 - \frac{NTY}{KL}) f_{MP}(F^{-1}_{MP,\lambda}(1 -\frac{NTY}{KL}))} - 2(1 - \sqrt{\frac{X}{Y}}) (- \frac{1}{2}) (\frac{X}{Y})^{-1 / 2} X Y^{-2} = -\frac{NT}{KL} F_{MP}^{-1}(1 -\frac{NTY}{KL}) - (1 - \sqrt{X / Y}) (\frac{X}{Y})^{-1 / 2} XY^{-2}$. Finally, since $X \leq Y$, this derivative is also negative, which in turn means that $Y=\Gamma$, with the same reasoning as the case where $X=\Delta$.
This proves that
    \begin{align} \label{IntegrUpper1}
    \mathbb{E}_{\mathbf{F}}\{\| \mathbf{D}\mathbf{E} - \mathbf{F}\|^{2}_{F}\}
 \geq   KL {\int_{(1- \sqrt{\frac{\delta \kappa}{\gamma}})^2}^{F^{-1}_{MP,\lambda}(1-T\frac{\gamma}{R})}}{x f_{MP,\lambda}(x) \:dx}
    \end{align}
for the case of $ X \leq Y$. As a result, this proves the lower bound in \eqref{lowerboundAverageAsymptotic}, and thus the entire lower bound (converse) of Theorem~\ref{asymptotic-capacity}, over all schemes for which $|\mathcal{R}_{\mathcal{P}}|  \geq |\mathcal{C}_{\mathcal{P}} |, \forall \mathcal{P} \in \mathcal{C}$. 

Now for the case of $ Y \leq X$, similarly we have $r=\frac{X}{Y}$, and now with $\sum^{m}_{i=1} \alpha_i Y = NT$, we see that
\begin{align*}
    \mathbb{E}_{\mathbf{F}}\{\sum^{m}_{i=1} \sum^{r_{\mathcal{P}_i}}_{j= (1-\alpha_i) r_{\mathcal{P}_i}+1} \sigma^{2}_j(\mathbf{F}_{\mathcal{P}_i})\} 
&\overset{(a,b,c,d)}{=}Y   \sum^{m}_{i=1} \mathbb{E}_{\mathbf{F}}\Big \{\sum^{\alpha_i Y}_{j=1} \int^{\infty}_{-\infty}(1 / \delta x)\mathbf{1}(  \frac{ \sigma^{2}_{j+ (1-\alpha_i) X} (\mathbf{F}_{\mathcal{P}_i}) }{Y} \leq x + \delta x)\\ &-\mathbf{1}(  \frac{ \sigma^{2}_{j+ (1-\alpha_i) X} (\mathbf{F}_{\mathcal{P}_i}) }{Y} \leq x ) x dx\Big \}
\\
&\overset{(e)}{=}Y X  \sum^{m}_{i=1}  \int^{F^{-1}_{MP,\lambda}(1 - \frac{Y}{X}\alpha_i)}_{(1 - \sqrt{r})^{2}}(1 / \delta x)[F_{MP}(x + \delta x) - F_{MP,\lambda}(x)] x dx
\\
&\overset{(f)}{=}Y X\sum^{m}_{i=1}  \int^{F^{-1}_{MP,\lambda}(1 -\frac{Y}{X}\alpha_i)}_{(1 - \sqrt{r})^{2}}f_{MP}(x ) x dx
\\
&\overset{(g)}{\geq }Y X m \int^{F^{-1}_{MP,\lambda}(1 -\frac{NT}{mX})}_{(1 - \sqrt{r})^{2}}f_{MP}(x ) x dx
\\
&\overset{(h)}{= }KL \int^{F^{-1}_{MP,\lambda}(1 -\frac{NTY}{KL}))}_{(1 - \sqrt{r})^{2}}f_{MP}(x ) x dx
\\
&\overset{(i)}{\geq  } KL\int^{F^{-1}_{MP,\lambda}(1 -\frac{NT\Gamma}{KL})}_{(1 - \sqrt{\frac{\Delta}{\Gamma}})^{2}}f_{MP}(x ) x dx
\end{align*}
where the first equality is proven the same way as the equalities $(a,b,c,d)$ in the previous case of $X \leq Y$, and where here $(e)$ is proven similar to equality $(f)$ of the case $\Delta > \Gamma$ in the proof of Lemma~\ref{MP-relation-Lemma}. Furthermore, here, $(f)$ follows from the fact that $f_{MP}$ is the derivative of $F_{MP}$. To establish $(g)$, we first consider  $g(\alpha_i)\triangleq \int^{F^{-1}_{MP,\lambda}(1 -\alpha_i)}_{(1 - \sqrt{r})^{2}} x f_{MP}(x ) dx$ subject to $\sum^{m}_{i=1} \alpha_i = \frac{NT}{Y}$, and then note that $g(\alpha_i)$ is a monotonic decreasing convex function of $\alpha_i$, in the range $\alpha_i \in \mathbb{R}$, thus allowing us to conclude that its Lagrange optimizer function  will be $\mathcal{L}(\alpha_1, \alpha_2,\hdots, \alpha_m) = Y \sum^{m}_{i=1} g(\alpha_i) - \lambda(\sum^{m}_{i=1} \alpha_i - \frac{NT}{Y})$ where to ensure  $\frac{\partial \mathcal{L}}{\partial \alpha_i} = 0, \forall i \in [m]$, we have $\frac{\partial g(\alpha_i)}{\partial \alpha_i} = F^{-1}_{MP,\lambda}(1 - \frac{NT}{m})= \lambda, \forall \alpha_i$, which means that $\alpha_1 =\alpha_2= \hdots = \alpha_m =\alpha$, which is then combined with the optimization constraints to yield that $\alpha= \frac{NT}{mY}$. Subsequently, $(h)$ results after noting that $m= \frac{KL}{XY}$ after recalling Lemma \ref{Asymptotic-Coverage} which tells us that for any asymptotic scheme with  normalized error, the representative supports must cover the entire $\mathbf{F}$. Finally, $(i)$ follows directly as inequality $(i)$ of the case of $X \leq Y$. As a result, this proves the lower bound in \eqref{lowerboundAverageAsymptotic}, and thus the entire lower bound (converse) of Theorem~\ref{asymptotic-capacity}, over all schemes for which $|\mathcal{R}_{\mathcal{P}}|  \leq |\mathcal{C}_{\mathcal{P}} |, \forall \mathcal{P} \in \mathcal{C}$. 

\bibliographystyle{ieeetr}
\bibliography{ref}

\begin{thebibliography}{10}

\bibitem{dean2008mapreduce}
J.~Dean and S.~Ghemawat, ``Mapreduce: simplified data processing on large clusters,'' {\em Communications of the ACM}, vol.~51, no.~1, pp.~107--113, 2008.

\bibitem{zaharia2010spark}
M.~Zaharia, M.~Chowdhury, M.~J. Franklin, S.~Shenker, and I.~Stoica, ``Spark: Cluster computing with working sets,'' in {\em 2nd USENIX Workshop on Hot Topics in Cloud Computing (HotCloud 10)}, 2010.

\bibitem{li2017scalable}
S.~Li, Q.~Yu, M.~A. Maddah-Ali, and A.~S. Avestimehr, ``A scalable framework for wireless distributed computing,'' {\em IEEE/ACM Transactions on Networking}, vol.~25, no.~5, pp.~2643--2654, 2017.

\bibitem{haddadpour2019trading}
F.~Haddadpour, M.~M. Kamani, M.~Mahdavi, and V.~Cadambe, ``Trading redundancy for communication: Speeding up distributed \text{SGD} for non-convex optimization,'' in {\em International Conference on Machine Learning}, pp.~2545--2554, PMLR, 2019.

\bibitem{yang2020coded}
C.-S. Yang, R.~Pedarsani, and A.~S. Avestimehr, ``Coded computing in unknown environment via online learning,'' in {\em 2020 IEEE International Symposium on Information Theory (ISIT)}, pp.~185--190, IEEE, 2020.

\bibitem{charalambides2021numerically}
N.~Charalambides, H.~Mahdavifar, and A.~O. Hero~III, ``Numerically stable binary coded computations,'' {\em arXiv preprint arXiv:2109.10484}, 2021.

\bibitem{soleymani2021analog}
M.~Soleymani, H.~Mahdavifar, and A.~S. Avestimehr, ``Analog \text{Lagrange} coded computing,'' {\em IEEE Journal on Selected Areas in Information Theory}, vol.~2, no.~1, pp.~283--295, 2021.

\bibitem{Brunero1}
F.~Brunero, K.~Wan, G.~Caire, and P.~Elia, ``Coded distributed computing for sparse functions with structured support,'' in {\em 2023 IEEE Information Theory Workshop (ITW)}, pp.~474--479, 2023.

\bibitem{Parrinello1}
E.~Parrinello, E.~Lampiris, and P.~Elia, ``Coded distributed computing with node cooperation substantially increases speedup factors,'' in {\em 2018 IEEE International Symposium on Information Theory (ISIT)}, pp.~1291--1295, 2018.

\bibitem{Reza1}
M.~R. Deylam~Salehi and D.~Malak, ``An achievable low complexity encoding scheme for coloring cyclic graphs,'' in {\em 2023 59th Annual Allerton Conference on Communication, Control, and Computing (Allerton)}, pp.~1--8, Sep. 2023.

\bibitem{Brunero2}
F.~Brunero and P.~Elia, ``Multi-access distributed computing,'' {\em IEEE Transactions on Information Theory}, vol.~Early Access, pp.~1--1, 2024.

\bibitem{Nezhad2023}
T.~Jahani-Nezhad, M.~A. Maddah-Ali, S.~Li, and G.~Caire, ``Swiftagg+: Achieving asymptotically optimal communication loads in secure aggregation for federated learning,'' {\em IEEE Journal on Selected Areas in Communications}, vol.~41, no.~4, pp.~977--989, 2023.

\bibitem{soleymani2020distributed}
M.~Soleymani and H.~Mahdavifar, ``Distributed multi-user secret sharing,'' {\em IEEE Transactions on Information Theory}, vol.~67, no.~1, pp.~164--178, 2020.

\bibitem{khalesi2021capacity}
A.~Khalesi, M.~Mirmohseni, and M.~A. Maddah-Ali, ``The capacity region of distributed multi-user secret sharing,'' {\em IEEE Journal on Selected Areas in Information Theory}, vol.~2, no.~3, pp.~1057--1071, 2021.

\bibitem{soleymani2020privacy}
M.~Soleymani, H.~Mahdavifar, and A.~S. Avestimehr, ``Privacy-preserving distributed learning in the analog domain,'' {\em arXiv preprint arXiv:2007.08803}, 2020.

\bibitem{soleymani2021list}
M.~Soleymani, R.~E. Ali, H.~Mahdavifar, and A.~S. Avestimehr, ``List-decodable coded computing: Breaking the adversarial toleration barrier,'' {\em IEEE Journal on Selected Areas in Information Theory}, vol.~2, no.~3, pp.~867--878, 2021.

\bibitem{bitar2022adaptive}
R.~Bitar, M.~Xhemrishi, and A.~Wachter-Zeh, ``Adaptive private distributed matrix multiplication,'' {\em IEEE Transactions on Information Theory}, vol.~68, no.~4, pp.~2653--2673, 2022.

\bibitem{yang2021coded}
C.-S. Yang and A.~S. Avestimehr, ``Coded computing for secure boolean computations,'' {\em IEEE Journal on Selected Areas in Information Theory}, vol.~2, no.~1, pp.~326--337, 2021.

\bibitem{yu2020coded}
Q.~Yu and A.~S. Avestimehr, ``Coded computing for resilient, secure, and privacy-preserving distributed matrix multiplication,'' {\em IEEE Transactions on Communications}, vol.~69, no.~1, pp.~59--72, 2020.

\bibitem{ehteram2023trainedmpc}
H.~Ehteram, M.~A. Maddah-Ali, and M.~Mirmohseni, ``Trained-{MPC}: A private inference by training-based multiparty computation,'' in {\em MLSys 2023 Workshop on Resource-Constrained Learning in Wireless Networks}, 2023.

\bibitem{AVEST1}
J.~So, R.~E. Ali, B.~G{\"u}ler, J.~Jiao, and A.~S. Avestimehr, ``Securing secure aggregation: Mitigating multi-round privacy leakage in federated learning,'' in {\em Proceedings of the AAAI Conference on Artificial Intelligence}, vol.~37, pp.~9864--9873, 2023.

\bibitem{raviv2020gradient}
N.~Raviv, I.~Tamo, R.~Tandon, and A.~G. Dimakis, ``Gradient coding from cyclic \text{MDS} codes and expander graphs,'' {\em IEEE Transactions on Information Theory}, vol.~66, no.~12, pp.~7475--7489, 2020.

\bibitem{lee2017speeding}
K.~Lee, M.~Lam, R.~Pedarsani, D.~Papailiopoulos, and K.~Ramchandran, ``Speeding up distributed machine learning using codes,'' {\em IEEE Transactions on Information Theory}, vol.~64, no.~3, pp.~1514--1529, 2017.

\bibitem{egger2022efficient}
M.~Egger, R.~Bitar, A.~Wachter-Zeh, and D.~G{\"u}nd{\"u}z, ``Efficient distributed machine learning via combinatorial multi-armed bandits,'' {\em arXiv preprint arXiv:2202.08302}, 2022.

\bibitem{kai1}
K.~Wan, H.~Sun, M.~Ji, and G.~Caire, ``Distributed linearly separable computation,'' {\em IEEE Transactions on Information Theory}, vol.~68, no.~2, pp.~1259--1278, 2022.

\bibitem{yu2020straggler}
Q.~Yu, M.~A. Maddah-Ali, and A.~S. Avestimehr, ``Straggler mitigation in distributed matrix multiplication: Fundamental limits and optimal coding,'' {\em IEEE Transactions on Information Theory}, vol.~66, no.~3, pp.~1920--1933, 2020.

\bibitem{yu2017polynomial}
Q.~Yu, M.~Maddah-Ali, and S.~Avestimehr, ``Polynomial codes: an optimal design for high-dimensional coded matrix multiplication,'' {\em Advances in Neural Information Processing Systems}, vol.~30, 2017.

\bibitem{jia2021cross}
Z.~Jia and S.~A. Jafar, ``Cross subspace alignment codes for coded distributed batch computation,'' {\em IEEE Transactions on Information Theory}, vol.~67, no.~5, pp.~2821--2846, 2021.

\bibitem{ng2020survey}
J.~S. Ng, W.~Y.~B. Lim, N.~C. Luong, Z.~Xiong, A.~Asheralieva, D.~Niyato, C.~Leung, and C.~Miao, ``A comprehensive survey on coded distributed computing: Fundamentals, challenges, and networking applications,'' {\em IEEE Communications Surveys \& Tutorials}, vol.~23, no.~3, pp.~1800--1837, 2021.

\bibitem{CIT-103}
S.~Li and S.~Avestimehr, {\em Coded Computing: Mitigating Fundamental Bottlenecks in Large-Scale Distributed Computing and Machine Learning}, vol.~17.
\newblock now Publishers Inc, 2020.

\bibitem{Parrinello5}
E.~Parrinello, A.~Bazco-Nogueras, and P.~Elia, ``Fundamental limits of topology-aware shared-cache networks,'' {\em IEEE Transactions on Information Theory}, vol.~70, no.~4, pp.~2538--2565, 2024.

\bibitem{Lampiris}
E.~Lampiris and P.~Elia, ``Full coded caching gains for cache-less users,'' {\em IEEE Transactions on Information Theory}, vol.~66, no.~12, pp.~7635--7651, 2020.

\bibitem{verbraeken2020survey}
J.~Verbraeken, M.~Wolting, J.~Katzy, J.~Kloppenburg, T.~Verbelen, and J.~S. Rellermeyer, ``A survey on distributed machine learning,'' {\em ACM Computing Surveys (CSUR)}, vol.~53, no.~2, pp.~1--33, 2020.

\bibitem{ulukus2022private}
S.~Ulukus, S.~Avestimehr, M.~Gastpar, S.~Jafar, R.~Tandon, and C.~Tian, ``Private retrieval, computing and learning: Recent progress and future challenges,'' {\em IEEE Journal on Selected Areas in Communications}, 2022.

\bibitem{wang2018fundamental}
S.~Wang, J.~Liu, N.~Shroff, and P.~Yang, ``Fundamental limits of coded linear transform,'' {\em arXiv preprint arXiv:1804.09791}, 2018.

\bibitem{li2017fundamental}
S.~Li, M.~A. Maddah-Ali, Q.~Yu, and A.~S. Avestimehr, ``A fundamental tradeoff between computation and communication in distributed computing,'' {\em IEEE Transactions on Information Theory}, vol.~64, no.~1, pp.~109--128, 2017.

\bibitem{wan2022cache}
K.~Wan, H.~Sun, M.~Ji, D.~Tuninetti, and G.~Caire, ``Cache-aided matrix multiplication retrieval,'' {\em IEEE Transactions on Information Theory}, vol.~68, no.~7, pp.~4301--4319, 2022.

\bibitem{yu2017polynomial2}
Q.~Yu, M.~Maddah-Ali, and S.~Avestimehr, ``Polynomial codes: an optimal design for high-dimensional coded matrix multiplication,'' {\em Advances in Neural Information Processing Systems}, vol.~30, 2017.

\bibitem{dutta2019optimal}
S.~Dutta, M.~Fahim, F.~Haddadpour, H.~Jeong, V.~Cadambe, and P.~Grover, ``On the optimal recovery threshold of coded matrix multiplication,'' {\em IEEE Transactions on Information Theory}, vol.~66, no.~1, pp.~278--301, 2019.

\bibitem{reisizadeh2021codedreduce}
A.~Reisizadeh, S.~Prakash, R.~Pedarsani, and A.~S. Avestimehr, ``Codedreduce: A fast and robust framework for gradient aggregation in distributed learning,'' {\em IEEE/ACM Transactions on Networking}, 2021.

\bibitem{woolsey2021new}
N.~Woolsey, R.-R. Chen, and M.~Ji, ``A new combinatorial coded design for heterogeneous distributed computing,'' {\em IEEE Transactions on Communications}, vol.~69, no.~9, pp.~5672--5685, 2021.

\bibitem{woolsey2021coded}
N.~Woolsey, R.-R. Chen, and M.~Ji, ``Coded elastic computing on machines with heterogeneous storage and computation speed,'' {\em IEEE Transactions on Communications}, vol.~69, no.~5, pp.~2894--2908, 2021.

\bibitem{woolsey2021practical}
N.~Woolsey, J.~Kliewer, R.-R. Chen, and M.~Ji, ``A practical algorithm design and evaluation for heterogeneous elastic computing with stragglers,'' in {\em 2021 IEEE Global Communications Conference (GLOBECOM)}, pp.~1--6, IEEE, 2021.

\bibitem{chen2021distributed}
M.~Chen, D.~G{\"u}nd{\"u}z, K.~Huang, W.~Saad, M.~Bennis, A.~V. Feljan, and H.~V. Poor, ``Distributed learning in wireless networks: Recent progress and future challenges,'' {\em IEEE Journal on Selected Areas in Communications}, vol.~39, no.~12, pp.~3579--3605, 2021.

\bibitem{wang2021price}
J.~Wang, Z.~Jia, and S.~A. Jafar, ``Price of precision in coded distributed matrix multiplication: A dimensional analysis,'' in {\em 2021 IEEE Information Theory Workshop (ITW)}, pp.~1--6, IEEE, 2021.

\bibitem{ozfatura2021coded}
E.~Ozfatura, S.~Ulukus, and D.~G{\"u}nd{\"u}z, ``Coded distributed computing with partial recovery,'' {\em IEEE Transactions on Information Theory}, vol.~68, no.~3, pp.~1945--1959, 2022.

\bibitem{Malak}
D.~Malak and M.~Médard, ``A distributed computationally aware quantizer design via hyper binning,'' {\em IEEE Transactions on Signal Processing}, vol.~71, pp.~76--91, 2023.

\bibitem{woodruff2014sketching}
D.~P. Woodruff {\em et~al.}, ``Sketching as a tool for numerical linear algebra,'' {\em Foundations and Trends{\textregistered} in Theoretical Computer Science}, vol.~10, no.~1--2, pp.~1--157, 2014.

\bibitem{jahani2021codedsketch}
T.~Jahani-Nezhad and M.~A. Maddah-Ali, ``Codedsketch: A coding scheme for distributed computation of approximated matrix multiplication,'' {\em IEEE Transactions on Information Theory}, vol.~67, no.~6, pp.~4185--4196, 2021.

\bibitem{RaviApproximated}
W.-T. Chang and R.~Tandon, ``Random sampling for distributed coded matrix multiplication,'' in {\em ICASSP 2019 - 2019 IEEE International Conference on Acoustics, Speech and Signal Processing (ICASSP)}, pp.~8187--8191, 2019.

\bibitem{CharalambidesApproximated}
N.~Charalambides, M.~Pilanci, and A.~O. Hero, ``Approximate weighted \text{CR} coded matrix multiplication,'' in {\em ICASSP 2021 - 2021 IEEE International Conference on Acoustics, Speech and Signal Processing (ICASSP)}, pp.~5095--5099, 2021.

\bibitem{RamchandranApproximated}
V.~Gupta, S.~Wang, T.~Courtade, and K.~Ramchandran, ``Oversketch: Approximate matrix multiplication for the cloud,'' in {\em 2018 IEEE International Conference on Big Data (Big Data)}, pp.~298--304, 2018.

\bibitem{StarkApproximated}
N.~S. Ferdinand and S.~C. Draper, ``Anytime coding for distributed computation,'' in {\em 2016 54th Annual Allerton Conference on Communication, Control, and Computing (Allerton)}, pp.~954--960, 2016.

\bibitem{ZhuRamchandranApproximated}
J.~Zhu, Y.~Pu, V.~Gupta, C.~Tomlin, and K.~Ramchandran, ``A sequential approximation framework for coded distributed optimization,'' in {\em 2017 55th Annual Allerton Conference on Communication, Control, and Computing (Allerton)}, pp.~1240--1247, 2017.

\bibitem{TayyebehBerrutMaddah-Ali}
T.~Jahani-Nezhad and M.~A. Maddah-Ali, ``Berrut approximated coded computing: Straggler resistance beyond polynomial computing,'' {\em IEEE Transactions on Pattern Analysis and Machine Intelligence}, vol.~45, no.~1, pp.~111--122, 2023.

\bibitem{CadambeApproximated}
H.~Jeong, A.~Devulapalli, V.~R. Cadambe, and F.~P. Calmon, ``$\epsilon$-approximate coded matrix multiplication is nearly twice as efficient as exact multiplication,'' {\em IEEE Journal on Selected Areas in Information Theory}, vol.~2, no.~3, pp.~845--854, 2021.

\bibitem{KianiStarkApproximated}
S.~Kiani and S.~C. Draper, ``Successive approximation coding for distributed matrix multiplication,'' {\em IEEE Journal on Selected Areas in Information Theory}, vol.~3, no.~2, pp.~286--305, 2022.

\bibitem{NarayananKrishna}
R.~Ji, A.~Heidarzadeh, and K.~R. Narayanan, ``Sparse random \text{Khatri-Rao} product codes for distributed matrix multiplication,'' in {\em 2022 IEEE Information Theory Workshop (ITW)}, pp.~416--421, 2022.

\bibitem{RashmiApproximated}
M.~Rudow, N.~Charalambides, A.~O. Hero, and K.~Rashmi, ``Compression-informed coded computing,'' in {\em 2023 IEEE International Symposium on Information Theory (ISIT)}, pp.~2177--2182, 2023.

\bibitem{ZhuJinggeApproximatedLearning}
N.~Agrawal, Y.~Qiu, M.~Frey, I.~Bjelakovic, S.~Maghsudi, S.~Stanczak, and J.~Zhu, ``A learning-based approach to approximate coded computation,'' in {\em 2022 IEEE Information Theory Workshop (ITW)}, pp.~600--605, 2022.

\bibitem{JahaniNezhadApproximated}
T.~Jahani-Nezhad and M.~A. Maddah-Ali, ``Optimal communication-computation trade-off in heterogeneous gradient coding,'' {\em IEEE Journal on Selected Areas in Information Theory}, vol.~2, no.~3, pp.~1002--1011, 2021.

\bibitem{MaddahAliApproximated}
T.~Jahani-Nezhad and M.~A. Maddah-Ali, ``Codedsketch: Coded distributed computation of approximated matrix multiplication,'' in {\em 2019 IEEE International Symposium on Information Theory (ISIT)}, pp.~2489--2493, 2019.

\bibitem{tandon2017gradient}
R.~Tandon, Q.~Lei, A.~G. Dimakis, and N.~Karampatziakis, ``Gradient coding: Avoiding stragglers in distributed learning,'' in {\em International Conference on Machine Learning}, pp.~3368--3376, PMLR, 2017.

\bibitem{ye2018communication}
M.~Ye and E.~Abbe, ``Communication-computation efficient gradient coding,'' in {\em International Conference on Machine Learning}, pp.~5610--5619, PMLR, 2018.

\bibitem{halbawi2018improving}
W.~Halbawi, N.~Azizan, F.~Salehi, and B.~Hassibi, ``Improving distributed gradient descent using \text{Reed-Solomon} codes,'' in {\em 2018 IEEE International Symposium on Information Theory (ISIT)}, pp.~2027--2031, IEEE, 2018.

\bibitem{dutta2016short}
S.~Dutta, V.~Cadambe, and P.~Grover, ``Short-dot: Computing large linear transforms distributedly using coded short dot products,'' {\em Advances In Neural Information Processing Systems}, vol.~29, 2016.

\bibitem{ramamoorthy2019universally}
A.~Ramamoorthy, L.~Tang, and P.~O. Vontobel, ``Universally decodable matrices for distributed matrix-vector multiplication,'' in {\em 2019 IEEE International Symposium on Information Theory (ISIT)}, pp.~1777--1781, 2019.

\bibitem{das2019distributed}
A.~B. Das and A.~Ramamoorthy, ``Distributed matrix-vector multiplication: A convolutional coding approach,'' in {\em 2019 IEEE International Symposium on Information Theory (ISIT)}, pp.~3022--3026, IEEE, 2019.

\bibitem{haddadpour2018codes}
F.~Haddadpour and V.~R. Cadambe, ``Codes for distributed finite alphabet matrix-vector multiplication,'' in {\em 2018 IEEE International Symposium on Information Theory (ISIT)}, pp.~1625--1629, IEEE, 2018.

\bibitem{wang2018coded}
S.~Wang, J.~Liu, and N.~Shroff, ``Coded sparse matrix multiplication,'' in {\em International Conference on Machine Learning}, pp.~5152--5160, PMLR, 2018.

\bibitem{ramamoorthy2020straggler}
A.~Ramamoorthy, A.~B. Das, and L.~Tang, ``Straggler-resistant distributed matrix computation via coding theory: Removing a bottleneck in large-scale data processing,'' {\em IEEE Signal Processing Magazine}, vol.~37, no.~3, pp.~136--145, 2020.

\bibitem{zinkevich2010parallelized}
M.~Zinkevich, M.~Weimer, L.~Li, and A.~Smola, ``Parallelized stochastic gradient descent,'' in {\em Advances in Neural Information Processing Systems}, vol.~23, 2010.

\bibitem{chilimbi2014project}
T.~Chilimbi, Y.~Suzue, J.~Apacible, and K.~Kalyanaraman, ``Project {A}dam: Building an efficient and scalable deep learning training system,'' in {\em 11th USENIX Symposium on Operating Systems Design and Implementation (OSDI 14)}, pp.~571--582, 2014.

\bibitem{kai2}
K.~Wan, H.~Sun, M.~Ji, and G.~Caire, ``On the tradeoff between computation and communication costs for distributed linearly separable computation,'' {\em IEEE Transactions on Communications}, vol.~69, no.~11, pp.~7390--7405, 2021.

\bibitem{kai3}
K.~Wan, H.~Sun, M.~Ji, and G.~Caire, ``On secure distributed linearly separable computation,'' {\em IEEE Journal on Selected Areas in Communications}, vol.~40, no.~3, pp.~912--926, 2022.

\bibitem{khalesi4}
A.~Khalesi and P.~Elia, ``Multi-user linearly-separable distributed computing,'' {\em IEEE Transactions on Information Theory}, vol.~69, no.~10, pp.~6314--6339, 2023.

\bibitem{makkonen2022analog}
O.~Makkonen and C.~Hollanti, ``Analog secure distributed matrix multiplication over complex numbers,'' in {\em 2022 IEEE International Symposium on Information Theory (ISIT)}, pp.~1211--1216, IEEE, 2022.

\bibitem{khalesi3}
A.~Khalesi and P.~Elia, ``Multi-user linearly separable computation: A coding theoretic approach,'' in {\em 2022 IEEE Information Theory Workshop (ITW)}, pp.~428--433, 2022.

\bibitem{khalesi5}
A.~Khalesi, S.~Daei, M.~Kountouris, and P.~Elia, ``Multi-user distributed computing via compressed sensing,'' in {\em 2023 IEEE Information Theory Workshop (ITW)}, pp.~509--514, 2023.

\bibitem{gribonval2010dictionary}
R.~Gribonval and K.~Schnass, ``Dictionary identification—sparse matrix-factorization via l-1 -minimization,'' {\em IEEE Transactions on Information Theory}, vol.~56, no.~7, pp.~3523--3539, 2010.

\bibitem{zheng2021identifiability}
L.~Zheng, E.~Riccietti, and R.~Gribonval, ``Identifiability in two-layer sparse matrix factorization,'' {\em arXiv preprint arXiv:2110.01235}, 2021.

\bibitem{zheng2023efficient}
L.~Zheng, E.~Riccietti, and R.~Gribonval, ``Efficient identification of butterfly sparse matrix factorizations,'' {\em SIAM Journal on Mathematics of Data Science}, vol.~5, no.~1, pp.~22--49, 2023.

\bibitem{le2023spurious}
Q.-T. Le, E.~Riccietti, and R.~Gribonval, ``Spurious valleys, np-hardness, and tractability of sparse matrix factorization with fixed support,'' {\em SIAM Journal on Matrix Analysis and Applications}, vol.~44, no.~2, pp.~503--529, 2023.

\bibitem{eckart1936approximation}
C.~Eckart and G.~Young, ``The approximation of one matrix by another of lower rank,'' {\em Psychometrika}, vol.~1, no.~3, pp.~211--218, 1936.

\bibitem{van1996matrix}
C.~F. Van~Loan and G.~Golub, ``Matrix computations (\text{Johns Hopkins} studies in mathematical sciences),'' {\em Matrix Computations}, vol.~5, 1996.

\bibitem{dao2019learning}
T.~Dao, A.~Gu, M.~Eichhorn, A.~Rudra, and C.~R{\'e}, ``Learning fast algorithms for linear transforms using butterfly factorizations,'' in {\em International conference on machine learning}, pp.~1517--1527, PMLR, 2019.

\bibitem{hackbusch1999sparse}
W.~Hackbusch, ``A sparse matrix arithmetic based on-matrices. \text{Part I}: Introduction to-matrices,'' {\em Computing}, vol.~62, no.~2, pp.~89--108, 1999.

\bibitem{johnson1990matrix}
C.~R. Johnson, ``Matrix completion problems: a survey,'' in {\em Matrix Theory and Applications}, vol.~40, pp.~171--198, 1990.

\bibitem{ardila2010tilings}
F.~Ardila and R.~P. Stanley, ``Tilings,'' {\em The Mathematical Intelligencer}, vol.~32, no.~4, pp.~32--43, 2010.

\bibitem{tao2012topics}
T.~Tao, {\em Topics in random matrix theory}, vol.~132.
\newblock American Mathematical Soc., 2012.

\bibitem{adams2022tiling}
C.~Adams, {\em The tiling book: An introduction to the mathematical theory of tilings}, vol.~142.
\newblock American Mathematical Society, 2022.

\bibitem{kishore2017literature}
N.~Kishore~Kumar and J.~Schneider, ``Literature survey on low rank approximation of matrices,'' {\em Linear and Multilinear Algebra}, vol.~65, no.~11, pp.~2212--2244, 2017.

\bibitem{marchenko1967distribution}
V.~A. Marchenko and L.~A. Pastur, ``Distribution of eigenvalues for some sets of random matrices,'' {\em Matematicheskii Sbornik}, vol.~114, no.~4, pp.~507--536, 1967.

\end{thebibliography}
\end{document}